\documentclass[aip,pof,reprint,onecolumn]{revtex4-1} 
\usepackage{graphicx}
\usepackage{amsmath}
\usepackage{multirow}

\usepackage[usenames,dvipsnames,svgnames,table]{xcolor} 
\usepackage[colorlinks=true,
            linkcolor=blue,
            urlcolor=blue,
            citecolor=BrickRed]{hyperref}

\usepackage{lineno}
\usepackage{xcolor}
\usepackage{soul}
\newcommand{\hltOn}{\color{black}} 
\newcommand{\hltOff}{\color{Black}}

\setstcolor{red}

\begin{document}

\title{Frozen propagation of Reynolds force vector from high-fidelity data into Reynolds-averaged simulations of secondary flows}

\author{Ali Amarloo}
\author{Pourya Forooghi}%
\author{Mahdi Abkar}
\email{abkar@mpe.au.dk}
\affiliation{%
Department of Mechanical and Production Engineering, Aarhus University, 8200 Aarhus N, Denmark
}%

\date{\today}

\begin{abstract}
Successful propagation of information from high-fidelity sources (i.e., direct numerical simulations and large-eddy simulations) into Reynolds-averaged Navier-Stokes (RANS) equations plays an important role in the emerging field of data-driven RANS modeling. Small errors carried in high-fidelity data can propagate amplified errors into the mean flow field, and higher Reynolds numbers worsen the error propagation. \hltOn In this study, we compare a series of propagation methods for two cases of Prandtl’s secondary flows of the second kind: square-duct flow at a low Reynolds number and roughness-induced secondary flow at a very high Reynolds number. \hltOff We show that frozen treatments result in less error propagation than the implicit treatment of Reynolds stress tensor (RST), and for cases with very high Reynolds numbers, explicit and implicit treatments are not recommended.  Inspired by the obtained results, \hltOn  we introduce the frozen treatment to the propagation of Reynolds force vector (RFV), which leads to less error propagation. Specifically, for both cases at low and high Reynolds numbers, propagation of RFV results in one order of magnitude lower error compared to RST propagation. \hltOff In the frozen treatment method, three different eddy-viscosity models are used to evaluate the effect of turbulent diffusion on error propagation. \hltOn We show that, regardless of the baseline model, the frozen treatment of RFV results in less error propagation. We combined one extra correction term for turbulent kinetic energy with the frozen treatment of RFV, which makes our propagation technique capable of reproducing both velocity and turbulent kinetic energy fields similar to high-fidelity data. \hltOff
\end{abstract}

\maketitle


\section{\label{sec:intro}Introduction}

In the simulation of turbulent flows, the physics of turbulence is commonly treated in three different levels. In direct numerical simulations (DNS), the whole size-spectrum of eddies is simulated. In large-eddy simulations (LES), the small eddies are filtered, and their effect on the large eddies is modelled. In Reynolds-averaged Navier-Stokes (RANS) simulations, the whole physics of turbulence is modelled by a Reynolds stress tensor (RST) model. Because of the high computational costs associated with high-fidelity methods (i.e., DNS and LES), the low-fidelity method (i.e., RANS) is mainly used to solve problems in industrial applications especially with complex geometries and/or at high Reynolds numbers \cite{slotnick2014cfd}.

Despite the popularity of the RANS simulations, it is well established that the RANS simulations may result in inaccurate solutions in several cases like flows with variation in the Reynolds normal stresses (e.g., square duct flow), or with strong adverse pressure gradients \cite{bush2019}. Duraisamy \textit{et al.} \cite{duraisamy2019turbulence} classified the uncertainties of the RANS method into 4 layers: (1) uncertainties because of ensemble-averaging of Navier-Stokes equations, (2) uncertainties caused by the representative model of the Reynolds stress, (3) uncertainties because of the functional form within the model, and (4) uncertainties induced by coefficients used in the functions. In this regard, the uncertainty quantification of RANS models is comprehensively reviewed in Ref.~\onlinecite{xiao2019quantification}. Considering the second, the third, and the fourth layers of uncertainties, a proper Reynolds stress model is essential for obtaining trustworthy results from the RANS simulations. 
To this end, one of the first steps for reducing the error associated with the Reynolds stress model is to benefit from the available high-fidelity data (e.g., from DNS and LES) of canonical cases (e.g., channel flow, periodic hills, square duct, etc.).
The first question in using the high-fidelity data is: Is it possible to reconstruct the mean velocity field by the propagation of information from high-fidelity data into the RANS equations? (the term \textit{propagation} is readapted from Ref.~\onlinecite{wu2019reynolds}). This question is the main concern of this study, and therefore, a brief review of studies focused on this question is presented here.

In one of the first studies about the propagation of high-fidelity data, Thompson \textit{et al.} \cite{thompson2016methodology} solved the RANS equations by using the Reynolds stress statistics obtained from the DNS. In other words, they propagated the DNS Reynolds stress into the RANS equations to calculate the mean velocity field. They applied this method to plane channel flow with a range of Reynolds numbers, and they reported that the calculated mean velocity from RANS has an unsatisfactory agreement with the calculated mean velocity from DNS. The discrepancy between these two mean velocities is referred to as error propagation. They used this error propagation technique as a quantified methodology for evaluating the statistical error associated with the accuracy of the DNS data. In a follow-up study, \citet{andrade2018analysis} used the same technique on both channel flow and pipe flow. They also reported the error propagation and the fact that a longer averaging time for the calculation of DNS statistics will result in lower error propagation.

In a similar study, Poroseva \textit{et al.} \cite{poroseva2016accuracy} calculated all the terms required in the RANS equations from the DNS statistics except the molecular diffusion. Therefore, they solved a diffusion equation for the mean velocity field with a known source term (including all the calculated statistics from DNS). As expected, the resulting velocity is not completely compatible with the DNS velocity field. They tried different numerical methods for solving the RANS equation with no success; therefore, the authors concluded that the uncertainty of DNS-informed RANS simulation is rooted in the inaccuracies of the DNS data.

\citet{wu2019reynolds} discussed the inconsistency of resulted mean velocity from RANS, after propagation of DNS data, with the mean velocity of DNS as ``ill-conditioning of RANS equations". They studied the error propagation of the RST calculated from DNS into the RANS equations more systematically by using the local ill-conditioning number. They concluded that ``RANS equations with explicit data-driven Reynolds stress closure can be ill-conditioned" \cite{wu2019reynolds}. In an approach that they termed the explicit method, they injected the DNS Reynolds stress explicitly into the RANS equations; as opposed to the implicit method in which they treated the linear part of the Reynolds stress as a turbulent diffusion term in the RANS equations. Besides using the second-order statistics of DNS (i.e., the RST), they used the first-order statistics (i.e., the mean velocity field) for the calculation of turbulent diffusivity required by the implicit treatment. They examined the capability of local conditioning number on turbulent channel flow and two more complex flows (the periodic hills and the square duct) in evaluating the error propagation. They showed that implicit treatment of Reynolds stress decreased both the local conditioning number and the error propagation. Guo \textit{et al.} \cite{guo2021computing} also studied the implicit treatment of the RST in the case of periodic hills. They tried two methods for the calculation of the turbulent viscosity for the implicit treatment of the linear part of RST. In one method, they used the mean velocity calculated from DNS (similar to Ref.~\onlinecite{wu2019reynolds}), and in the second method, they used an auxiliary RANS model (i.e., Spalart-Allmaras \cite{spalart1992one}). They reported that the second method resulted in lower error propagation. 

The above-mentioned ill-conditioning issue needs to be addressed, in particular, in the context of data-driven RANS models, which have received increasing interest in recent years. It was noted by Duraisamy \textit{et al.} \cite{duraisamy2019turbulence} that the difference between the learning environment (information from high-fidelity data) and the propagation environment (the data-injected RANS simulation) can pose challenges because even if a well-trained data-driven technique can predict terms very similar to the high-fidelity data, obtaining the correct velocity field from that perfect prediction is still troublesome. In a recent comprehensive review of data-driven turbulence modeling techniques, \citet{sandberg2022machine} highlighted that successful training of RST based on high-fidelity data cannot guarantee a successful prediction of mean flow fields in RANS simulations. Therefore, the propagation method can play an important role in the development of data-driven models. For example, \citet{li2022data} developed a data-driven model to predict RST successfully but the resulting mean velocity domain is at best similar to the propagation results. They reported that one of the main reasons for this problem is using the explicit treatment of RST. In another attempt for a data-driven turbulence modeling, \citet{jiang2021interpretable} pointed out that the linear part of the predicted RST should be treated implicitly to avoid the ill-conditioning problem. \citet{cruz2019use} introduced a new propagation method for their data-driven model. In this method, they only used the first-order DNS statistics (i.e., mean velocity field), and they replaced that in RANS equations and calculated the term corresponding to the Reynolds stress (i.e., the divergence of RST, called Reynolds force vector (RFV)). They applied the propagation of both the RST and the RFV from the DNS of the square duct case explicitly into the RANS equations. They also trained neural networks to predict both the RSTs and the RFVs. The results showed that using the RFV decreases the error propagation in both cases of the DNS data and the predicted values. 

In a comprehensive study, \citet{brener2021conditioning} evaluated both the implicit and the explicit treatment methods for both RST and RFV in the turbulent channel flow, the square duct, and the periodic hills. They showed that the propagation of the RFV (both implicitly and explicitly) decreased the error propagation in their simulations. Also, they reported that implicit treatment did not induce a significant change in error propagation, but it helped with the robustness and convergence rate of simulations. They also reported that it was challenging to apply the explicit treatment in complex geometry (i.e., periodic hill); therefore, the implicit treatment was recommended.

In another study of data-driven RANS modeling, \citet{weatheritt2017development} introduced a novel hybrid framework for the implicit treatment of the RST. They calculated the mean velocity field, turbulent kinetic energy (TKE), and the RST from high-fidelity data. They used a linear eddy-viscosity model ($k-\omega SST$ \cite{menter1994two}) and solved the $\omega$ equation over frozen fields of the velocity and the TKE. Therefore, they can calculate the discrepancy of the high-fidelity RST with the baseline modeled RST. In the propagation process, they propagated the discrepancy of the RST into the RANS equation solved by the baseline model. They also reported the existence of error propagation but the performance of the framework was reasonably good enough to be used for training a data-driven model. In an extension to this propagation method, Schmelzer \textit{et al.}\cite{schmelzer2020discovery} introduced an extra corrective term to the TKE equation in the baseline model with which it is possible to also propagate the TKE (calculated from high-fidelity data) into the RANS simulation. They named this propagation framework as k-corrective frozen (KCF) RANS, and they reported a reasonably low error propagation; therefore, they used this framework for the discovery of algebraic Reynolds stress models. In a recent study by \citet{mandler2022frozen}, the k-corrective frozen treatment of RST was investigated in detail for a data-driven RST model. They showed that including the extra term for the correction of TKE can be important for cases with complex geometries.

As a problem in which commonly used linear eddy-viscosity RANS models inherently fall short, Prandtl’s secondary flow of the second kind (e.g. in square ducts) has been employed as a standard test case by many studies cited above \cite{brener2021conditioning,cruz2019use,wu2019reynolds}. However, while it has been shown that error propagation can be intensified by increasing the Reynolds number \cite{thompson2016methodology, poroseva2016accuracy, wu2019reynolds}, all these studies focused on flows with low/moderate Reynolds number. Motivated by this fact, in the present work, we introduce a new test case, that is a large-scale secondary flow induced by spanwise heterogeneous roughness at nominally infinite Reynolds number. It has been shown that longitudinal roughness patches in the half channel flow will create a large-scale secondary mean flow \cite{willingham2014turbulent}, and it is categorized as Prandtl's secondary flow of the second kind \cite{anderson2015numerical}. For this study, the LES technique is used to produce the high-fidelity data required for the propagation into the RANS equations\cite{forooghi2020roughness, amarloo2022secondary}.

\hltOn The aim of this work is to compare the performance of different propagation approaches, particularly in the simulation of Prandtl’s secondary flows of the second kind.
Inspired by the previously proposed propagation methods in the literature, we introduce a novel framework referred to as k-corrective frozen RFV (KCF-RFV), which uses the frozen treatment of RFV (instead of RST) with including the TKE correction. We systematically evaluate the performance of this approach, along with the ones previously proposed in the literature, on the case of secondary flows in a square duct at a low Reynolds number, and then, we repeat the same procedure for the prediction of roughness-induced secondary motions at a very high Reynolds number. \hltOff Therefore, the rest of the manuscript is organized as follows: In Section \ref{sec:method}, four different treatment techniques for the propagation of both the RST and RFV are described, and the high-fidelity data sources are presented. In Section \ref{sec:Results}, the results and the error propagation of all the propagation techniques are shown and compared. The paper is concluded with the final remarks presented in Section \ref{sec:Conclusion}.

\section{\label{sec:method}Methodology}
In this section, four different methods for the propagation of both RST and RFV values into the RANS equations will be presented. By using the Reynolds decomposition of velocity and pressure, the RANS equations for an incompressible steady flow are written \hltOn as \cite{pope2000turbulent} \hltOff
\begin{equation}
\label{eq:ransDC}
u_i = \overline{u}_i + u'_i, \hspace{0.2cm}  p = \overline{p} + p',   
\end{equation}
\begin{equation}
\label{eq:ransEq}
\partial_i \overline{u}_i = 0, \hspace{0.2cm} \partial_j (\overline{u}_i \overline{u}_j) = -\frac{1}{\rho}\partial_i \overline{P} + \partial_j \left(\nu \partial_j \overline{u}_i - a_{ij}\right),   
\end{equation}
where i = 1, 2, 3 are streamwise ($x$), spanwise ($y$), and wall-normal ($z$) directions, respectively, $u_i$ is the velocity, $p$ is the pressure, $\overline{u}_i$ and $\overline{p}$ are the mean velocity and the mean pressure, respectively (generally, the overbar indicates time-averaging), $u'_i$ and $p'$ are the fluctuations of the velocity and pressure,  respectively, $\rho$ is the fluid density, $\nu$ is the kinematic viscosity, $\overline{P} = \overline{p} + 1/3\rho\overline{u'_iu'_i}$ is the modified mean pressure, $ a_{ij} =\overline{u'_iu'_j} - 1/3\overline{u'_ku'_k} \delta_{ij}$ is the deviatoric anisotropic part of Reynolds stress, and $\delta_{ij}$ is the Kronecker delta.

By using the standard $k-\epsilon$\cite{launder1975progress} as a baseline RANS model, $a_{ij}$ can be modelled as $a^{BL}_{ij} = -\nu_t (\partial_i \overline{u}_j + \partial_j \overline{u}_i)$, where $\nu_t = C_\mu k^2/\epsilon$ is the turbulent viscosity, $k$ is the TKE, and $\epsilon$ is the dissipation of TKE. These two terms are modelled by two partial differential equations \hltOn as \cite{launder1975progress} \hltOff  
\begin{equation}
\label{eq:kEq}
    \partial_j (k \overline{u}_j) =\partial_j \left((\nu+\nu_t/\sigma_k) \partial_j k \right) + G_k - \epsilon ,
\end{equation}
\begin{equation}
\label{eq:epsEq}
    \partial_j (\epsilon \overline{u}_j) = \partial_j \left((\nu+\nu_t/\sigma_{\epsilon}) \partial_j \epsilon \right) +  C_{1\epsilon} \epsilon G_k /k - C_{2\epsilon} \epsilon^2/k, 
\end{equation}
where $G_k = - a_{ij} \partial_i \overline{u}_j$ is the production source of TKE by the RST (again, $a_{ij}$ is modelled by $a^{BL}_{ij}$), and constants of equations are considered as $C_\mu=0.09$, $\sigma_k=1.0$, $\sigma_{\epsilon}=1.3$, $C_{1\epsilon}=1.44$, $C_{2\epsilon}=1.92$ \cite{pope2000turbulent}.

Instead of using a baseline model and for the purpose of propagation, the values of the deviatoric anisotropic part of RST ($a_{ij}$) or the force vector corresponding to that ($t_i = \partial_j a_{ij}$) can be directly calculated from the high-fidelity data. The parameters that are calculated directly from the high-fidelity data or based on the mean velocity of the high-fidelity data are indicated by a superscript $^*$.

\begin{figure*}
\includegraphics[width=\textwidth]{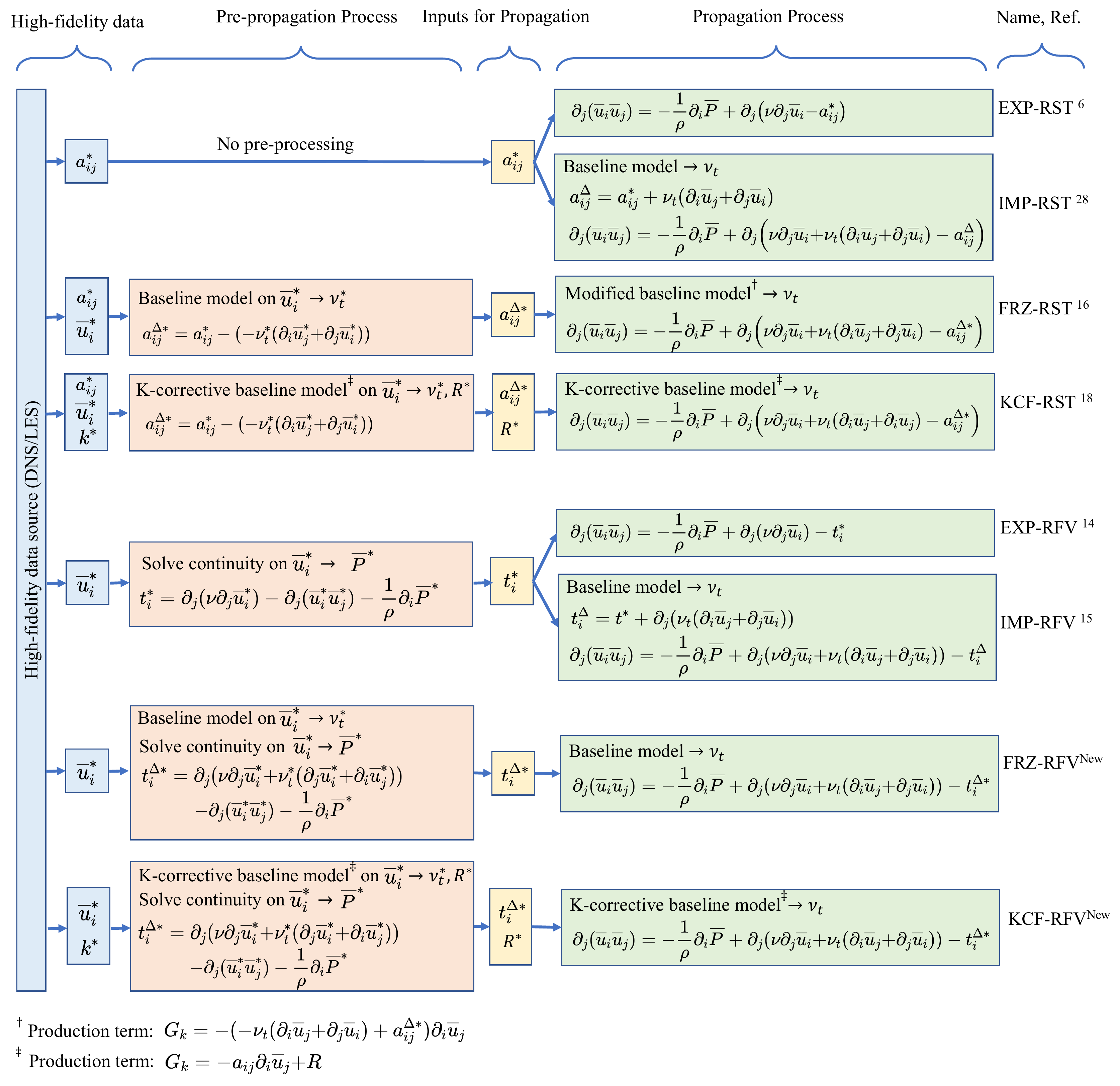}%
\caption{\label{fig:methods} The list of 4 different treatment methods of the propagation for both the RST and the RFV from the high fidelity data into the RANS equations.}%
\end{figure*}

The information from the high-fidelity data can be propagated into the RANS equations with 4 different treatment techniques which will be explained in this section. All of these methods are summarised and compared in Fig.~\ref{fig:methods}. The implementation of all these methods is applied in an open-source finite-volume code OpenFOAM\cite{weller1998tensorial} (v2112). For all of the methods that need a baseline model, as a default, we use standard $k-\epsilon$ \cite{launder1975progress}. For the analysis of sensitivity on the baseline model, we also use realizable $k-\epsilon$ \cite{shih1993realizable} and $k-\omega-SST$ \cite{menter1994two}, which all of them are already implemented inside the OpenFOAM.

\subsection{\label{sec:RST} Propagation of the Reynolds stress tensor (RST)}
The values of high-fidelity RST can be calculated from the second-order statistics of the high-fidelity source. These values can be propagated into the RANS equations with different treatment techniques, which will be explained in this subsection.
\subsubsection{\label{sec:EXPRST}Explicit treatment of the RST}
The values of the deviatoric anisotropic part of the RST can be calculated from the second-order statistics of high-fidelity data \hltOn as \cite{pope2000turbulent} \hltOff
\begin{equation}
    a_{ij}^* =\overline{u'_iu'_j}^* - 1/3\overline{u'_iu'_i}^* \delta_{ij}.
\end{equation}
In the explicit treatment, these values will be directly replaced in the RANS equations \hltOn as \cite{thompson2016methodology} \hltOff
\begin{equation}
\label{eq:EXPRST}
    \partial_j (\overline{u}_i \overline{u}_j) = -\frac{1}{\rho}\partial_i \overline{P} + \partial_j \left(\nu \partial_j \overline{u}_i - a^*_{ij}\right).
\end{equation}
This method was introduced by Thompson \textit{et al.} \cite{thompson2016methodology} and will be referred to as \textit{EXP-RST}.

\subsubsection{\label{sec:IMPRST}Implicit treatment of the RST}
In this approach, the turbulent diffusion term will be considered inside the RANS equations. Therefore, instead of the direct replacement of the $a^*_{ij}$ inside the RANS equations, the discrepancy of the $a^*_{ij}$ with a baseline-modelled value will be calculated at each iteration of the simulation \hltOn as \cite{wu2018physics} \hltOff
\begin{equation}
\label{eq:RSTdivison}
    a_{ij}^{\Delta} = a_{ij}^* - (-\nu_t (\partial_i \overline{u}_j + \partial_j \overline{u}_i)).
\end{equation}

The discrepancy term ($a_{ij}^{\Delta}$) will be placed explicitly into the RANS equations, and the linear part of the $a_{ij}^{*}$ will be treated as a turbulent diffusion term implicitly inside the RANS equations  \hltOn as \cite{wu2018physics} \hltOff
\begin{equation}
\label{eq:IMPRST}
    \partial_j (\overline{u}_i \overline{u}_j) = -\frac{1}{\rho}\partial_i \overline{P} + \partial_j \left(\nu \partial_j \overline{u}_i + \nu_t (\partial_i \overline{u}_j + \partial_j \overline{u}_i) - a^{\Delta}_{ij}\right).
\end{equation}
In this technique, which was proposed by Wu \textit{et al.}\cite{wu2018physics}, a proper value for the turbulent viscosity ($\nu_t$) must be considered. In our study, we use the standard $k-\epsilon$ model (Eqs.~(\ref{eq:kEq},\ref{eq:epsEq})) at each iteration only for predicting the values of $\nu_t$ dynamically. In other words, the new values of velocity in each iteration will lead to new values for $\nu_t$ and $a_{ij}^{\Delta}$. This method will be referred to as \textit{IMP-RST}. 

\subsubsection{\label{sec:FRZRST}Frozen treatment of the RST}
A novel implicit treatment of the RST was first proposed by \citet{weatheritt2017development}, and it is called \textit{frozen} treatment because the mean velocity from the high-fidelity data ($\overline{u}^*_i$) is kept frozen for the calculation of the $a_{ij}^{\Delta*}$. Here, the $a_{ij}^{\Delta*}$ will be calculated in a pre-processing step in which $\overline{u}^*_i$ is known and frozen, and only the equations of the baseline RANS model (Eqs.~(\ref{eq:kEq},\ref{eq:epsEq})) are going to be solved to find the value of $\nu^*_t$ (associated to $\overline{u}^*_i$) \hltOn as \cite{weatheritt2017development} \hltOff
\begin{equation}
    a_{ij}^{\Delta*} = a_{ij}^* - (-\nu^*_t (\partial_i \overline{u}^*_j + \partial_j \overline{u}^*_i)).
\end{equation}
After finding the $a_{ij}^{\Delta*}$, it will be inserted into the RANS equations \hltOn as \cite{weatheritt2017development} \hltOff
\begin{equation}
\label{eq:FRZRST}
    \partial_j (\overline{u}_i \overline{u}_j) = -\frac{1}{\rho}\partial_i \overline{P} + \partial_j \left(\nu \partial_j \overline{u}_i + \nu_t (\partial_i \overline{u}_j + \partial_j \overline{u}_i) - a^{\Delta*}_{ij}\right).
\end{equation}
It should be noted that in this method, the values of $a_{ij}^{\Delta*}$ are constant through the propagation process, but the values of $\nu_t$ are calculated by the baseline model (Eqs.~(\ref{eq:kEq},\ref{eq:epsEq})) at each iteration. Inside the baseline model, the production term is modified to be compatible with the pre-processing step \hltOn as \cite{schmelzer2020discovery} \hltOff
\begin{equation}
    G_k = - (-\nu_t (\partial_i \overline{u}_j + \partial_j \overline{u}_i) + a^{\Delta*}_{ij}) \partial_i \overline{u}_j.
\end{equation}
This method will be referred to as \textit{FRZ-RST}.

\subsubsection{\label{sec:KCFRST}K-corrective frozen treatment of the RST}
Schmelzer \textit{et al.}\cite{schmelzer2020discovery} proposed a modified version of the frozen treatment of the RST. The whole process of this technique is similar to the FRZ-RST method but the TKE equation. In this method, they calculated the values of the TKE from the high-fidelity data ($k^* = 0.5\overline{u'_iu'_i}^*$) instead of modeling the TKE by Eq.~(\ref{eq:kEq}). In the pre-processing step, it is only needed to solve the $\epsilon$ equation (Eq.~(\ref{eq:epsEq})) while $\overline{u}^*_i$ and $k^*$ are frozen to find the values of $\nu^*_t$. Inside the baseline model, they added one corrective term, named $R$, to correct the production term \hltOn as \cite{schmelzer2020discovery} \hltOff
\begin{equation}
\label{eq:RinG}
    G_k = - a_{ij} \partial_i \overline{u}_j + R.
\end{equation}
During the pre-processing step, the values of $R$ can be calculated because the values of the TKE are known and frozen, and the values of $\epsilon$ are modelled by considering the new production term (Eq.~(\ref{eq:RinG})) inside the dissipation equation (Eq.~(\ref{eq:epsEq})) \cite{schmelzer2020discovery},
\begin{equation}
\label{eq:rEq}
    R^* = \partial_j (k^* \overline{u}^*_j) - \partial_j \left((\nu+\nu^*_t/\sigma_k) \partial_j k^* \right) + a^*_{ij} \partial_i \overline{u}^*_j + \epsilon.
\end{equation}
After finding the $a_{ij}^{\Delta*}$ and $R^*$, they will be propagated to the RANS equations. The propagation of $a_{ij}^{\Delta*}$ will be done by the Eq.~(\ref{eq:FRZRST}). The propagation of the $R^*$ to the k-corrective baseline model will be included in the production term (Eq.~(\ref{eq:RinG})). Including the extra correction term ($R^*$) makes the RANS simulation possible to model the TKE as close as possible to the values from the high-fidelity data ($k^*$). This method will be referred to as \textit{KCF-RST}.

\subsection{\label{sec:RFV} Propagation of the Reynolds force vector (RFV)}
Instead of the RST, here the RFV will be calculated from the first-order statistics (i.e., mean velocity field) of the high-fidelity data and propagated into the RANS equations. Similar to the RST, the propagation of the RFV can also be done in four different treatment techniques.

\subsubsection{\label{sec:EXPRFV}Explicit treatment of the RFV}
This method was proposed by \citet{cruz2019use}, and the values of the RFV ($t_i = \partial_j a_{ij}$) will be calculated in the pre-processing step \hltOn as \cite{cruz2019use} \hltOff
\begin{equation}
\label{eq:preEXPRFV}
    t^*_i = - \partial_j (\overline{u}^*_i \overline{u}^*_j) + \partial_j (\nu \partial_j \overline{u}^*_i) -\frac{1}{\rho}\partial_i \overline{P}^*,
\end{equation}
while the values of the mean modified pressure ($\overline{P}^*$) can be calculated by solving the continuity equation over the known velocity field ($\overline{u}^*_i$). In the propagation process, the values of $t^*_i$ can be propagated by directly replacing them inside the RANS equations \hltOn as \cite{cruz2019use} \hltOff
\begin{equation}
\label{eq:EXPRFV}
    \partial_j (\overline{u}_i \overline{u}_j) = -\frac{1}{\rho}\partial_i \overline{P} + \partial_j (\nu \partial_j \overline{u}_i) - t^*_i.
\end{equation}
This method will be referred to as \textit{EXP-RFV}.

\subsubsection{\label{sec:IMPRFV}Implicit treatment of the RFV}
In this method, which was introduced by \citet{brener2021conditioning}, the values of $t^*_i$ will be calculated the same as EXP-RFV by Eq.~(\ref{eq:preEXPRFV}). These values are not going to be explicitly replaced in the RANS equations. A linear eddy-viscosity model will be used to provide a turbulent diffusion inside the RANS equations. The discrepancy of $t^*_i$ with the divergence of modelled RST will be added explicitly to the RANS equations \hltOn as \cite{brener2021conditioning} \hltOff
\begin{equation}
    t_{i}^{\Delta} = t_{i}^* - \partial_j(-\nu_t (\partial_i \overline{u}_j + \partial_j \overline{u}_i)),
\end{equation}
\begin{equation}
\label{eq:IMPRFV}
    \partial_j (\overline{u}_i \overline{u}_j) = -\frac{1}{\rho}\partial_i \overline{P} + \partial_j \left(\nu \partial_j \overline{u}_i + \nu_t (\partial_i \overline{u}_j + \partial_j \overline{u}_i)\right) - t^{\Delta}_i,
\end{equation}
where the value of the $t_{i}^{\Delta}$ will be calculated at each iteration of the solution. Similar to the IMP-RST, a proper value of $\nu_t$ is needed which is dynamically calculated by using the standard $k-\epsilon$ (Eqs.~(\ref{eq:kEq},\ref{eq:epsEq})) as a baseline model. This method will be referred to as \textit{IMP-RFV}.

\subsubsection{\label{sec:FRZRFV}Frozen treatment of the RFV}
Since \citet{brener2021conditioning} have shown that the propagation of the RFV instead of the RST (in both implicit and explicit treatments) will result in lower error propagation, we apply the frozen approach to the propagation of the RFV. In this method, the mean velocity field from the high-fidelity data ($u^*_i$) will be kept frozen in the pre-processing step, and the baseline model (Eqs.~(\ref{eq:kEq},\ref{eq:epsEq}) will be solved to obtain the $\nu_t^*$ associated with the $u^*_i$. Therefore, the values of the $t_{i}^{\Delta*}$ can be calculated,
\begin{equation}
\label{eq:preFRZRFV}
    t^{\Delta*}_i = - \partial_j (\overline{u}^*_i \overline{u}^*_j) + \partial_j (\nu \partial_j \overline{u}^*_i + \nu_t^* (\partial_j \overline{u}^*_i + \partial_i \overline{u}^*_j))-\frac{1}{\rho}\partial_i \overline{P}^*,
\end{equation}
while the values of the mean modified pressure ($\overline{P}^*$) can be calculated by solving the continuity equation over the known velocity field ($\overline{u}^*_i$). In the propagation process, the constant values of the $t_{i}^{\Delta*}$ will be injected into the RANS equations while the baseline model will dynamically predict $\nu_t$ at each iteration to solve the turbulent diffusion implicitly,
\begin{equation}
\label{eq:FRZRFV}
    \partial_j (\overline{u}_i \overline{u}_j) = -\frac{1}{\rho}\partial_i \overline{P} + \partial_j \left(\nu \partial_j \overline{u}_i + \nu_t (\partial_i \overline{u}_j + \partial_j \overline{u}_i)\right) - t^{\Delta*}_i.
\end{equation}
It should be noted that since in the pre-processing step only the first-order statistics of high-fidelity data are used, there is no information about the RST; therefore, the production term inside the TKE equation ($G_k$) will be calculated by the modelled RST ($a^{BL}_{ij}$). This method will be referred to as \textit{FRZ-RFV}.

\subsubsection{\label{sec:KCFRFV}K-corrective frozen treatment of the RFV}
Since the KCF-RST\cite{schmelzer2020discovery} method has the ability to model the TKE as close as possible to the calculated values from high-fidelity data ($k^*$), here we apply the k-corrective baseline model to the frozen treatment of the RFV. Therefore, the values of the $t_{i}^{\Delta*}$ will be calculated by Eq.~(\ref{eq:preFRZRFV}). Inside the baseline model, the production term of the TKE equation will be modified the same as Eq.~(\ref{eq:RinG}). Therefore, the value of $\nu_t^*$ and $R^*$ will be calculated based on the modelled $\epsilon, a_{ij}$ and the TKE from high-fidelity data ($k^*$),
\begin{equation}
\label{eq:r2Eq}
    R^* = \partial_j (k^* \overline{u}^*_j) - \partial_j \left((\nu+\nu^*_t/\sigma_k) \partial_j k^* \right) + a^{BL}_{ij} \partial_i \overline{u}^*_j + \epsilon.
\end{equation}

During the propagation process, the values of the $t_{i}^{\Delta*}$ will be propagated into RANS equations by using Eq.~(\ref{eq:FRZRFV}), and the values of the $R^*$ will be propagated into the baseline model (Eqs.~(\ref{eq:kEq},\ref{eq:epsEq})) by a modified production term (Eq.~(\ref{eq:RinG})). Therefore, this new method combines the ability to predict the values of TKE (similar to the KCF method proposed by Schmelzer \textit{et al.}\cite{schmelzer2020discovery}) with a lower error propagation associated with the implicit treatment of the RFV (proposed by \citet{brener2021conditioning}). This method will be referred to as \textit{KCF-RFV}.

\subsection{\label{sec:HFData} High-fidelity data}
In this study, we considered two cases of Prandtl’s secondary flows of the second kind \cite{nikitin2021prandtl}. One case is the secondary flows inside the square duct and the second one is the secondary flows induced by roughness heterogeneity in a half channel flow. It is well established in the literature\cite{nikitin2021prandtl} that linear eddy-viscosity models in the RANS simulations cannot capture these secondary flows that are induced by the anisotropy of the RST; therefore, a successful propagation of the high-fidelity data into the RANS simulations can provide a more promising precursor for a data-driven RANS model.

\begin{figure*}
\includegraphics[width=0.8\textwidth]{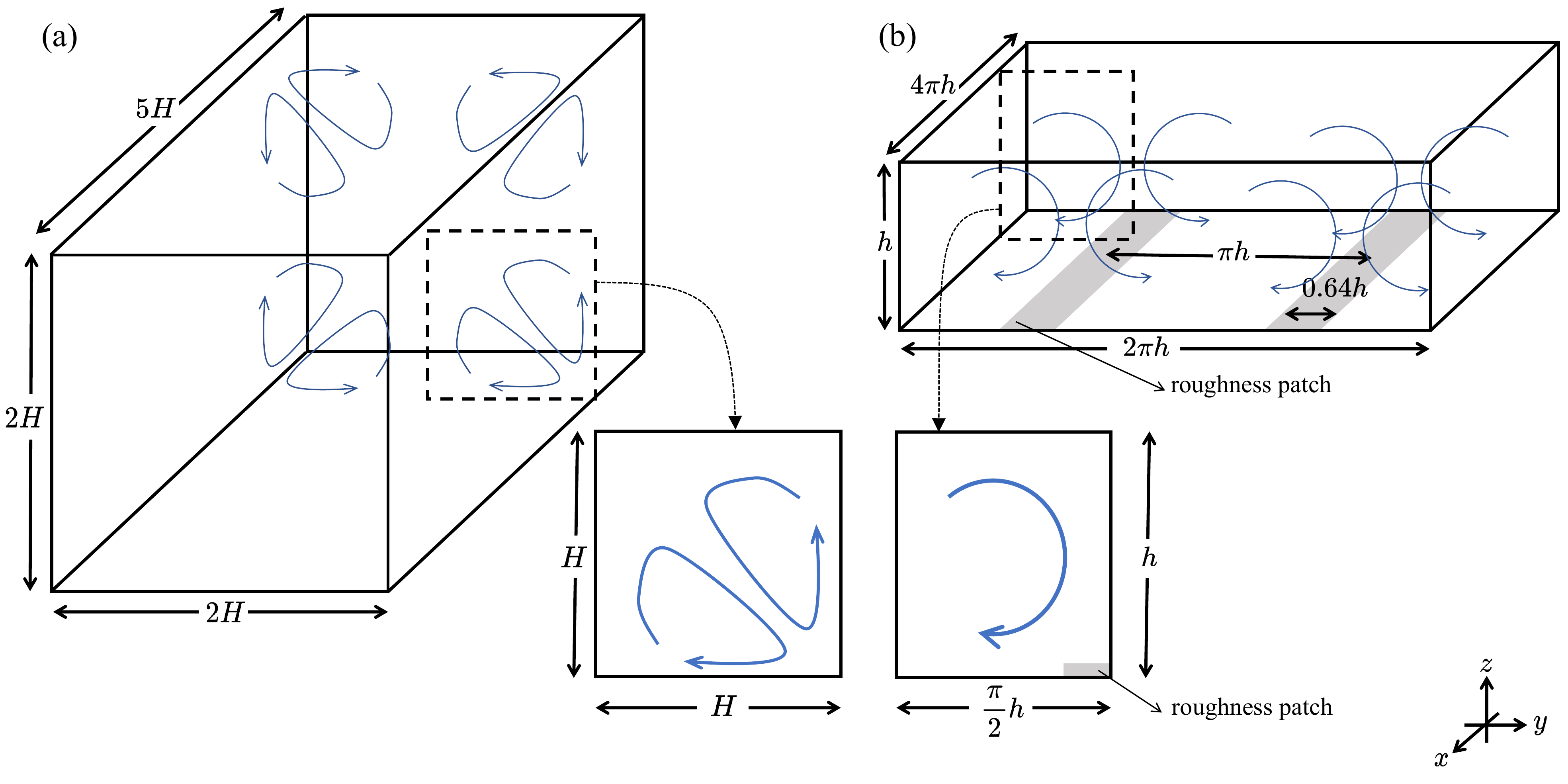}%
\caption{\label{fig:Geo} The schematic of the geometry for (a) the square-duct secondary flow, and (b) the roughness-induced secondary flows: 3D geometries for the DNS/LES, and 2D geometries for the RANS simulations.}%
\end{figure*}

First, all of the propagation methods (listed in Fig.~\ref{fig:methods}) are going to be evaluated in the case of the square-duct flow, depicted in the Fig.~\ref{fig:Geo}(a). In Fig.~\ref{fig:Geo}(a), the 3-dimensional geometry is used by Ref. \citet{pinelli2010reynolds} for the DNS, and here, for the RANS simulations only a 2-dimensional quarter of the domain is considered. The high-fidelity data for this case comes from the DNSs done by \citet{pinelli2010reynolds} and curated by McConkey \textit{et al.}\cite{mcconkey2021curated}. The bulk streamwise velocity is $U_b = 0.482m/s$, and the fluid viscosity is $\nu = 6.886\times10^{-5}m^2/s$. Because of the symmetry in the cross-section of the square duct, only a quarter of the domain is considered for the RANS simulations. The height and width of the square duct are $2H= 1m$. This will result in a bulk Reynolds number of $Re = 3500$. The RANS simulations are initialized by uniform conditions, and the periodic condition is considered for the streamwise direction. No-slip condition is used for the walls, and the rest of the boundaries are considered symmetric.

Secondly, the secondary flows induced by longitudinal roughness patches are considered to be evaluated. Fig.~\ref{fig:Geo}(b) shows the schematic of the geometry for a half channel with spanwise heterogeneous roughness. Here, we used an in-house pseudo-spectral finite-difference code for the LES of the case. The sub-grid scale stress tensor is modelled by the anisotropic minimum dissipation model \cite{rozema2015minimum,abkar2016minimum,abkar2017large}. This code has been repeatedly used for the simulation of turbulent boundary layer\cite{yang2018hierarchical,bastankhah2019multirotor,yang2020scaling,yang2022logarithmic,Eidi2022} and also for the simulation of the roughness-induced secondary flows \cite{forooghi2020roughness,amarloo2022secondary}.

Since the error propagation is amplified by increasing the Reynolds number\cite{wu2019reynolds}, the case of roughness-induced secondary flow has the nominally infinite Reynolds number. In our LESs, the friction velocity is $u_\tau = 0.4 m/s$, the height of the half channel is $h = 500 m$, the kinematic viscosity is $\nu = 1.5 \times 10^{-5} m^2/s$, high-roughness patch height $z_{o,H} = 0.5 m$, low-roughness patch height $z_{o,L} = 0.005 m$.
The friction Reynolds number is $Re_\tau = 1.3 \times 10^7$; therefore, the flow is not affected by the Reynolds number and is nominally referred to as infinite Reynolds number flow (see, e.g., Refs. \onlinecite{stevens2014large,yang2018numerical,yang2018hierarchical}). In Fig.~\ref{fig:Geo}(b), the 3-dimensional geometry is used for the LES, and for the RANS simulations only a 2-dimensional quarter of the domain is considered because of the symmetry in the spanwise direction. More details about the LESs can be found in Refs.  \onlinecite{forooghi2020roughness,amarloo2022secondary}. The RANS simulations are initialized by uniform conditions, and the periodic condition is considered for the streamwise direction. A rough-wall boundary condition with wall functions is used for the bottom wall, and the spanwise boundaries are considered symmetric.

\section{\label{sec:Results}Results and discussion}
In this section, the results of applying all of the 8 propagation methods (listed in Fig.~\ref{fig:methods}) are presented. First, the performance of new methods (FRZ-RFV and KCF-RFV) will be compared with that of previously introduced methods (EXP-RST, IMP-RST, FRZ-RST, KCF-RST, EXP-RFV, and IMP-RFV) in the canonical case of the square duct. Secondly, the case of roughness-induced secondary flow will be examined by all of the methods to evaluate the reliability of the propagation methods in a case with a very high Reynolds number. In the last subsection, the effect of using different baseline RANS models in the propagation process with frozen treatment is studied.

\subsection{\label{sec:SDresults} Square-duct secondary flow}
\begin{figure*}
\includegraphics[height=0.25\textwidth]{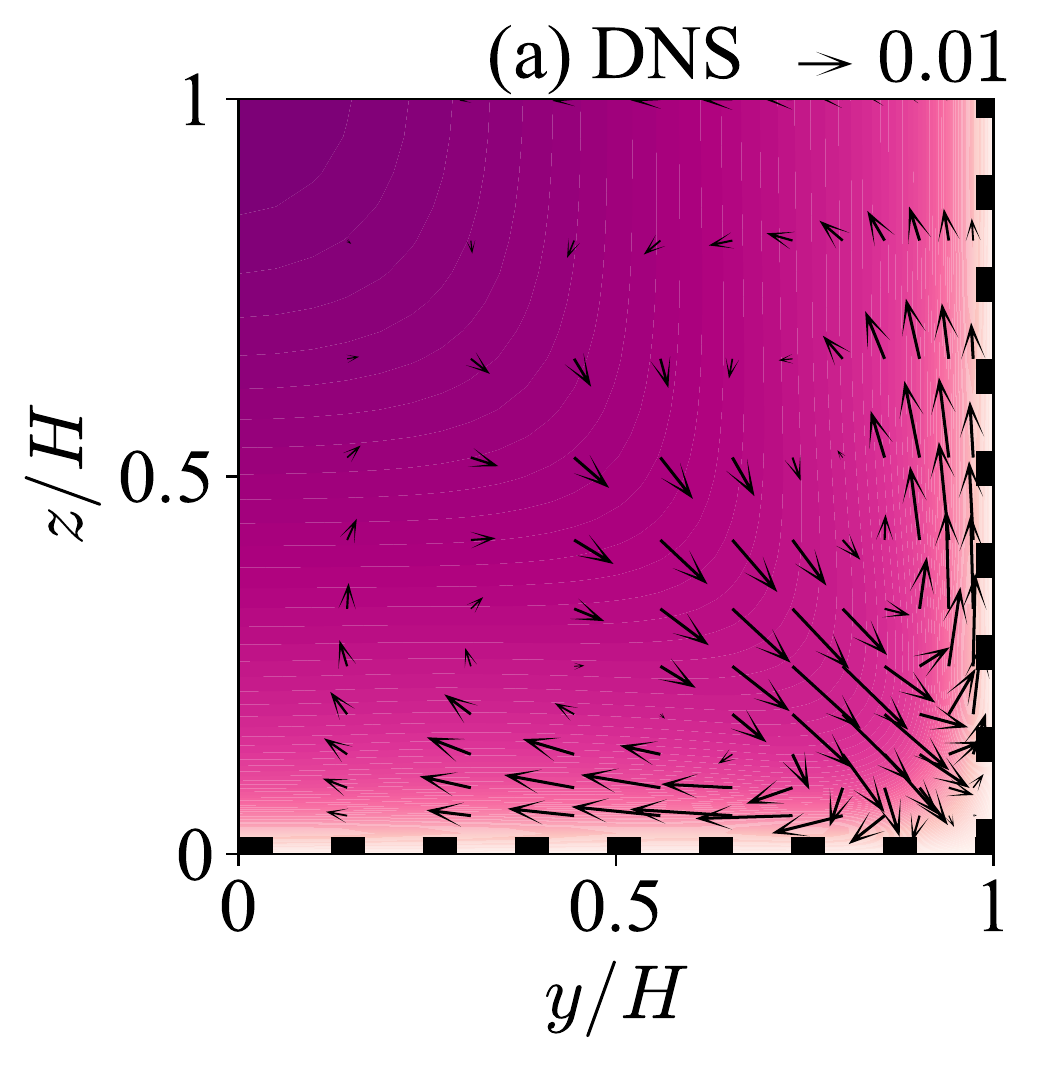}
\includegraphics[height=0.25\textwidth]{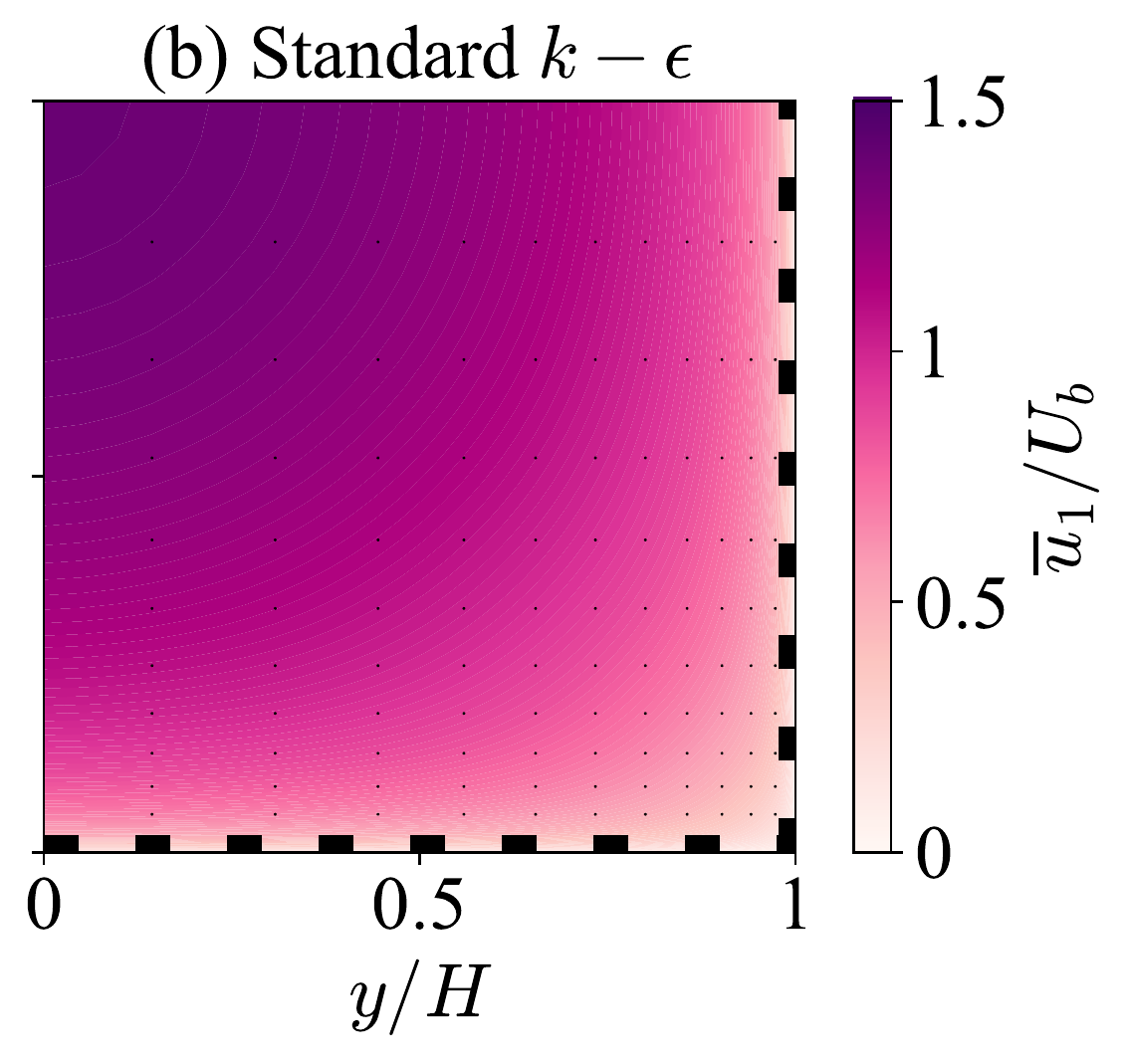}
\includegraphics[height=0.25\textwidth]{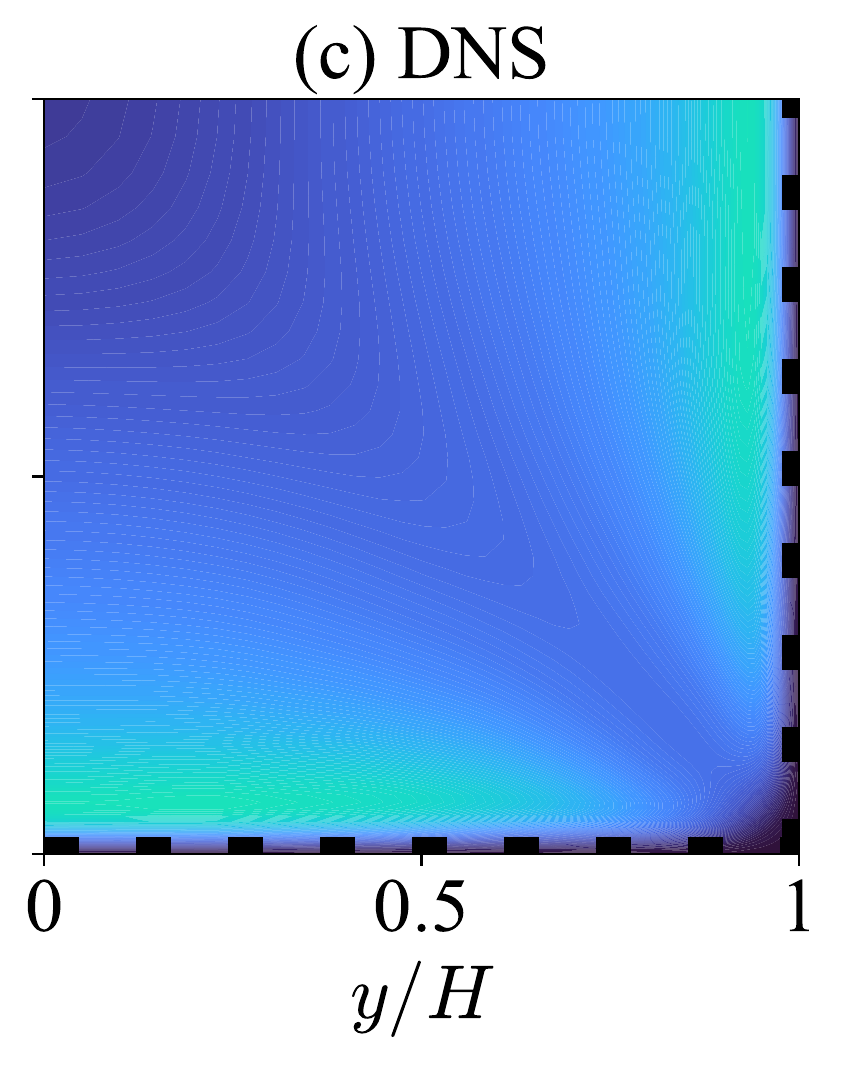}
\includegraphics[height=0.25\textwidth]{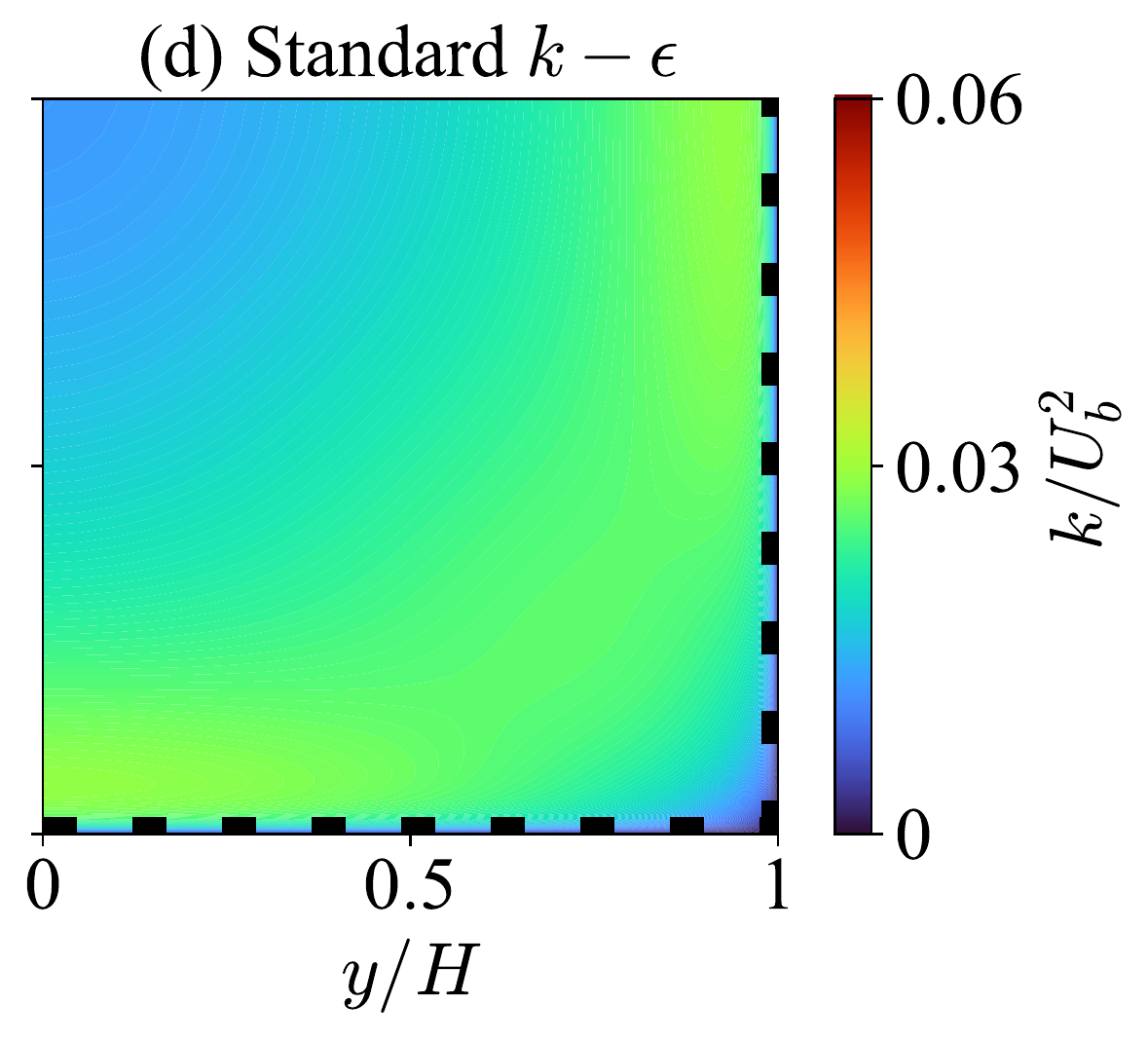}

\caption{\label{fig:SD_rolls} Reference values for the case square duct with $Re = 3500$: (a,c) DNS data \cite{mcconkey2021curated} and (b,d) RANS simulation with standard $k-\epsilon$. The contour plot shows: (a,b) the mean streamwise velocity and (c,d) the TKE. The vectors indicate the in-plane velocity normalized by $U_b$. The black dots inside (b) indicate the absolute zero values for the in-plane velocity. The dashed line indicates the walls of the square duct.}%
\end{figure*}

In Fig.~\ref{fig:SD_rolls}(a), the DNS mean velocity field of the square duct with $Re = 3500$ (readapted from Ref.~\onlinecite{mcconkey2021curated}) is shown, and also as a reference, the RANS simulation of the same case by the standard $k-\epsilon$ is reported in Fig.~\ref{fig:SD_rolls}(b). The contour plot shows the normalized mean streamwise velocity, and the vectors are showing the in-plane velocity. \hltOn Fig.~\ref{fig:SD_rolls}(b) shows that the the standard $k-\epsilon$ model, consistent with the previous studies \cite{nikitin2021prandtl}, is not capable of reproducing the secondary flows which are captured by DNS (Fig.~\ref{fig:SD_rolls}(a)). This error is due to the importance of modeling the normal components of the Reynolds stress tensor. \hltOff The TKE from the high-fidelity data\cite{mcconkey2021curated} and from the RANS simulation with standard  $k-\epsilon$ are reported in Figs.~\ref{fig:SD_rolls}(c,d) respectively.

\begin{figure*}
\includegraphics[height=0.25\textwidth]{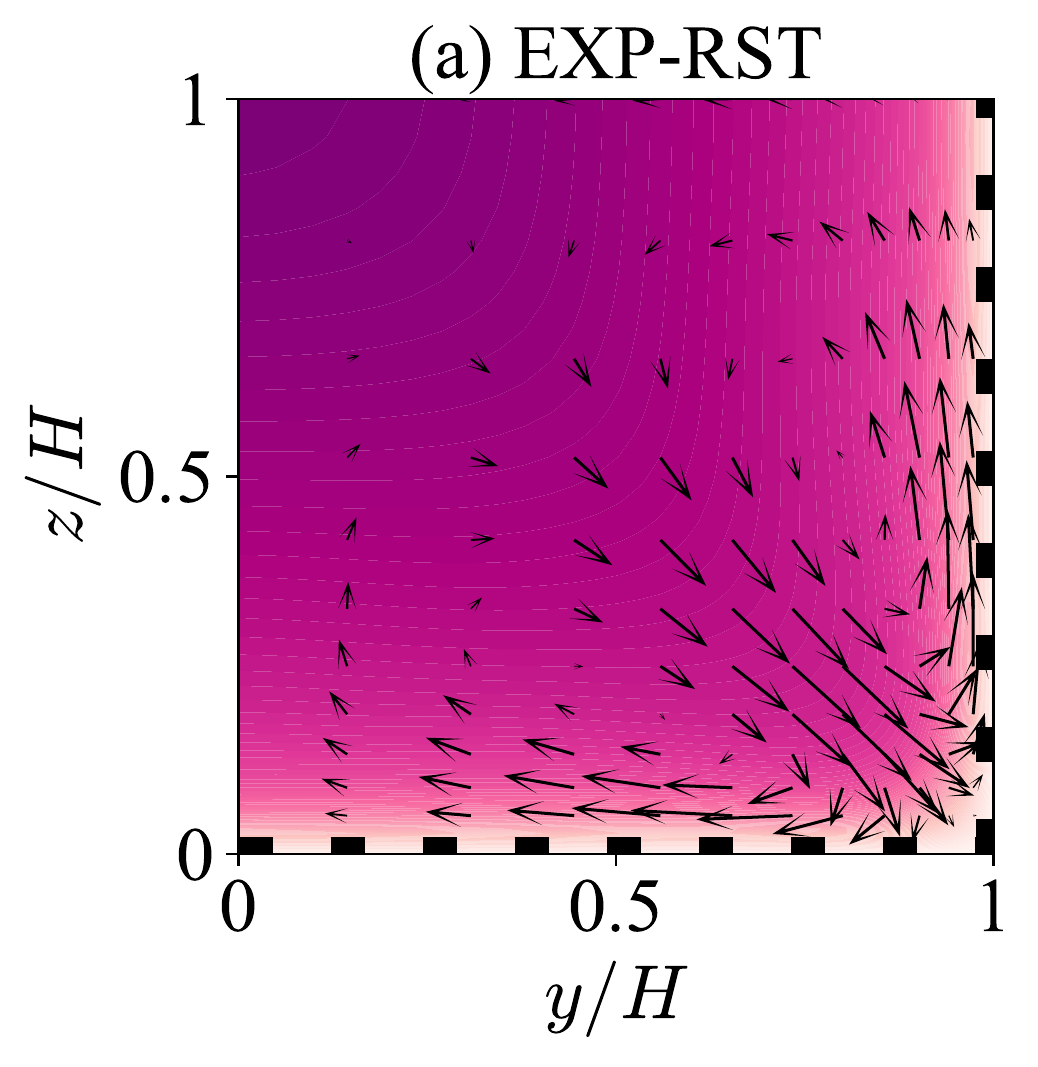}
\includegraphics[height=0.25\textwidth]{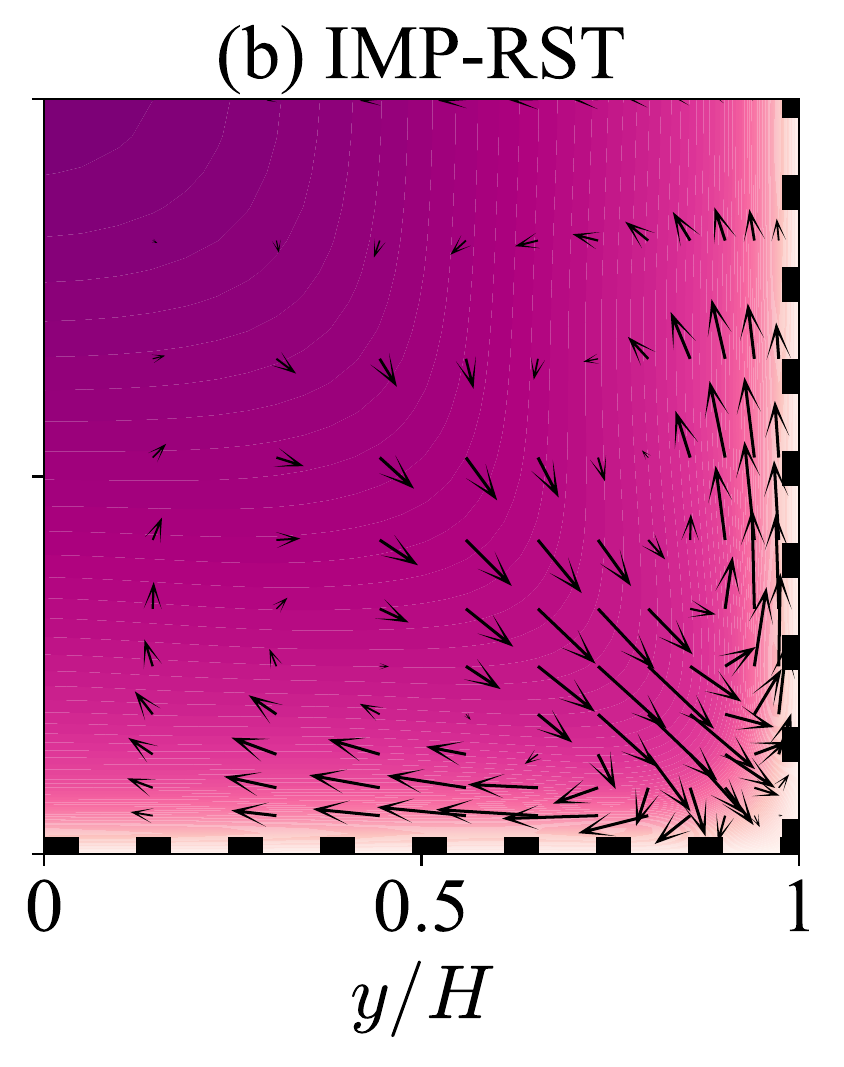}
\includegraphics[height=0.25\textwidth]{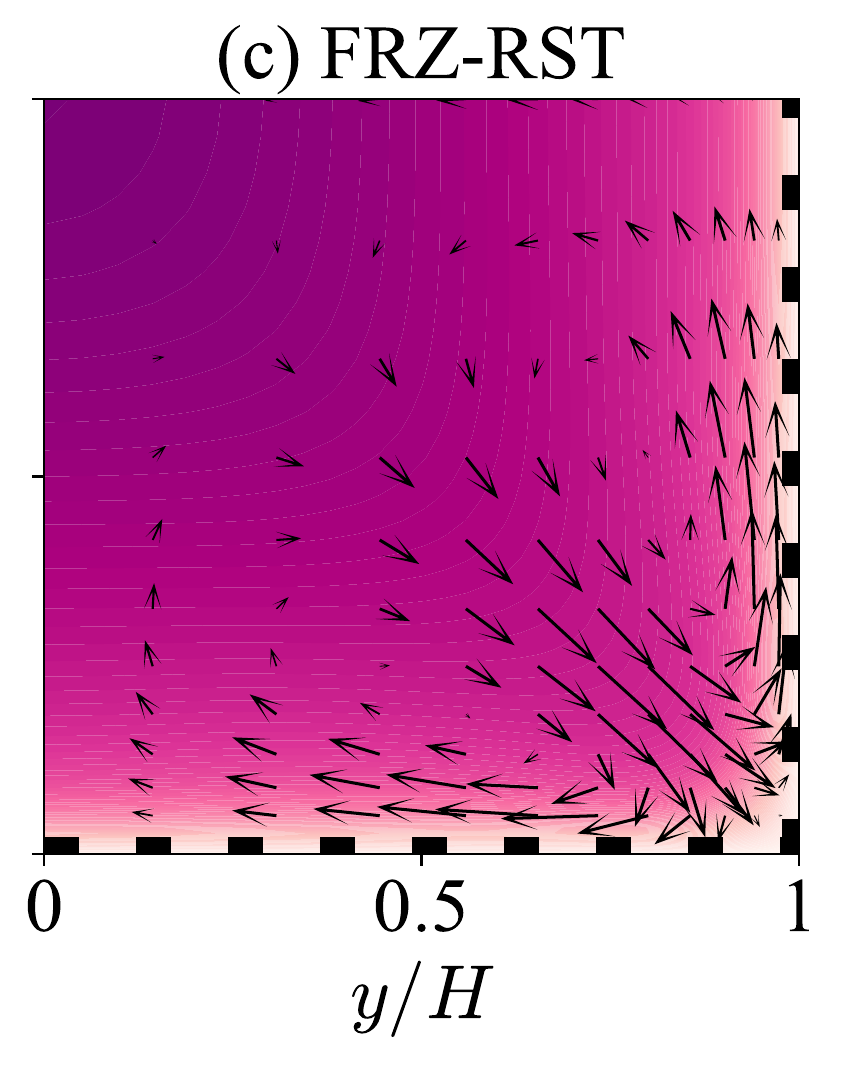}
\includegraphics[height=0.25\textwidth]{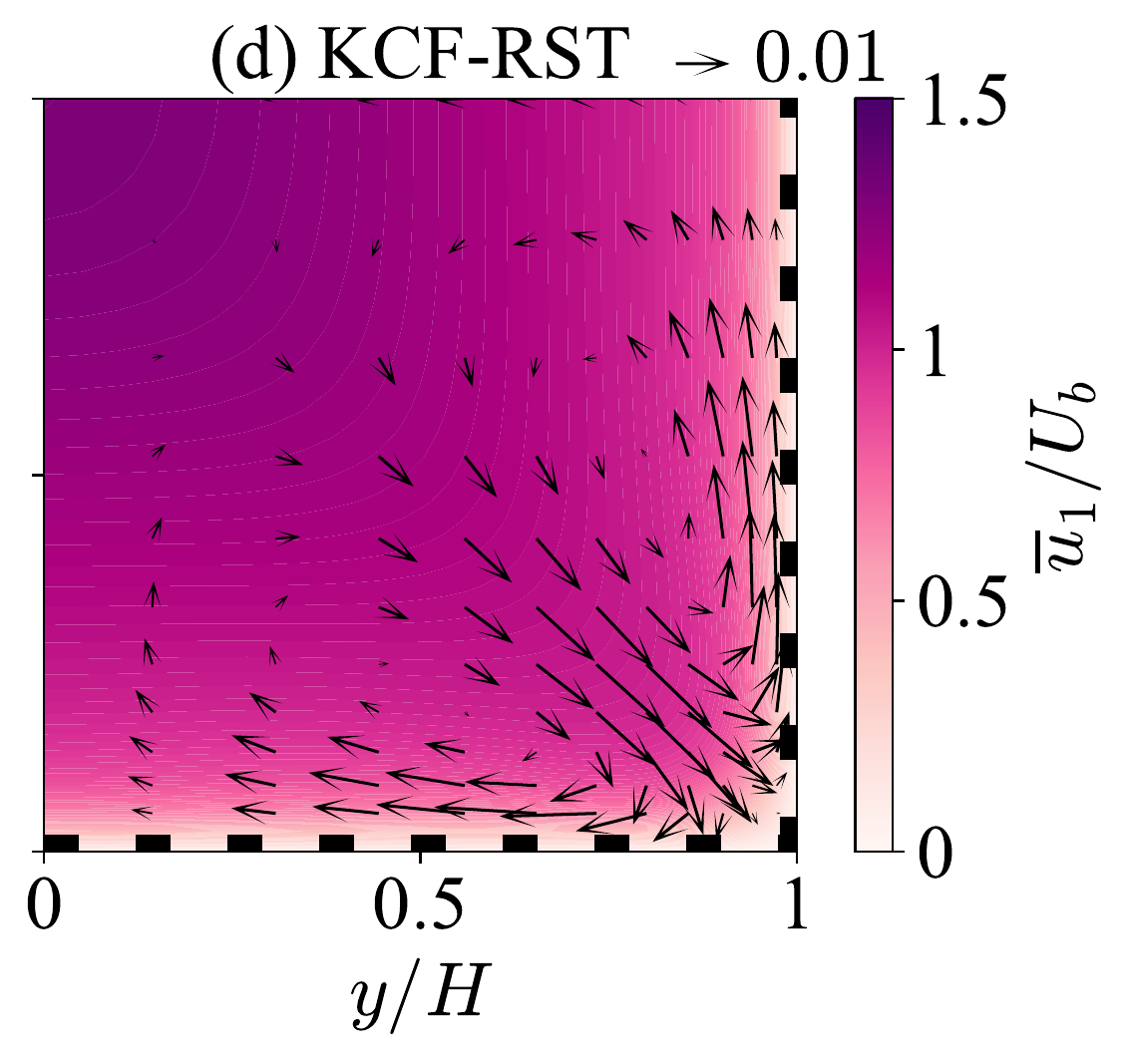}
\caption{\label{fig:SD_rolls_RST} Mean velocity field with the propagation of the RST into the RANS simulations for the case square duct with $Re = 3500$: (a) the EXP-RST, (b) the IMP-RST, (c) the FRZ-RST, (d) the KCF-RST. The contour shows the mean streamwise velocity, and the vectors indicate the in-plane velocity normalized by $U_b$. The dashed line indicates the walls of the square duct.}%
    \includegraphics[height=0.25\textwidth]{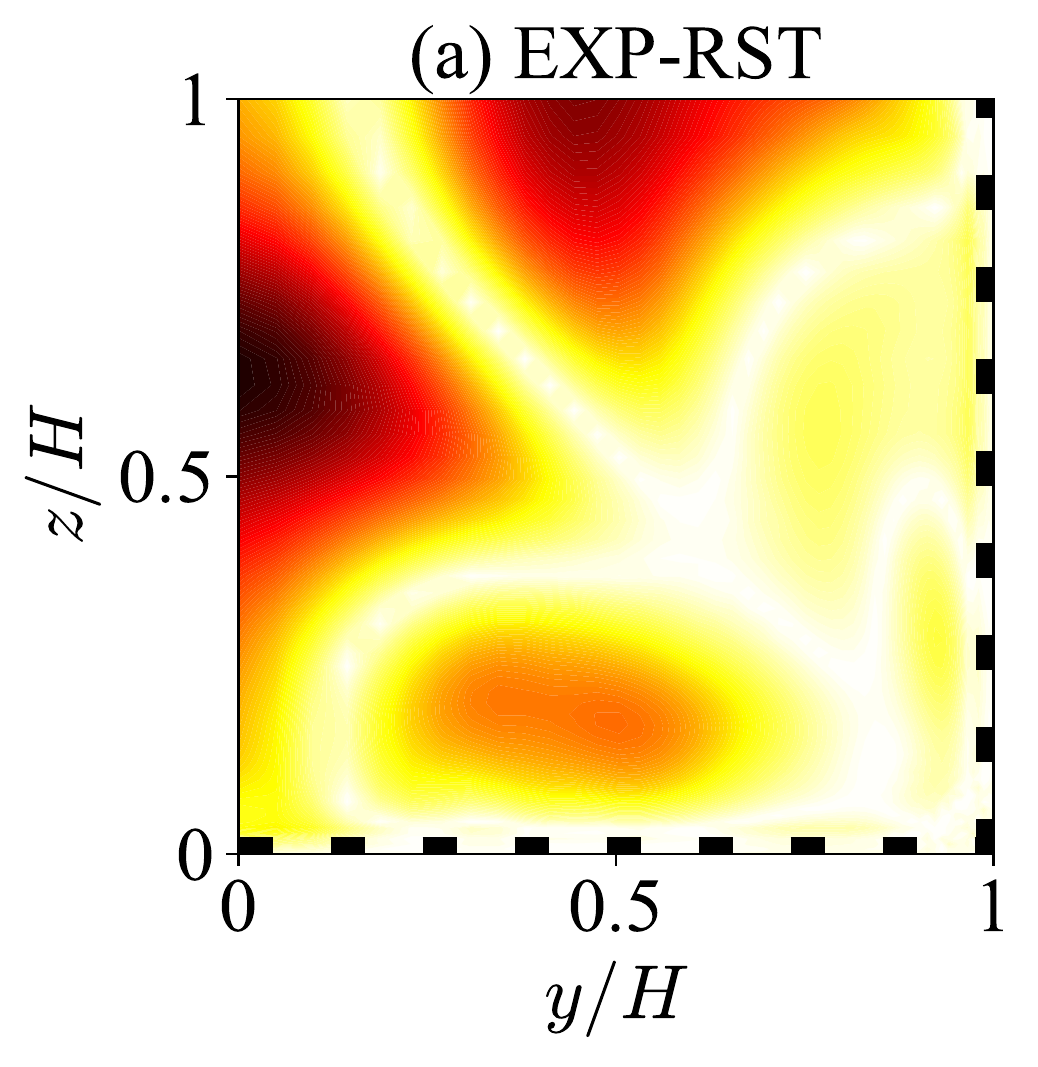}
    \includegraphics[height=0.25\textwidth]{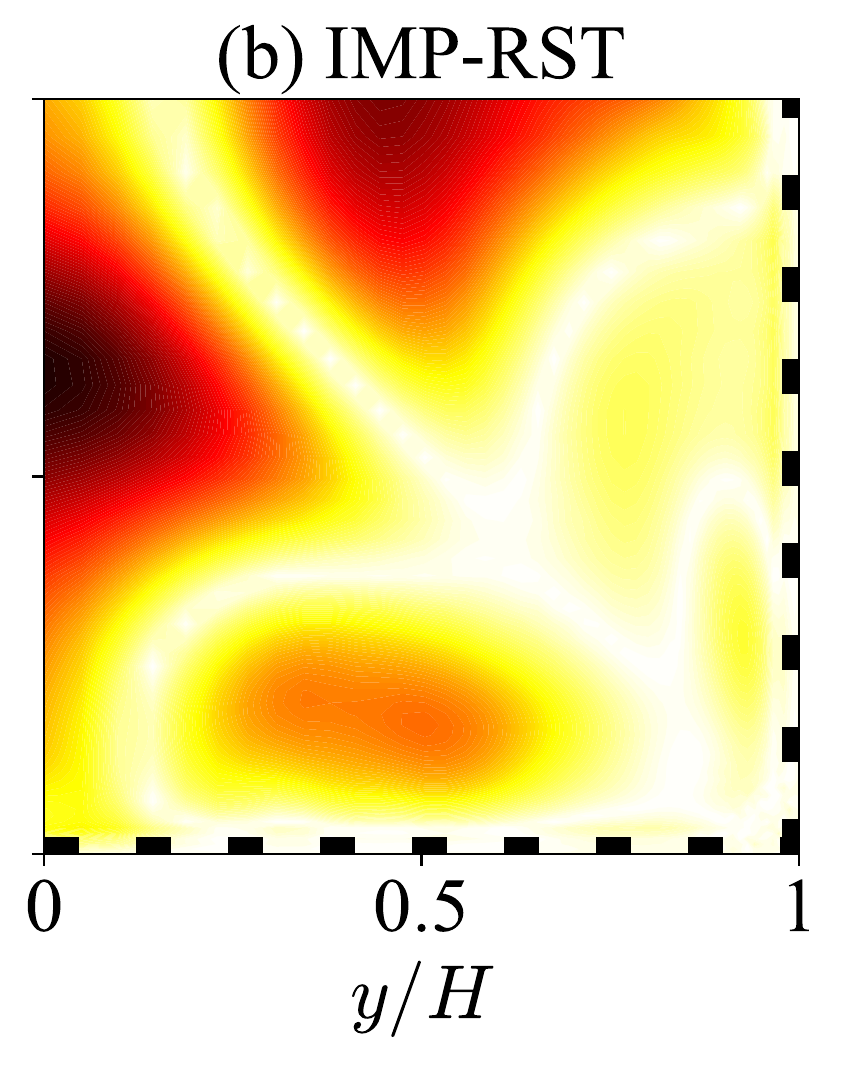}
    \includegraphics[height=0.25\textwidth]{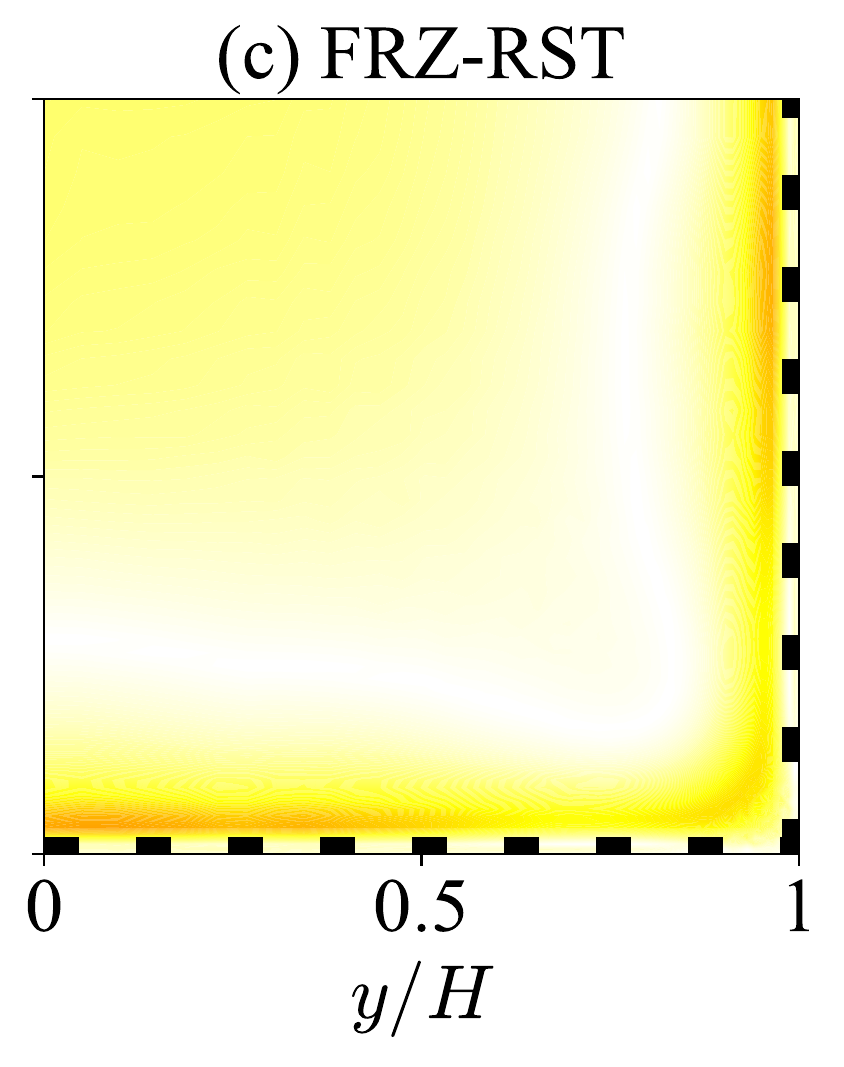}
    \includegraphics[height=0.25\textwidth]{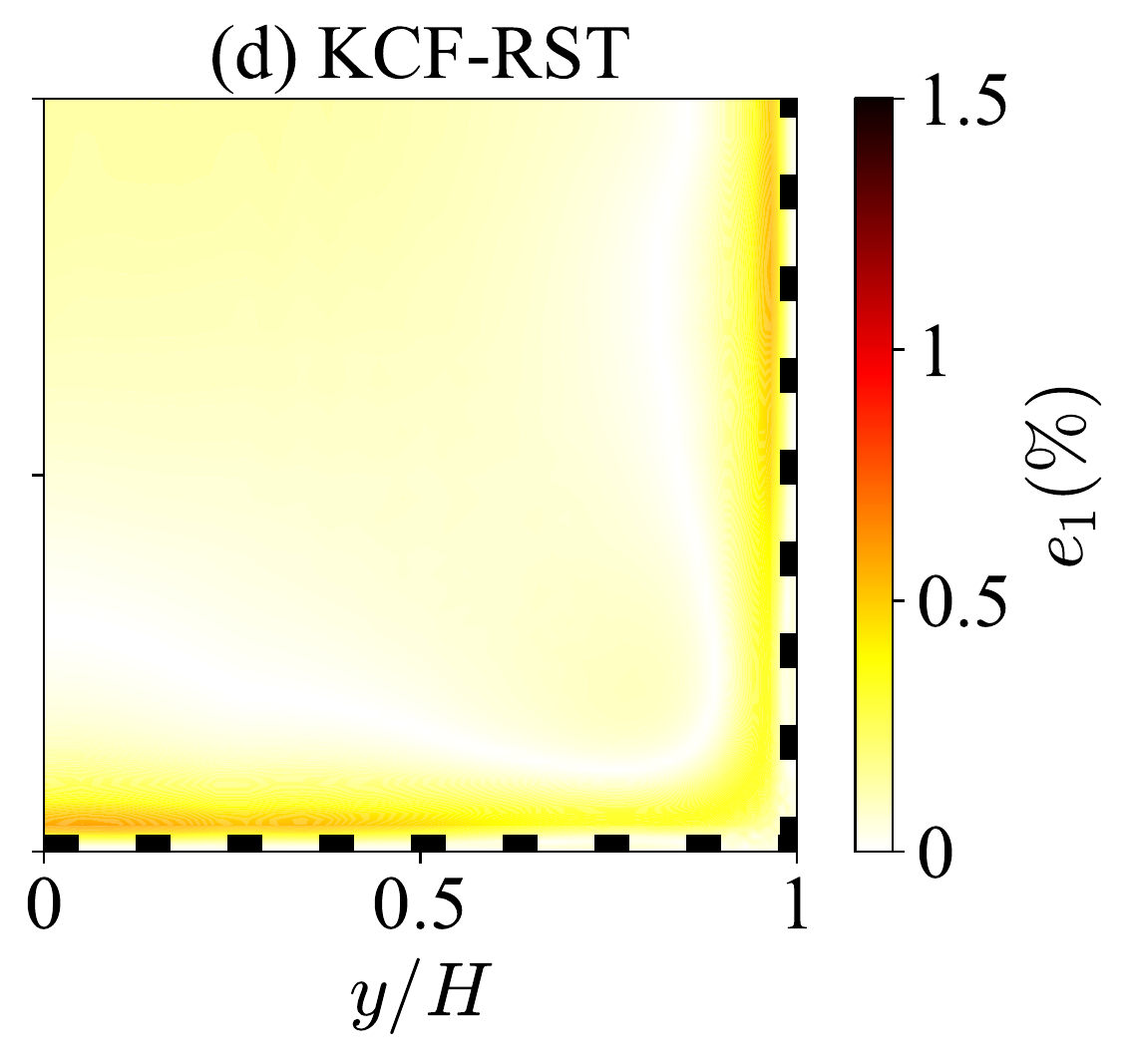}
\caption{\label{fig:SD_error_RST_u1} \hltOn The error propagation of the streamwise ($x$-direction) velocity by the propagation of the RST for the case of square duct with $Re = 3500$: (a)EXP-RST, (b)IMP-RST, (c)FRZ-RST, (d)KCF-RST. The dashed line indicates the walls of the square duct. \hltOff}%
\end{figure*}

First, the RST is propagated into the RANS simulations with different treatment techniques listed in Fig.~\ref{fig:methods}. Fig.~\ref{fig:SD_rolls_RST} shows the contours of streamwise mean velocity and the vectors of in-plane velocity. Fig.~\ref{fig:SD_rolls_RST} shows that the propagation of the RST, regardless of the treatment technique, has a good performance to reproduce the velocity field similar to the DNS data. For a better comparison, the normalized error values will be compared for the evaluation of the different treatment methods of the RST. The percentage of error is calculated based on the difference between the resulted mean velocity $\overline{u}_i$ and the high-fidelity mean velocity $\overline{u}^*_i$. The difference is normalized by the bulk magnitude of each velocity component. Therefore, the error for each velocity component is defined as 
\begin{equation}
\label{eq:e_error}
    e_{i} = \frac{|\overline{u}_i - \overline{u}^*_i|}{u^{ref}_i}\times100,~\text{where}~ u^{ref}_i = \frac{1}{S}\iint_S |\overline{u}^*_i| ds,
\end{equation}
where $u^{ref}_i$ is the bulk value of the absolute velocity which is calculated as the area-averaged of the velocity component magnitude in direction $i$ over the cross-section $S$, and the area-averaged error, $E_i$, is defined as 
\begin{equation}
\label{eq:E_error}
    E_i = \frac{1}{S}\iint_S e_{i} ds.
\end{equation}

The error propagation of the streamwise ($x$-direction) velocity is shown in Fig.~\ref{fig:SD_error_RST_u1} with 4 different treatment methods for the propagation of the RST. It shows that the implicit treatment (Fig.~\ref{fig:SD_error_RST_u1}(b)) does not significantly change the error propagation compared to the explicit treatment (Fig.~\ref{fig:SD_error_RST_u1}(a)) which was reported by the Ref.~\onlinecite{brener2021conditioning}. By comparing Fig.~\ref{fig:SD_error_RST_u1}(b) with Fig.~\ref{fig:SD_error_RST_u1}(c), it is clear that the error propagation is reduced by the frozen treatment of the RST. The comparison of FRZ-RST and KCF-RST in Fig.~\ref{fig:SD_error_RST_u1} shows that a k-corrective baseline does not change the performance of the frozen treatment technique on the reconstruction of the velocity domain.

\begin{figure*}
\includegraphics[height=0.25\textwidth]{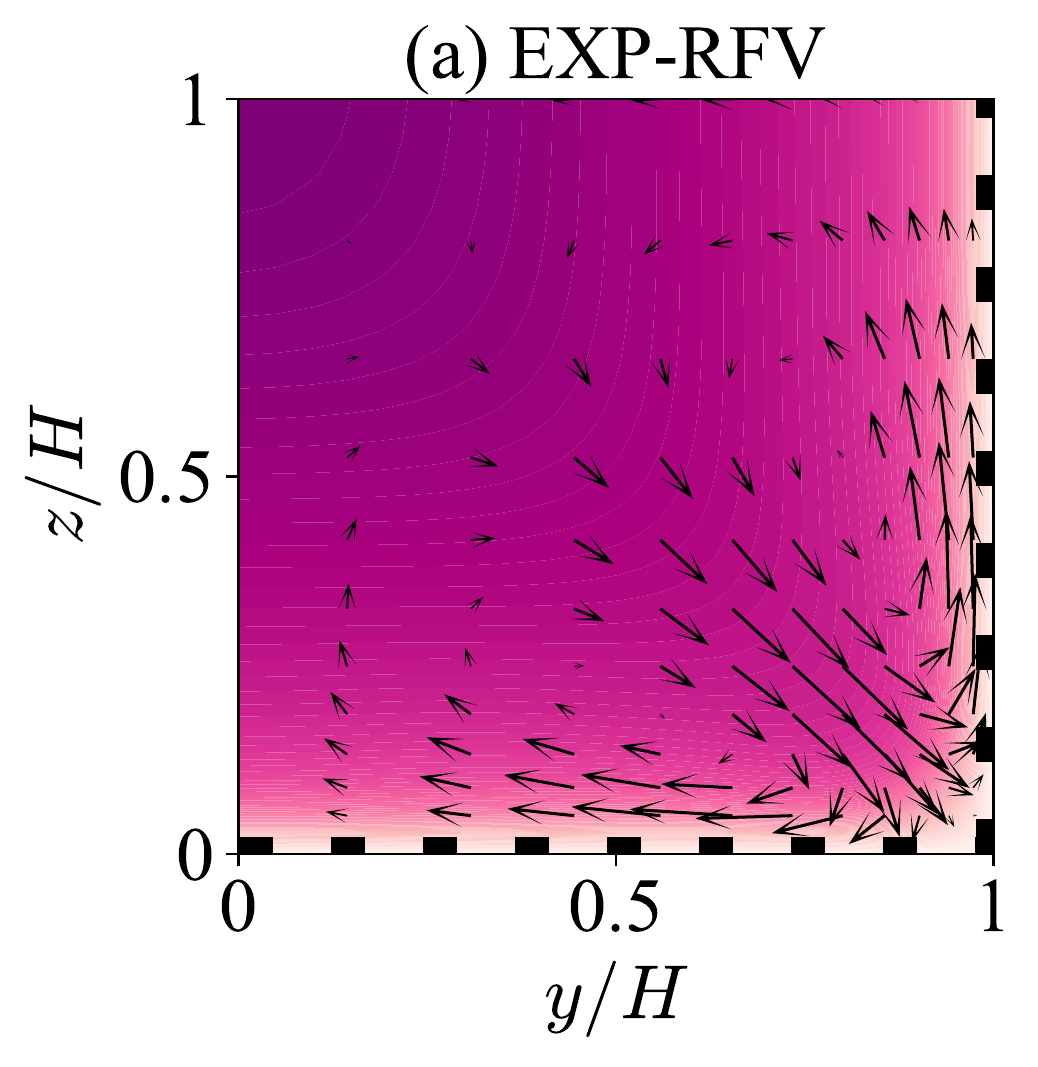}
\includegraphics[height=0.25\textwidth]{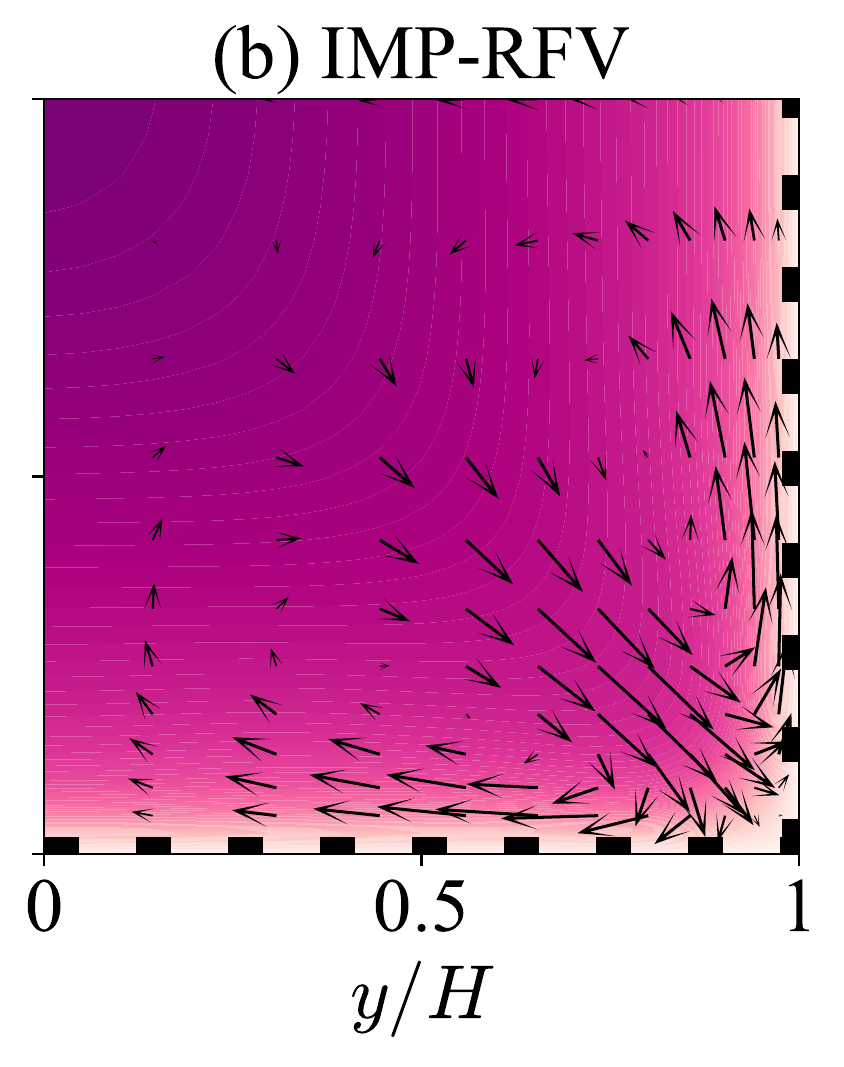}
\includegraphics[height=0.25\textwidth]{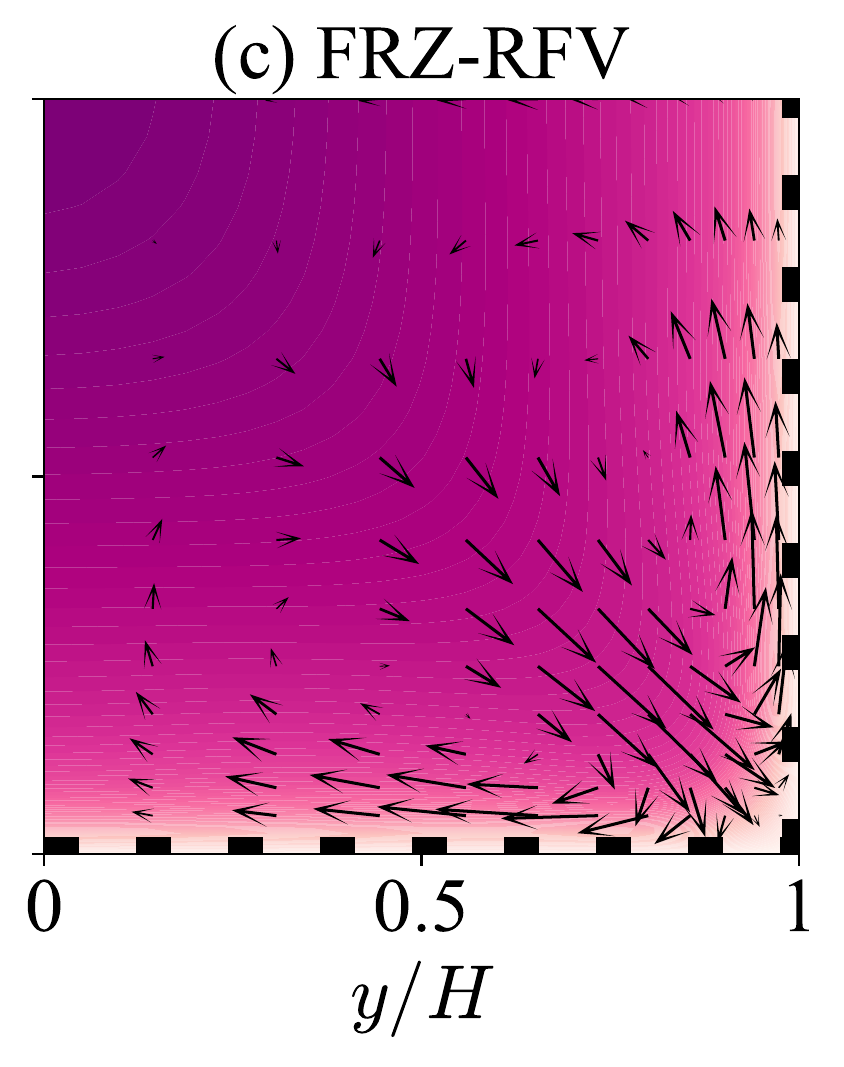}
\includegraphics[height=0.25\textwidth]{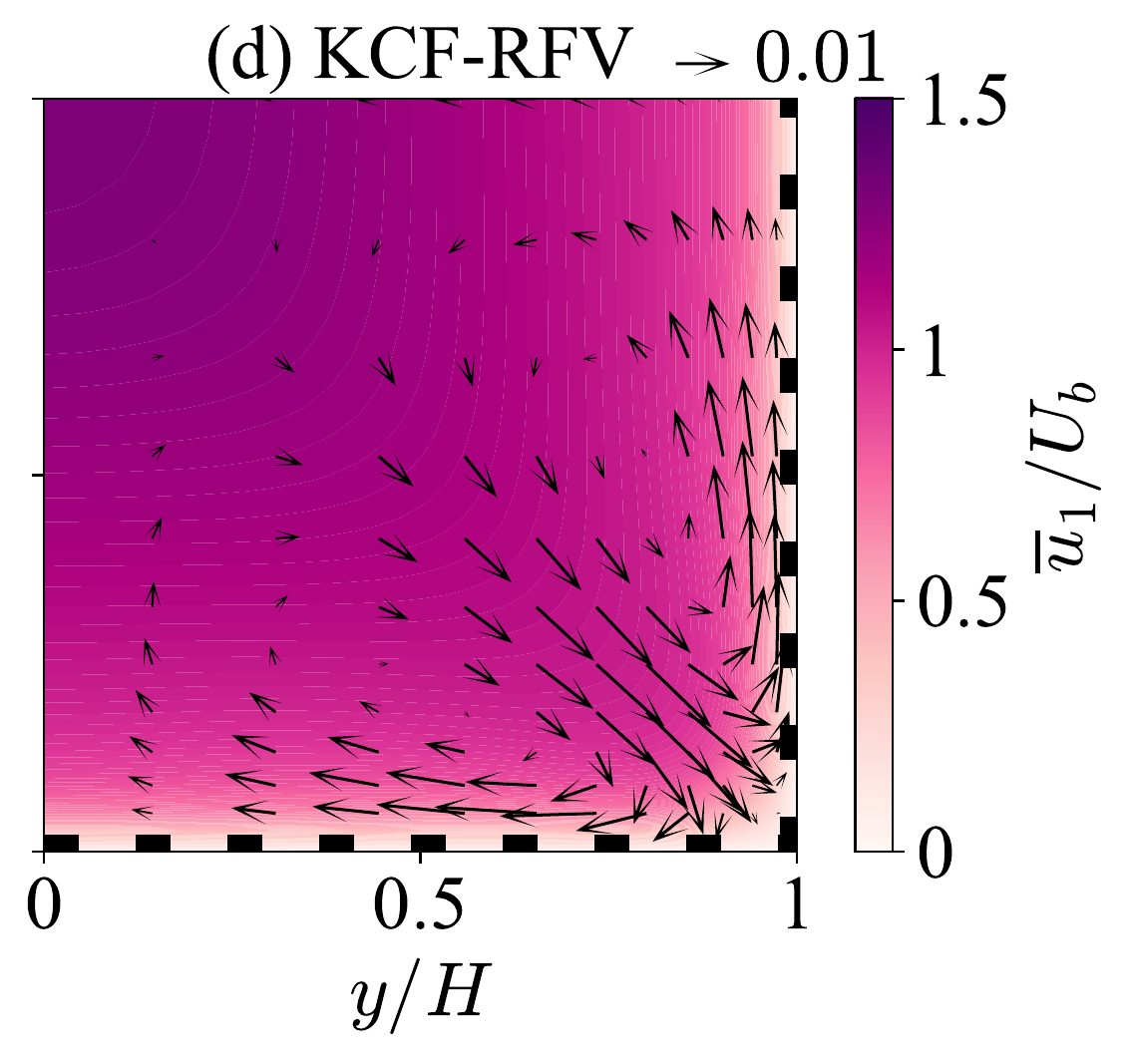}
\caption{\label{fig:SD_rolls_RFV} Mean velocity field with the propagation of the RFV into the RANS simulations for the case square duct with $Re = 3500$: (a) the EXP-RFV, (b) the IMP-RFV, (c) the FRZ-RFV, (d) the KCF-RFV. The contour shows the mean streamwise velocity, and the vectors indicate the in-plane velocity normalized by $U_b$. The dashed line indicates the walls of the square duct.}%
    \includegraphics[height=0.25\textwidth]{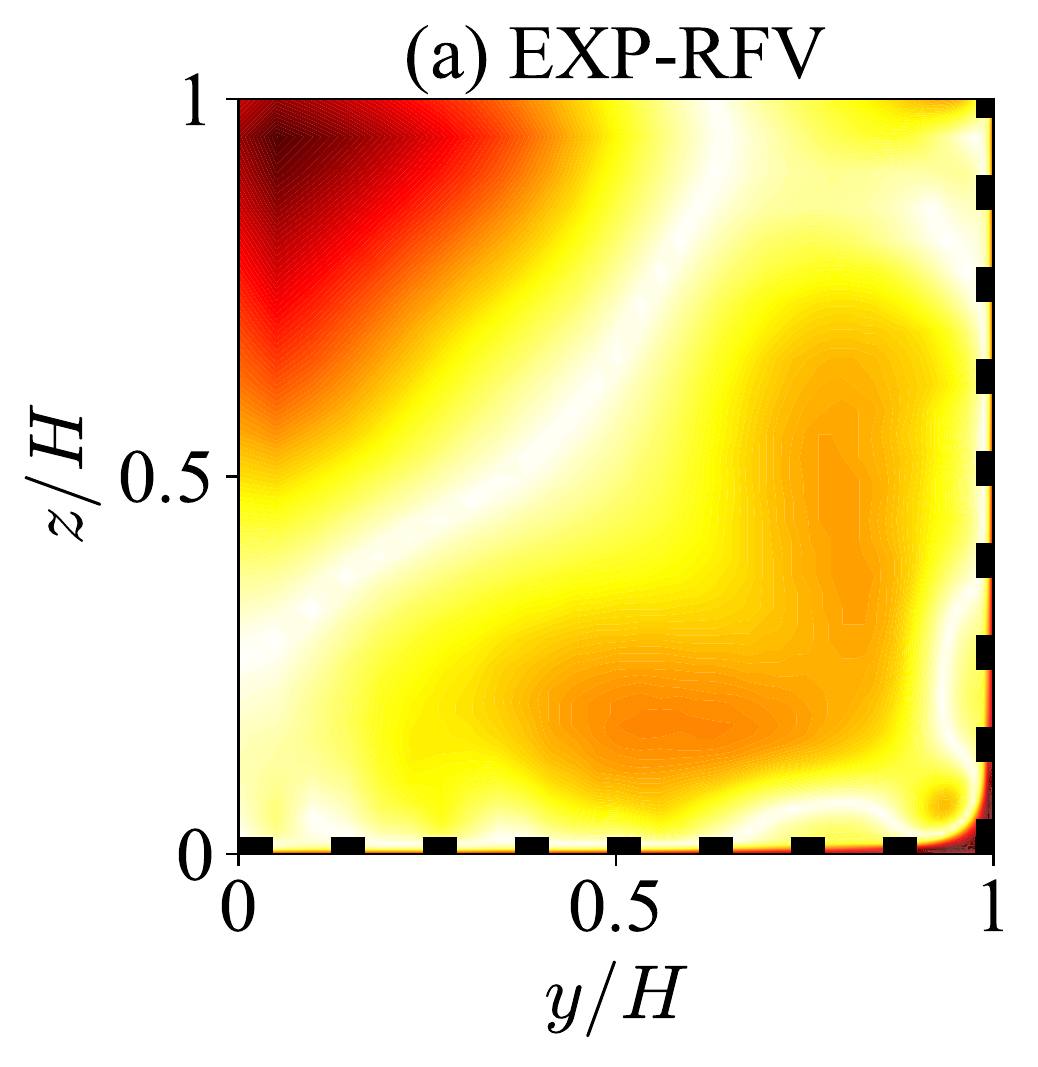}
    \includegraphics[height=0.25\textwidth]{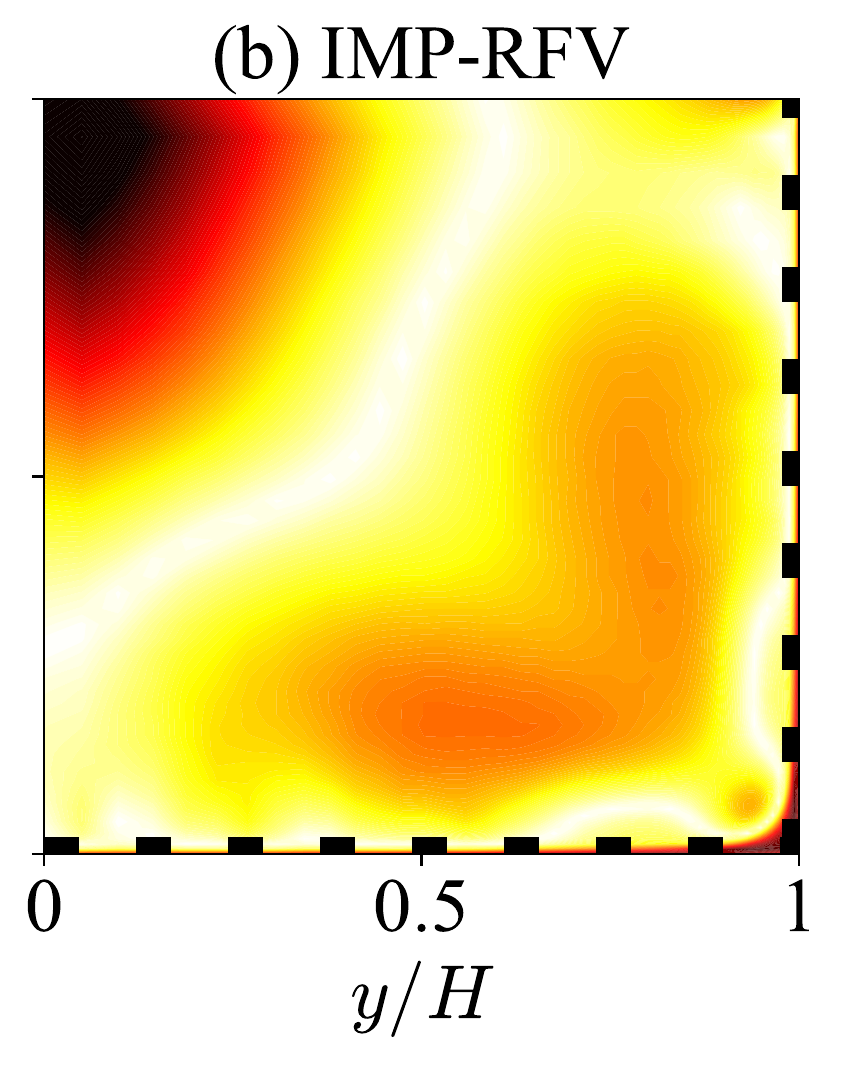}
    \includegraphics[height=0.25\textwidth]{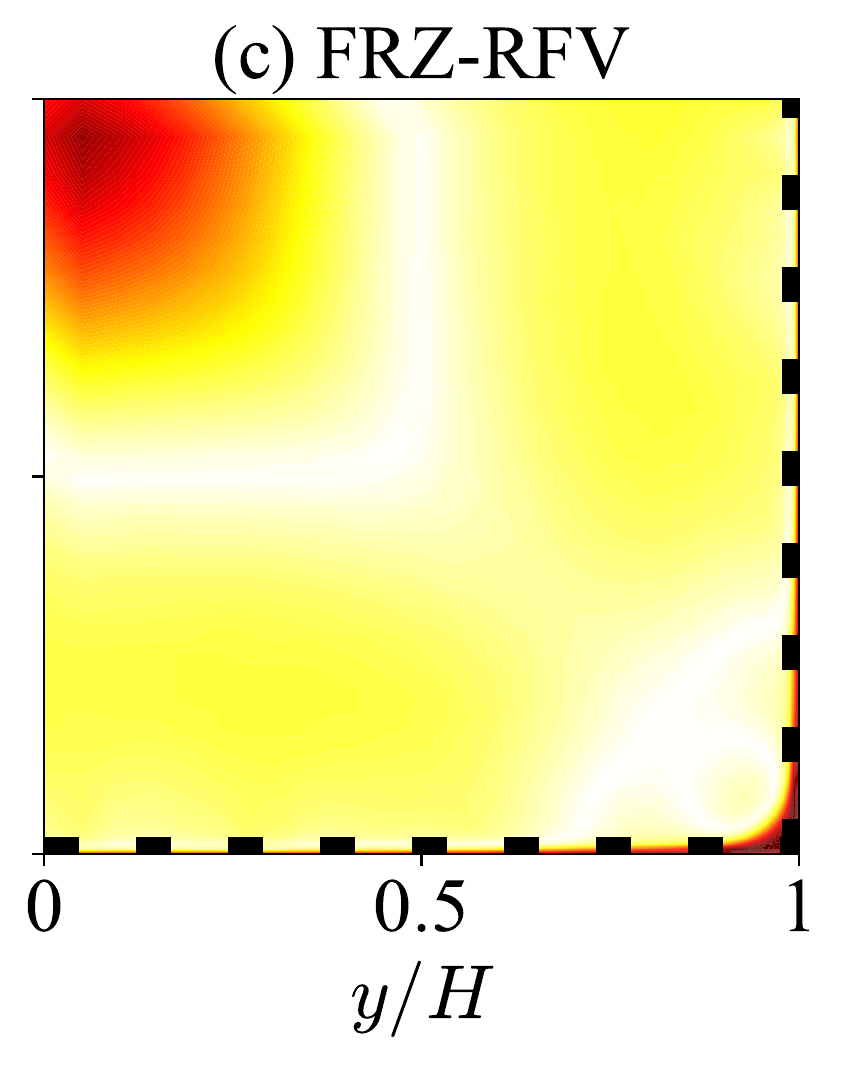}
    \includegraphics[height=0.25\textwidth]{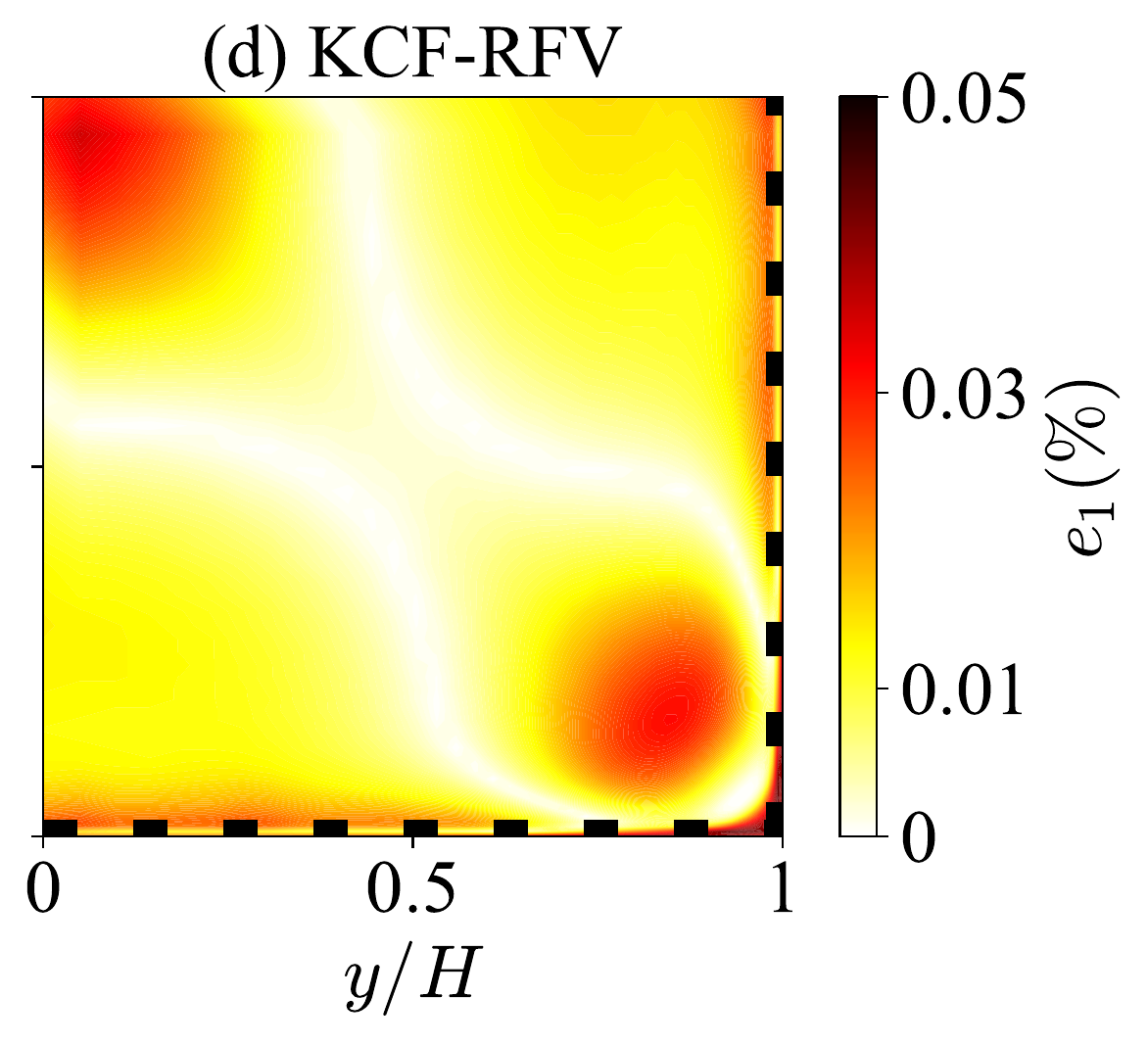}
\caption{\label{fig:SD_error_RFV_u1} \hltOn The error propagation of the streamwise ($x$-direction) velocity by the propagation of the RFV for the case of square duct with $Re = 3500$: (a)EXP-RFV, (b)IMP-RFV, (c)FRZ-RFV, (d)KCF-RFV. The dashed line indicates the walls of the square duct. \hltOff}%
\end{figure*}

Similar to the Fig.~\ref{fig:SD_rolls_RST}, Fig.~\ref{fig:SD_rolls_RFV} shows that the propagation of the RFV, regardless of the treatment technique, also shows a good performance to reproduce the velocity field similar to the DNS data. For a better comparison, Fig.~\ref{fig:SD_error_RFV_u1} shows the error propagation of streamwise velocity by propagating the RFV with 4 different treatment methods into the RANS equation. First, the comparison of the magnitude of error for streamwise velocity in Fig.~\ref{fig:SD_error_RST_u1} and Fig.~\ref{fig:SD_error_RFV_u1} shows that, regardless of the treatment method, the propagation of the RFV (errors between $0\%-0.05\%$) will result in a lower error propagation compared to the propagation of the RST (errors between $0\%-1.5\%$). \hltOn Note that it was shown before by \citet{brener2021conditioning} that the implicit treatment cannot induce a significant change for error propagation, which is also visible in our results. They also showed that the implicit propagation of the RFV is better than the implicit propagation of the RST; here, we combine the strength of frozen treatment and RFV propagation. \hltOff Fig.~\ref{fig:SD_error_RFV_u1} shows that a similar pattern to Fig.~\ref{fig:SD_error_RST_u1} (the propagation of RST) is happening for the propagation of the RFV. In other words, the propagation of the RFV results in lower error than the propagation of the RST, and the frozen treatment makes it even lower. 

Figs.~\ref{fig:SD_error_RST_u1} and \ref{fig:SD_error_RFV_u1} show that the pattern of error propagation changes when the frozen treatment is used. By using explicit and implicit treatments, the areas with high error values are mainly inside the domain, and by using the frozen treatment, near-boundary areas have higher error values. This change of pattern can be rooted in the treatment technique, where carried errors from the first order statistics (i.e., the mean velocity) of the high-fidelity source play an important role. It should be also noted that when only second-order statistics are used (Figs.~\ref{fig:SD_error_RST_u1} (a) and (b)) the pattern of error is not symmetric about the diagonal line, as opposed to the cases when first-order statistics are used in which the error pattern is symmetric about the diagonal line. This shows that the propagation of error from the high-fidelity source into RANS simulations is dependent on the statistical errors of high-fidelity data, as Thompson \textit{et al.} \cite{thompson2016methodology} pointed out. This is most likely due to the fact that integrating first-order statistics, which usually have lower statistical errors, into the propagation technique mitigates the error propagation while it can also change the distribution of error inside the domain.

For a better quantitative comparison of the methods, the area-averaged propagation error (defined in Eq.~(\ref{eq:E_error})) for each of the velocity components is calculated and presented in Table~\ref{tab:SD_errors}. For a better grasp of the error magnitudes, the errors of the RANS simulation with the standard $k-\epsilon$ is also reported in Table~\ref{tab:SD_errors} as a reference. In this simulation, the magnitudes of $E_2$ and $E_3$ are $100\%$, and they show the complete failure of the standard $k-\epsilon$ model for capturing secondary flows (as it is well established in the literature \cite{nikitin2021prandtl}). 
The values of $E_2$ and $E_3$ for different propagation techniques show that the explicit treatment of the RST ($E_2 = 3.754\%$) has a satisfying performance in reconstructing the secondary flow and the implicit treatment of the RST makes secondary flow error lower but the frozen treatment reduces the error magnitude to one order of value lower. All of the treatment techniques for the RFV propagation show similar performance in reconstructing the secondary flow but their error values are lower than that of the RST propagation techniques. 
It can be concluded from Table~\ref{tab:SD_errors} that the explicit treatment of the RST (EXP-RST) results in the highest magnitudes of propagation error. As remedies, the frozen treatment of the RST and the propagation of the RFV (instead of RST) can remarkably reduce the error propagation; therefore, the frozen treatment of the RFV results in the lowest magnitudes of the propagation error.

\begin{table}
\caption{\label{tab:SD_errors} \hltOn The area-averaged error propagation for the comparison of all the 8 methods of propagation (listed in Fig.~\ref{fig:methods}) for the square duct case with $Re=3500$. \hltOff}
\begin{ruledtabular}
\begin{tabular}{lccc}
Method&$E_{1}$ ($\%$)&$E_{2}$ ($\%$)& $E_{3}$ ($\%$)\\
\hline
Standard $k-\epsilon$   & 6.845     & 100.000   & 100.000\\
EXP-RST         & 0.415     & 3.754     & 3.850\\
IMP-RST         & 0.416     & 3.404     & 3.501\\
FRZ-RST         & 0.141     & 0.741     & 0.803\\	
KCF-RST         & 0.106     & 0.638     & 0.656\\
EXP-RFV         & 0.014	    & 0.209	    & 0.221\\
IMP-RFV         & 0.015	    & 0.212	    & 0.220\\
FRZ-RFV         & 0.009	    & 0.199	    & 0.203\\
KCF-RFV         & 0.011	    & 0.234	    & 0.231\\

\end{tabular}
\end{ruledtabular}
\end{table}

\begin{figure*}
    \includegraphics[height=0.25\textwidth]{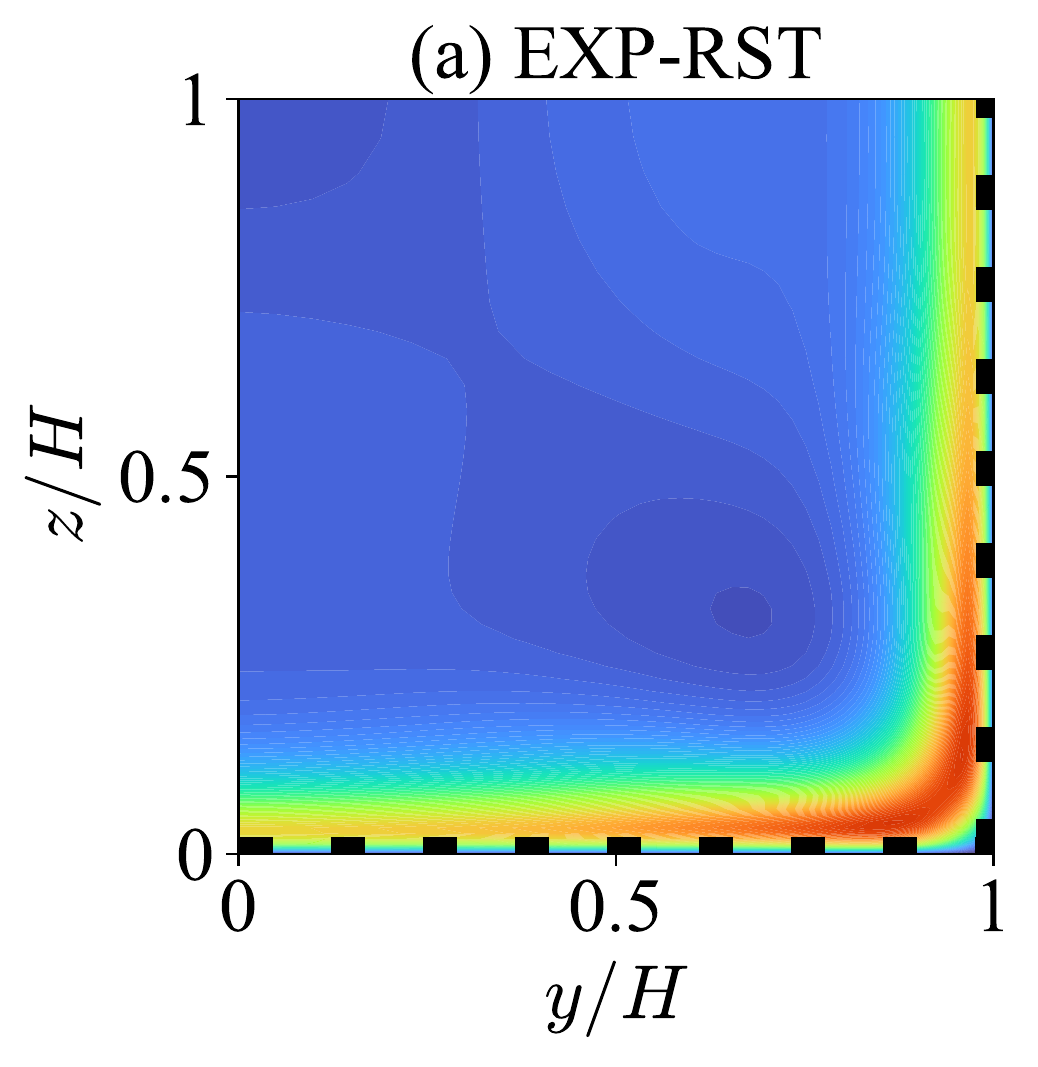}
    \includegraphics[height=0.25\textwidth]{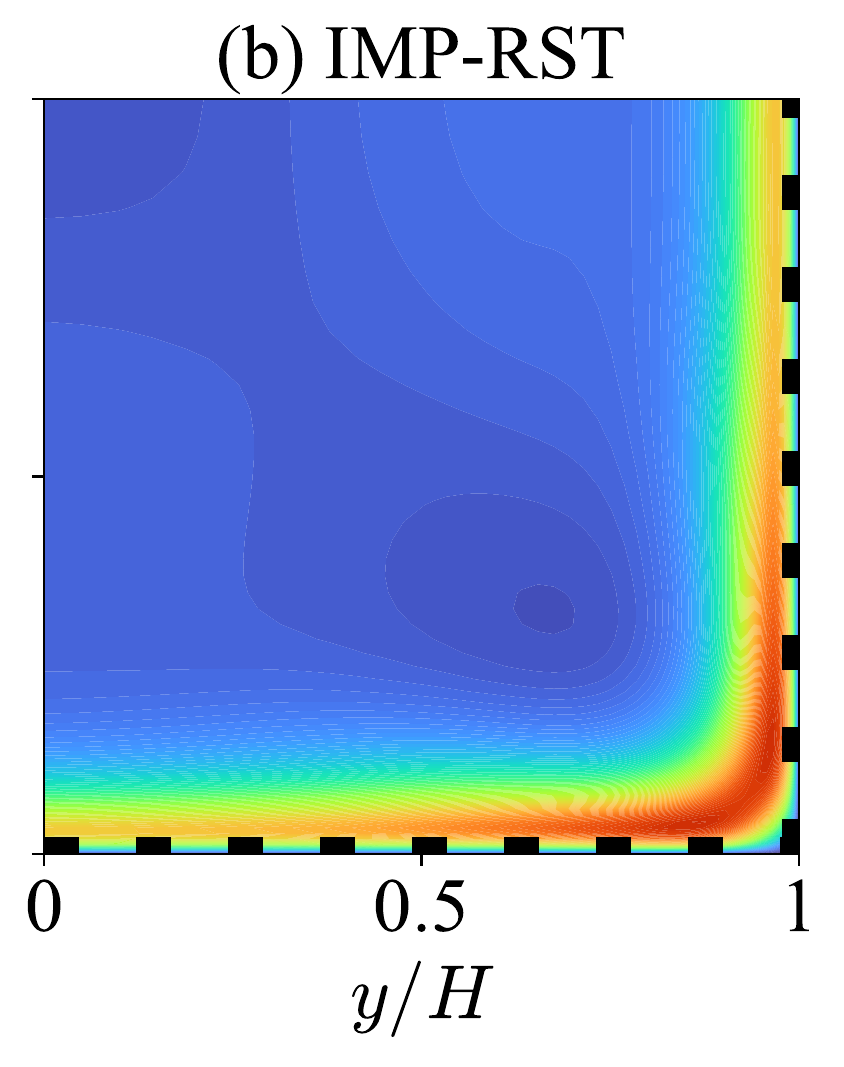}
    \includegraphics[height=0.25\textwidth]{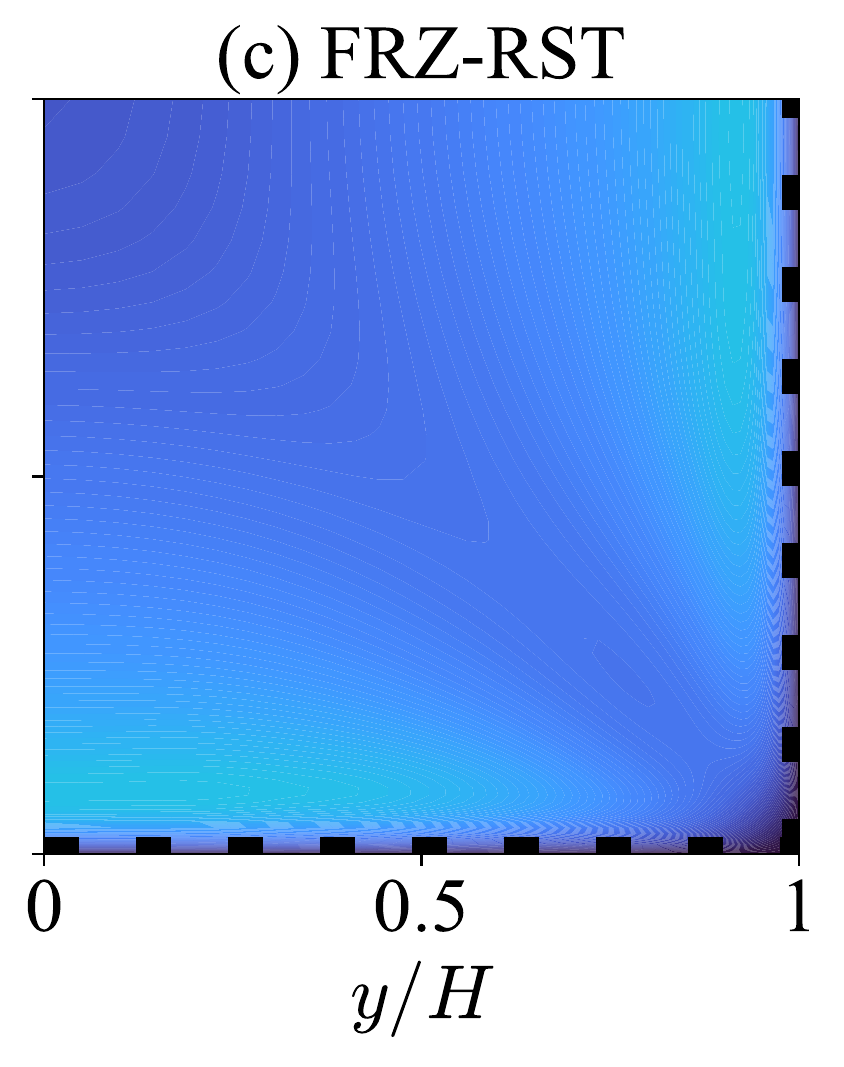}
    \includegraphics[height=0.25\textwidth]{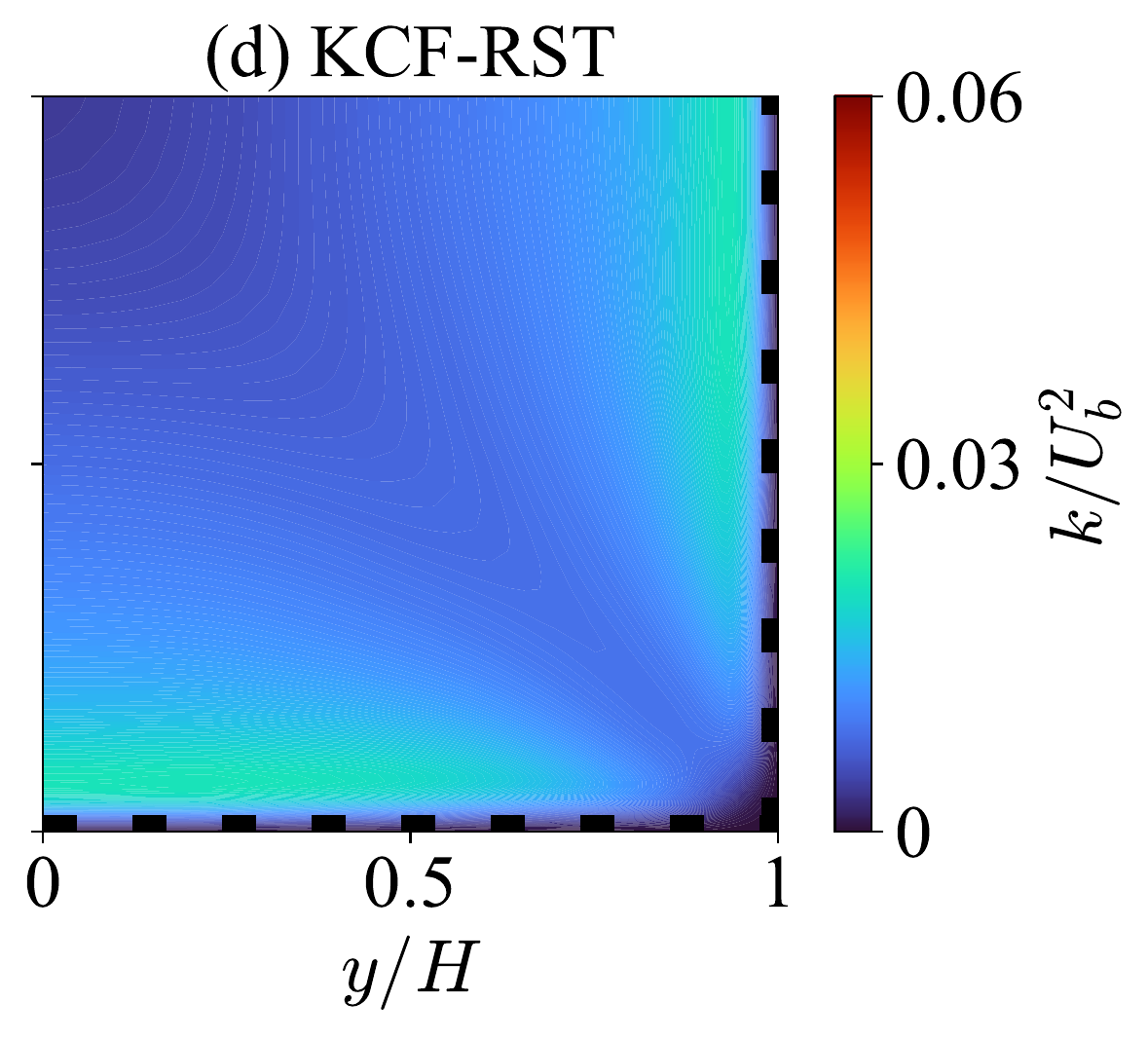}
\caption{\label{fig:SD_tke_rst} Contours of TKE after the propagation of the RST into RANS simulation of the case of square duct with $Re = 3500$: (a)EXP-RST, (b)IMP-RST, (c)FRZ-RST, (d)KCF-RST. The dashed line indicates the walls of the square duct.}%
    \includegraphics[height=0.25\textwidth]{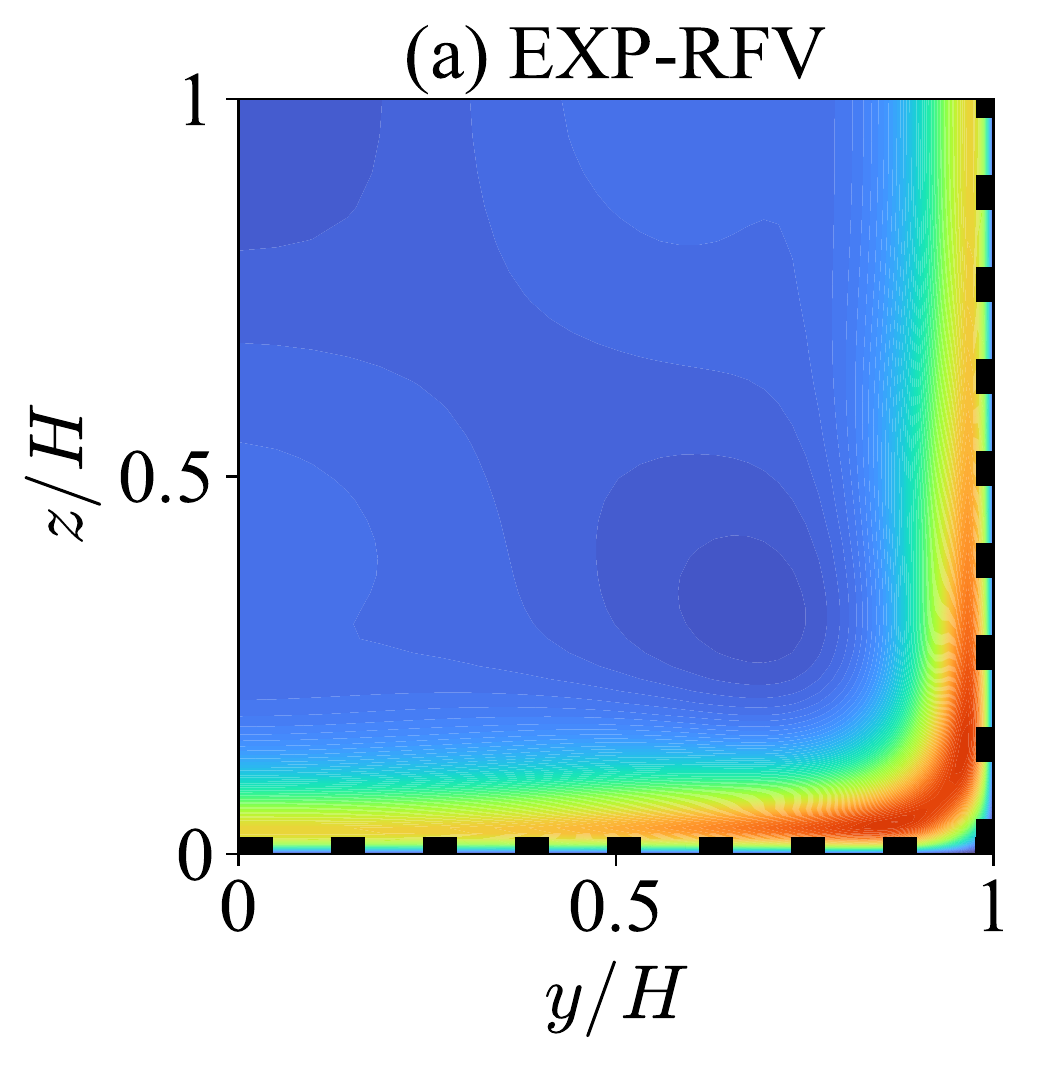}
    \includegraphics[height=0.25\textwidth]{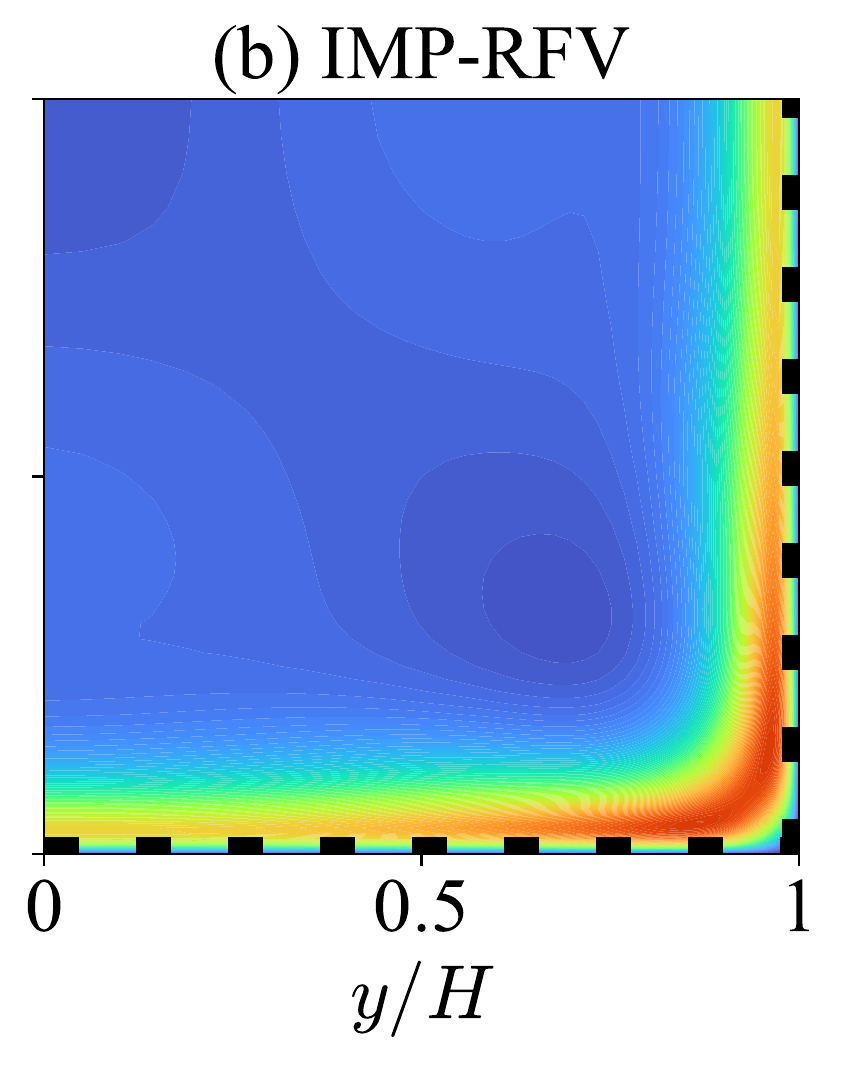}
    \includegraphics[height=0.25\textwidth]{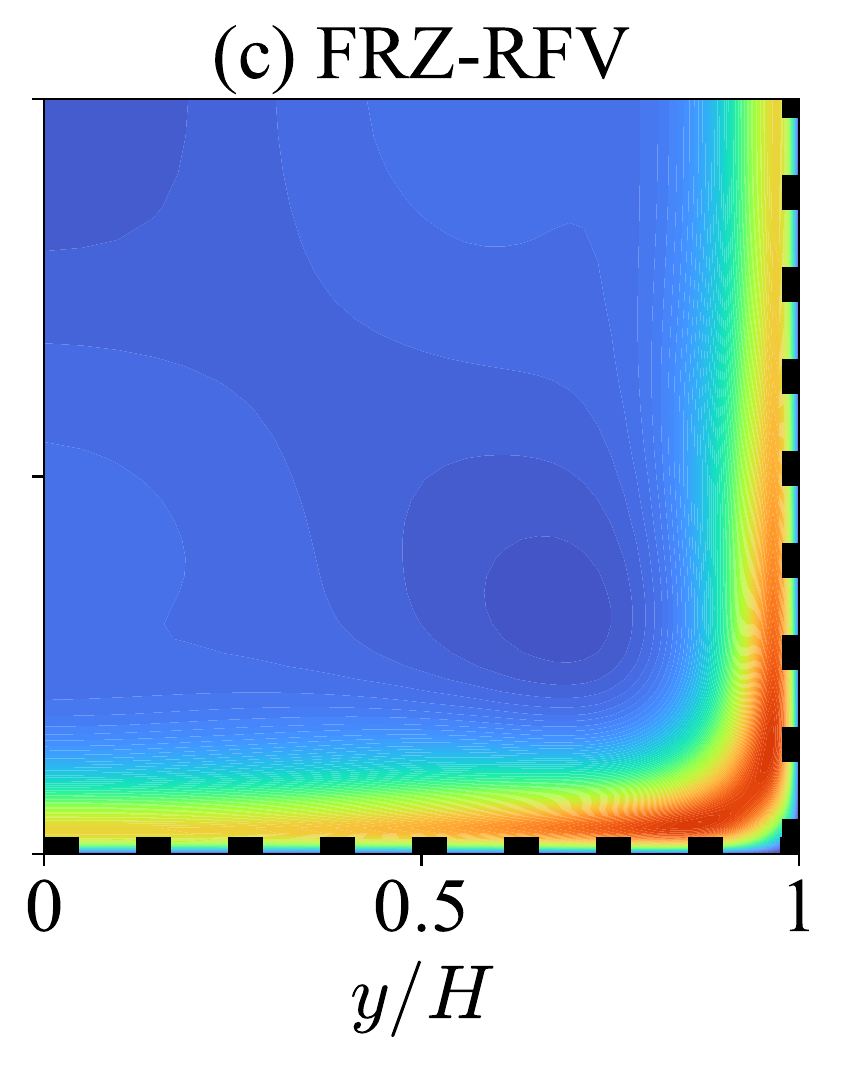}
    \includegraphics[height=0.25\textwidth]{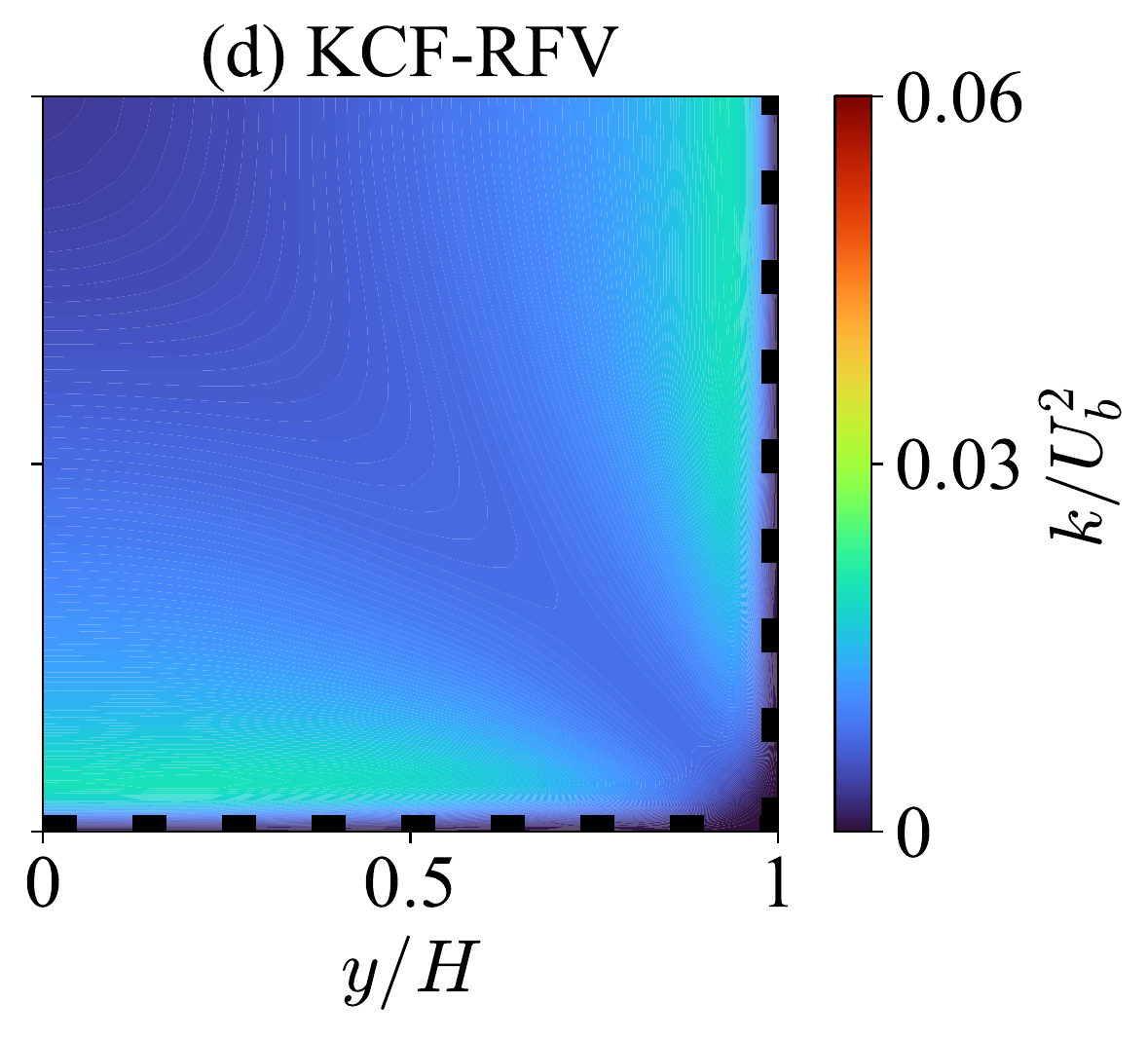}
\caption{\label{fig:SD_tke_rfv} Contours of TKE after the propagation of the RFV into RANS simulation of the case of square duct with $Re = 3500$: (a)EXP-RFV, (b)IMP-RFV, (c)FRZ-RFV, (d)KCF-RFV. The dashed line indicates the walls of the square duct.}%
\end{figure*}

Since the only difference between the frozen treatment and k-corrective frozen treatment is the extra correction term inside the baseline model for better modeling of the TKE, the values of the TKE after the propagation of the RST and the RFV are presented in Figs.~\ref{fig:SD_tke_rst} and~\ref{fig:SD_tke_rfv} respectively. Figs.~\ref{fig:SD_tke_rst}(a), \ref{fig:SD_tke_rst}(b), \ref{fig:SD_tke_rfv}(a), and \ref{fig:SD_tke_rfv}(b) show similar contours of the TKE because the standard $k-\epsilon$ is used for the calculation of the TKE over similar mean velocity domain. Fig.~\ref{fig:SD_tke_rst}(c) shows that including the $a_{ij}^{\Delta*}$ inside the $k-\epsilon$ can improve the prediction of the TKE within the FRZ-RST technique. Fig.~\ref{fig:SD_tke_rfv}(c) shows that the FRZ-RFV technique, similar to implicit and explicit treatments, overestimates the values of the TKE because there is no correction term inside the TKE equation. Figs.~\ref{fig:SD_tke_rst}(d) and ~\ref{fig:SD_tke_rfv}(d) show that including the extra correction term $R^*$ makes the KCF-RST and KCF-RFV methods capable of reproducing the TKE values similar to the high-fidelity data (Fig.~\ref{fig:SD_rolls}(c)). 

\subsection{\label{sec:RIresults} Roughness-induced secondary Flow}
\begin{figure*}
    \includegraphics[height=0.24\linewidth]{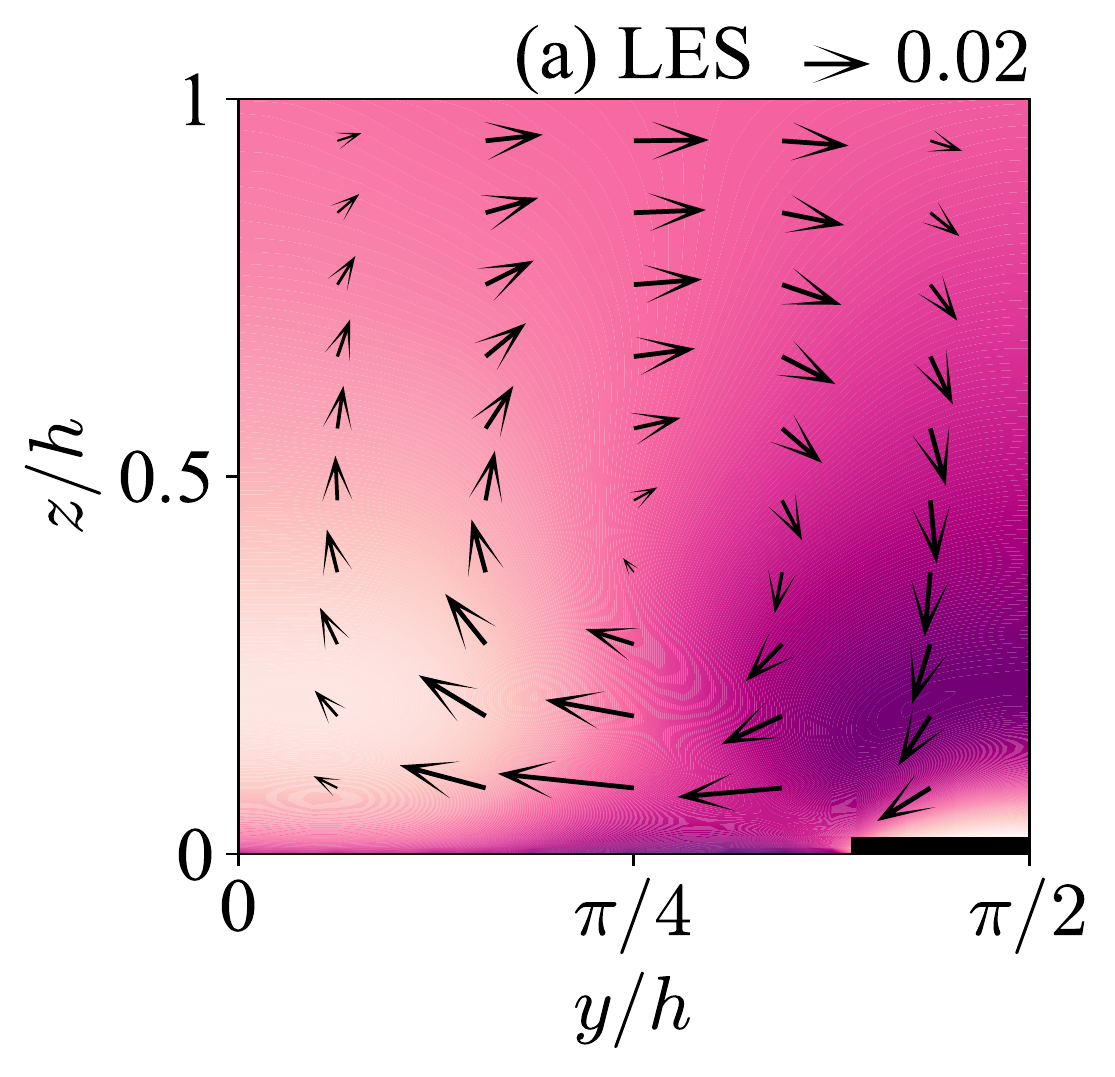}
    \includegraphics[height=0.24\linewidth]{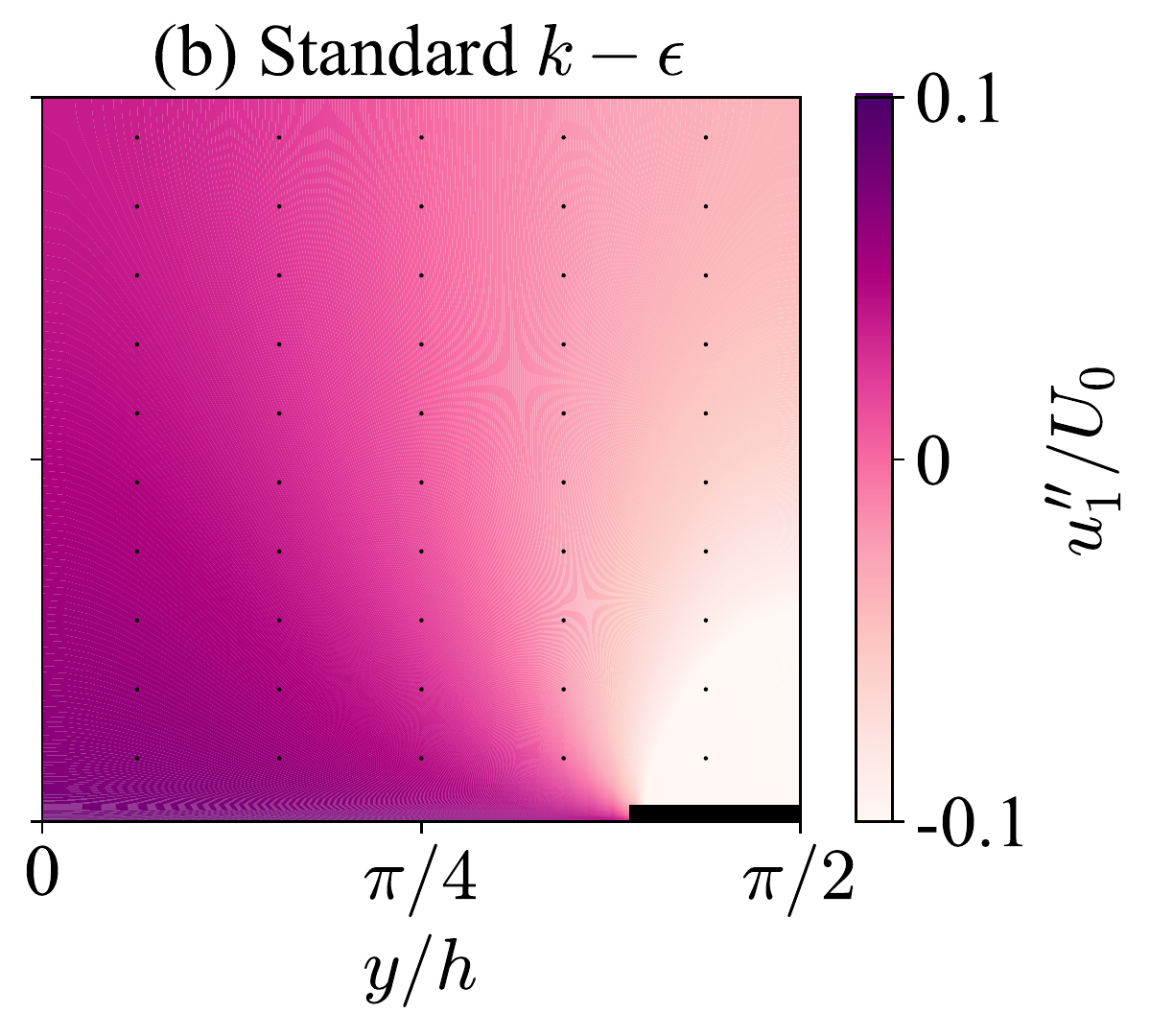}
    \includegraphics[height=0.24\textwidth]{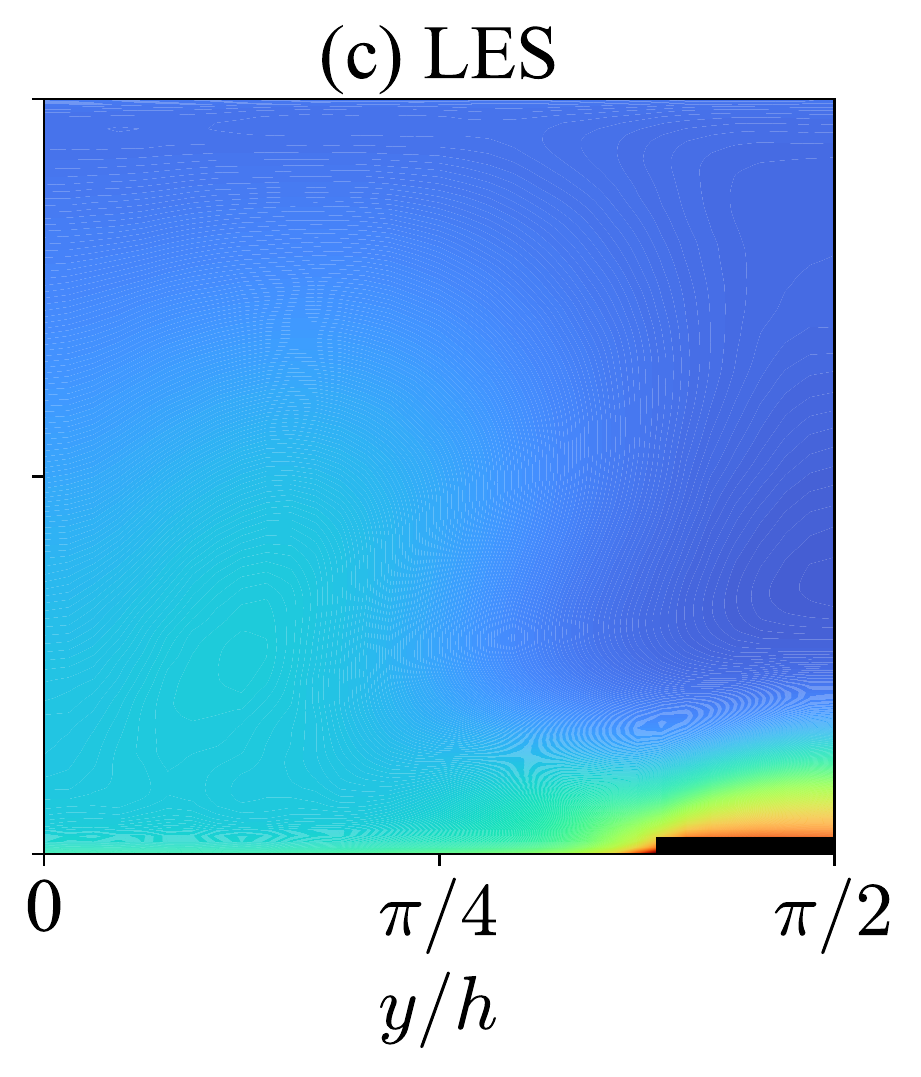}
    \includegraphics[height=0.24\textwidth]{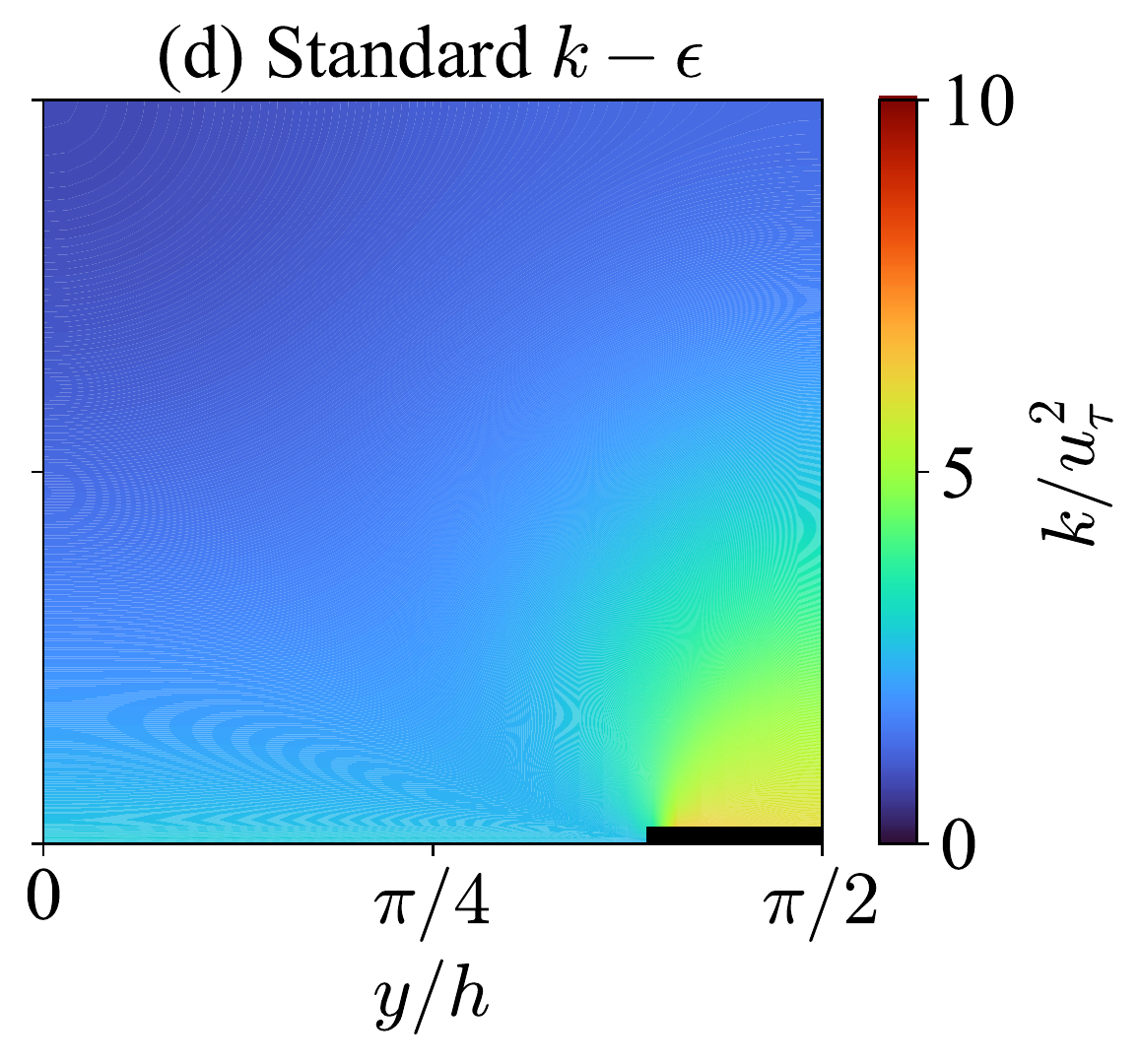}
\caption{\label{fig:RI_rolls_imp} Reference values for the case of roughness-induced secondary flow: (a,c) LES high-fidelity data and (b,d) RANS simulation with standard $k-\epsilon$. The contour plot shows: (a,b) the normalized streamwise dispersive velocity and (c,d) the TKE. The vectors indicate the in-plane velocity normalized by $U_0$. The black dots inside (b) indicate the absolute zero values for the in-plane velocity. The solid black line indicates the high-roughness patch.}
\end{figure*}

In this section, the results of the propagation of LES data for a nominally infinite Reynolds number case of roughness-induced secondary flow into the RANS simulation are presented. The mean velocity field of the LES is shown in Fig.~\ref{fig:RI_rolls_imp}(a). The contour plot shows the dispersive streamwise velocity normalized by $U_0$, which is the mean streamwise velocity horizontally averaged at $z/h=1$. The dispersive velocity is defined \hltOn as \cite{amarloo2022secondary} \hltOff 

\begin{equation}
\label{eq:dsprs}
    u''_i = \overline{u}_i - \langle \overline{u}_i \rangle,
\end{equation}
where $u''_i$ is the dispersive velocity, and the angle brackets indicate the averaged value in the $y$-direction. Fig.~\ref{fig:RI_rolls_imp}(b) shows the result of using RANS simulation with standard $k-\epsilon$ as a reference. As expected\cite{nikitin2021prandtl}, it shows that linear eddy-viscosity models cannot capture the secondary flows induced by roughness patches. \hltOn Similar to the square duct case (depicted in Fig.~\ref{fig:SD_rolls}(b) and (b)), this error is associated with the modelling of normal components of the Reynolds stress tensor. \hltOff Also, the values of the TKE obtained from the high-fidelity data (i.e., LES) and from the RANS simulation with standard $k-\epsilon$ are reported in Figs.~\ref{fig:RI_rolls_imp}(c) and~\ref{fig:RI_rolls_imp}(d) respectively.

All of the 8 propagation methods listed in Fig.~\ref{fig:methods} were applied for this case, and the explicit treatment of both the RST and the RFV (i.e. EXP-RST and EXP-RFV) failed in convergence. For a better understanding of the reason behind the failure, we should indicate that because of the nominally infinite Reynolds number, the molecular diffusivity is negligible compared to the turbulent diffusivity. Therefore, the existence of a turbulent diffusion term inside the equations (i.e., the implicit treatment of the Reynolds stress) is necessary for a case with a very high Reynolds number. The implicit treatments of both the RST and the RFV (i.e., IMP-RST and IMP-RFV) were converged but they resulted in a high amount of error propagation and failed in the reconstruction of the secondary flows. We hypothesize that the reason behind these high values of error is rooted in the calculation of the turbulent viscosity. In the implicit treatment, the RST will be split into two parts: a parallel tensor and an orthogonal tensor to the strain-rate tensor (Eq.~\ref{eq:RSTdivison}). The parallel tensor will be treated as a diffusion term but the orthogonal tensor will be treated explicitly. The division of the RST will add a degree of freedom in the strength of the turbulent viscosity which can lead to intense error propagation. In the frozen treatment method, as a remedy, the turbulent viscosity can be determined based on the high-fidelity velocity during the pre-propagation process to limit the magnitudes of both parallel and orthogonal tensors.

\begin{figure}
\includegraphics[width=0.45\textwidth]{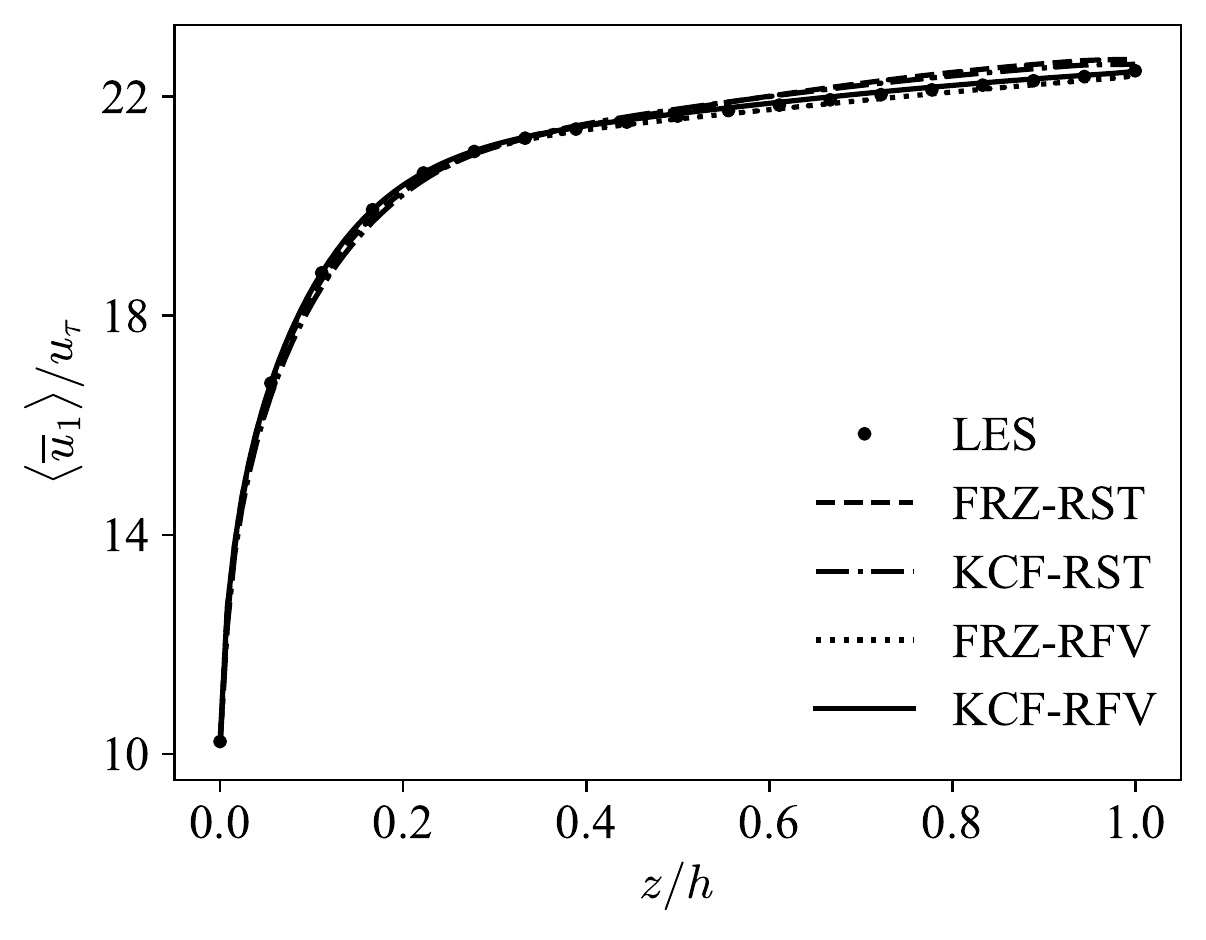}%
\caption{\label{fig:RI_uMean_frz} Normalized spanwise-averaged mean streamwise velocity for the case of roughness-induced secondary flow.}%
\end{figure}

\begin{figure*}
\includegraphics[height=0.24\linewidth]{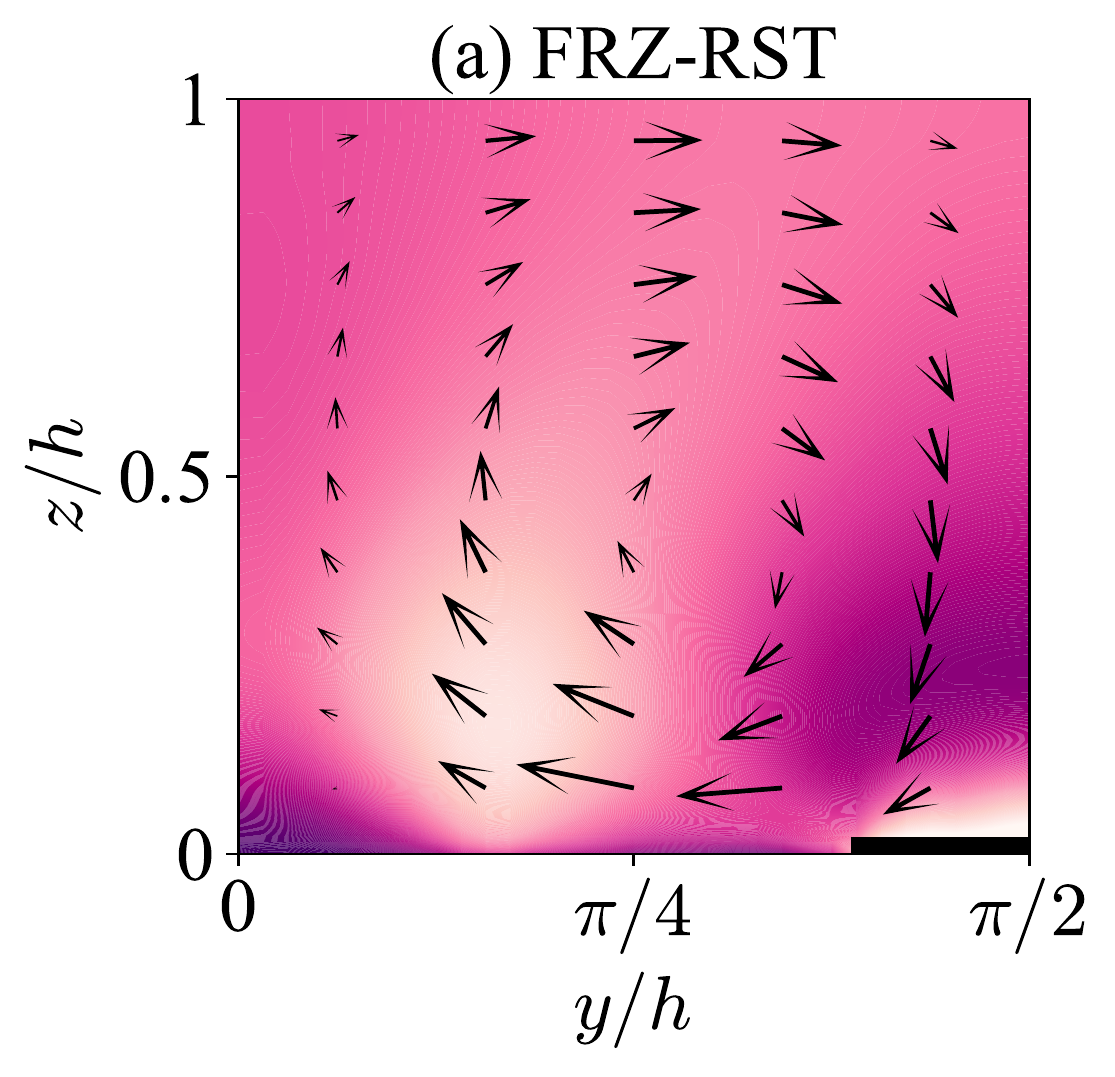}
\includegraphics[height=0.24\linewidth]{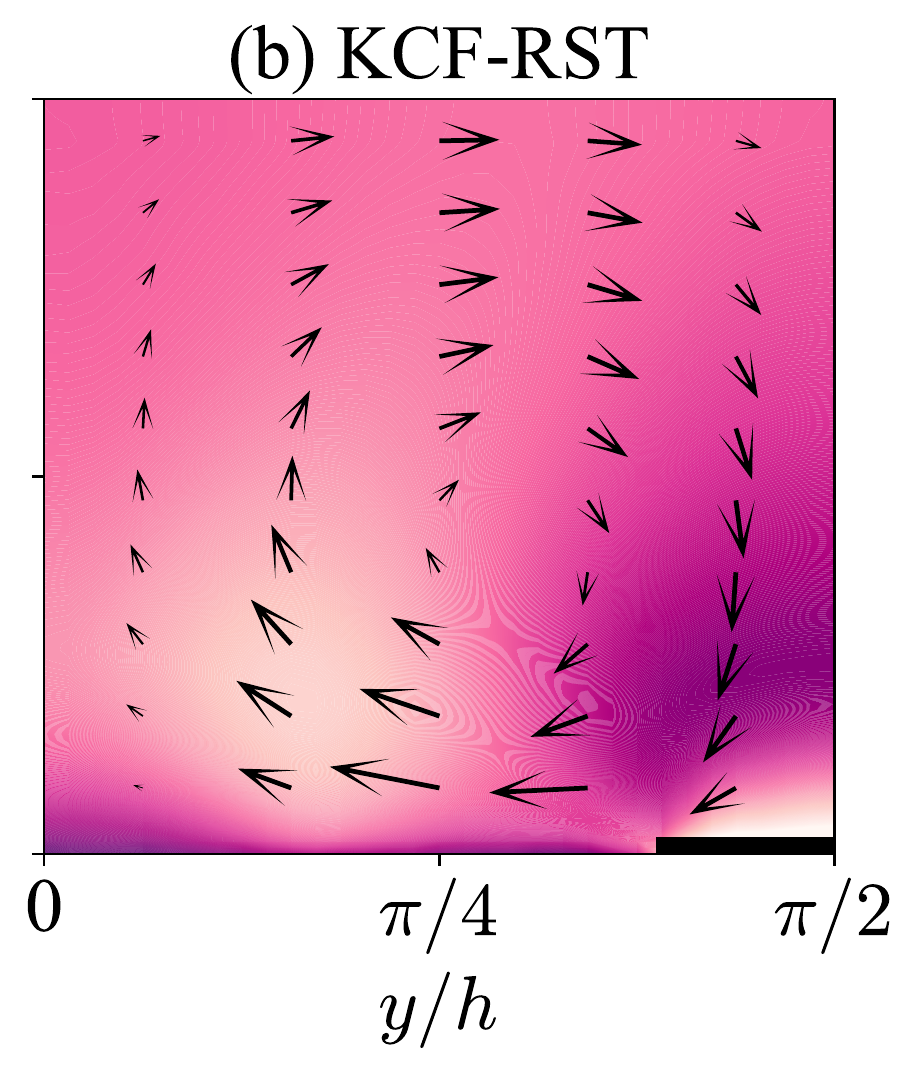}
\includegraphics[height=0.24\linewidth]{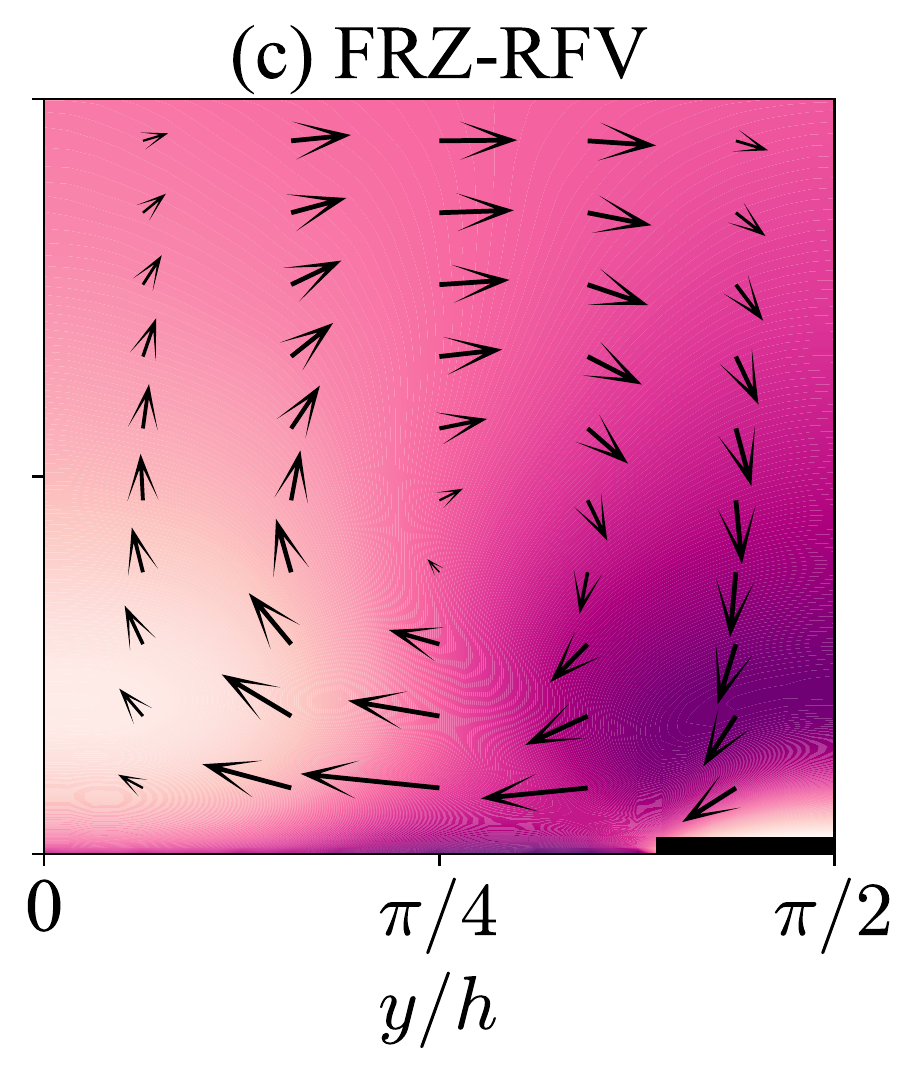}%
\includegraphics[height=0.24\linewidth]{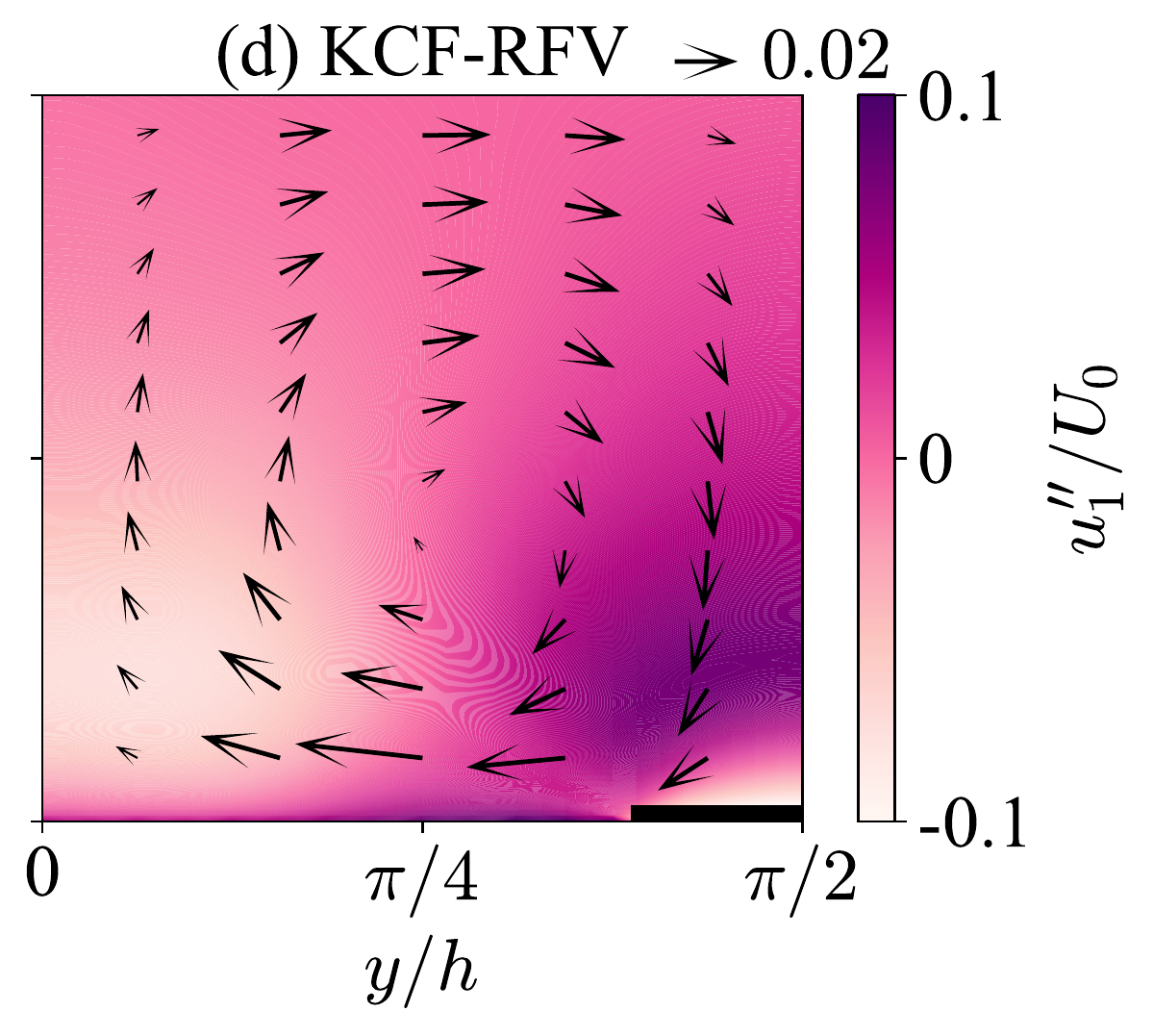}

\caption{\label{fig:RI_rolls_frz} Contour plots of the normalized streamwise dispersive velocity, and the vectors of the in-plane velocity normalized by $U_0$. The frozen treatment of the RST and RFV: (a) FRZ-RST, (b) KCF-RST, (c) FRZ-RFV, (d) KCF-RFV. The solid black line indicates the high-roughness patch.}%
\end{figure*}

Fig.~\ref{fig:RI_uMean_frz} shows the mean spanwise-averaged streamwise velocity after the propagation of data by the frozen treatments of both the RFV and RST. Fig.~\ref{fig:RI_uMean_frz} shows that the frozen treatments have a remarkable capability for the successful propagation of information from high-fidelity data sources in this case with very high Reynolds numbers.
For a better illustration of the roughness-induced secondary flows, Fig.~\ref{fig:RI_rolls_frz} shows the contours of normalized streamwise dispersive velocity and vectors of the in-plane motions by the frozen treatment of both the RST and the RFV. The comparison of Fig.~\ref{fig:RI_rolls_frz} with the high-fidelity data (Fig.~\ref{fig:RI_rolls_imp}(a)) shows that the frozen treatments of both the RST and the RFV are capable of reconstructing the secondary flows induced by roughness.  Also, the frozen treatments of the RFV (i.e. FRZ-RFV and KCF-RFV) show relatively better agreement in terms of mean dispersive streamwise velocity. 


\begin{figure*}
\includegraphics[height=0.25\textwidth]{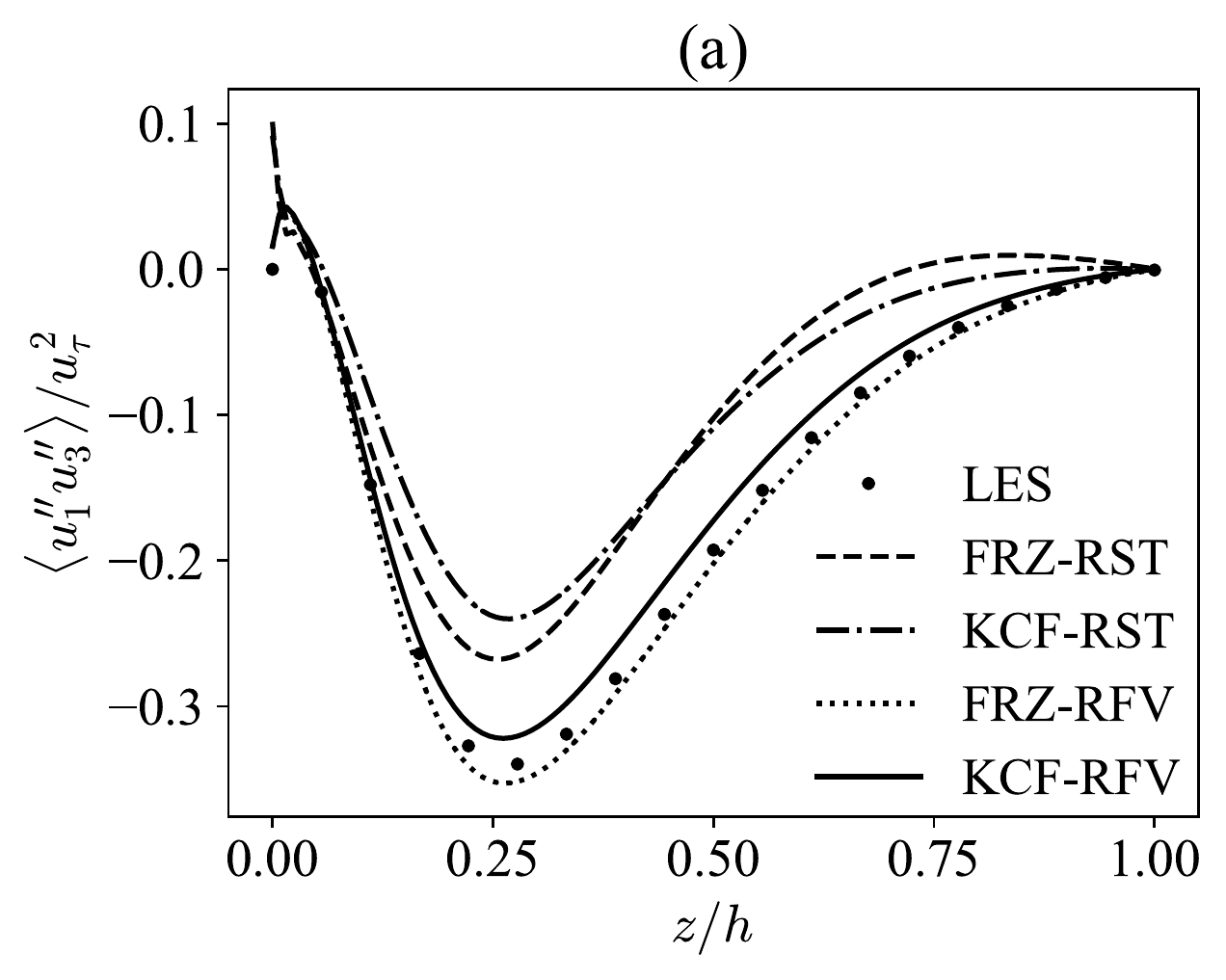}
\includegraphics[height=0.25\textwidth]{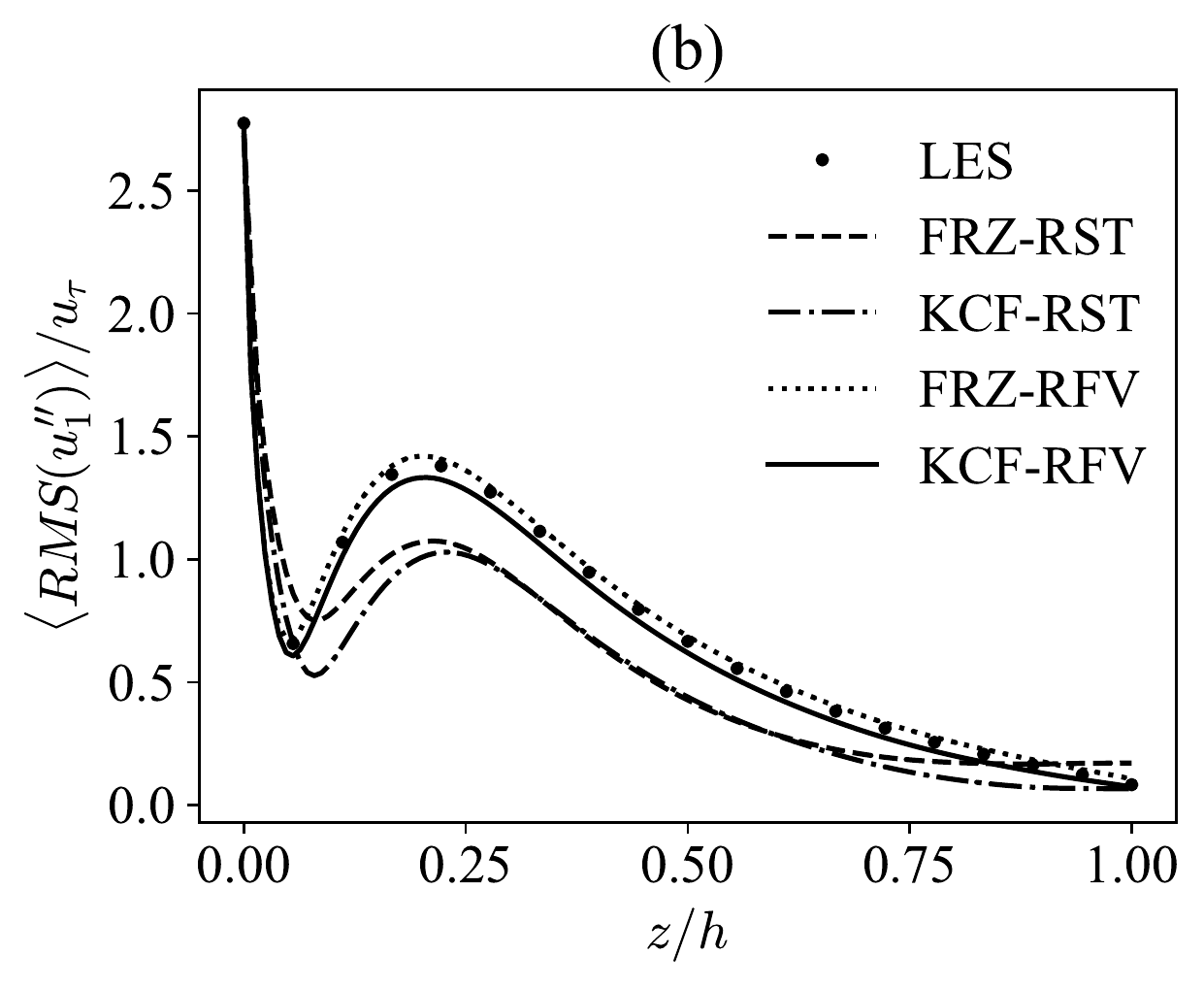}
\includegraphics[height=0.25\textwidth]{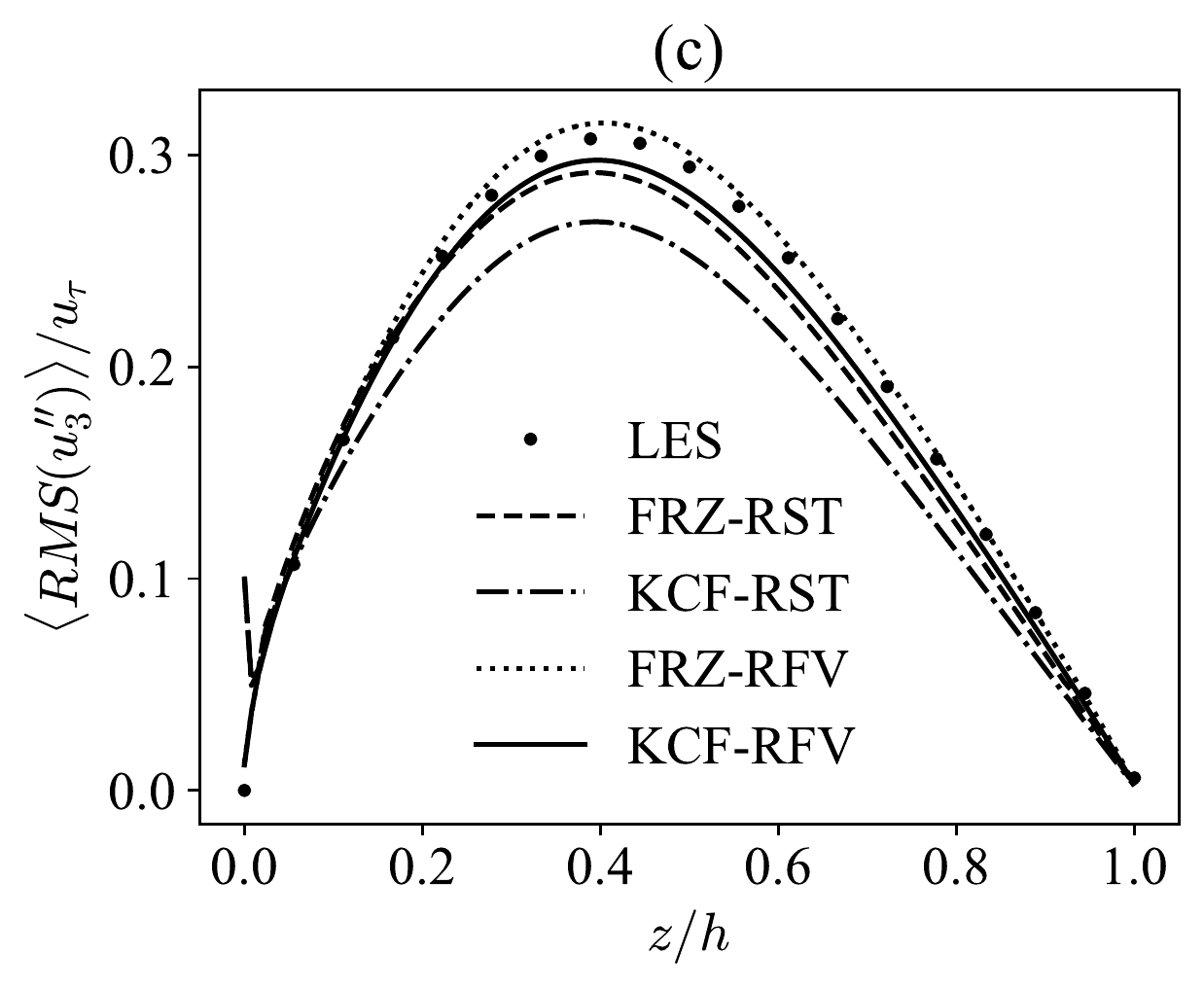}
\caption{\label{fig:RI_dsp_frz} Normalized spanwise-averaged (a) dispersive wall-normal momentum flux, (b) RMS of the dispersive streamwise velocity (c) RMS of the dispersive wall-normal velocity.}%
\end{figure*}

For a better comparison of the frozen methods on reconstructing of the secondary flows, Fig.~\ref{fig:RI_dsp_frz}(a) shows the spanwise-averaged dispersive momentum flux normalized by bottom-wall shear stress ($\tau_{wall} = u_{\tau}^2$), and Figs.~\ref{fig:RI_dsp_frz}(b) and \ref{fig:RI_dsp_frz}(c) show the root mean square (RMS) of streamwise and wall-normal dispersive velocity. Fig.~\ref{fig:RI_dsp_frz} can be used to evaluate the capability of the propagation methods in the reconstructing both shape and strength of the secondary flows, and it shows that the frozen treatments of the RFV have better results in this respect. 

For a more quantitative comparison of all the methods, the area-averaged errors (calculated by Eq.~(\ref{eq:E_error})) are reported in Table~\ref{tab:RO_errors}. The error values associated with the RANS simulation with the standard $k-\epsilon$ model are also reported, and the error values for in-plane velocity components are $100\%$ because the linear eddy-viscosity model completely fails in capturing the secondary flows as expected\cite{nikitin2021prandtl}. A comparison between Table~\ref{tab:SD_errors} and Table~\ref{tab:RO_errors} shows that the error values in reproducing the secondary flow are higher in the case of a very high Reynolds number. Also, the propagation of the RFV results in lower values of error than the propagation of the RST in both cases of low Reynolds number and high Reynolds number; therefore in both cases, the combination of the RFV propagation and the frozen treatment results in the lowest error propagation.   

\begin{table}
\caption{\label{tab:RO_errors} \hltOn The area-averaged error propagation for the comparison of the frozen propagation methods (listed in Fig.~\ref{fig:methods}) for the roughness-induced secondary flow. \hltOff}
\begin{ruledtabular}
\begin{tabular}{lccc}
Method&$E_{1}$ ($\%$)&$E_{2}$ ($\%$)& $E_{3}$ ($\%$)\\
\hline
Standard $k-\epsilon$   & 10.197     & 100.000   & 100.000\\
FRZ-RST         & 2.362      & 18.195    & 21.076\\
KCF-RST         & 1.846     & 19.165    & 22.096\\
FRZ-RFV         & 0.269     & 3.243     & 3.248\\
KCF-RFV         & 0.229     & 3.929     & 4.352\\

\end{tabular}
\end{ruledtabular}
\end{table}

\begin{figure*}
\includegraphics[height=0.24\textwidth]{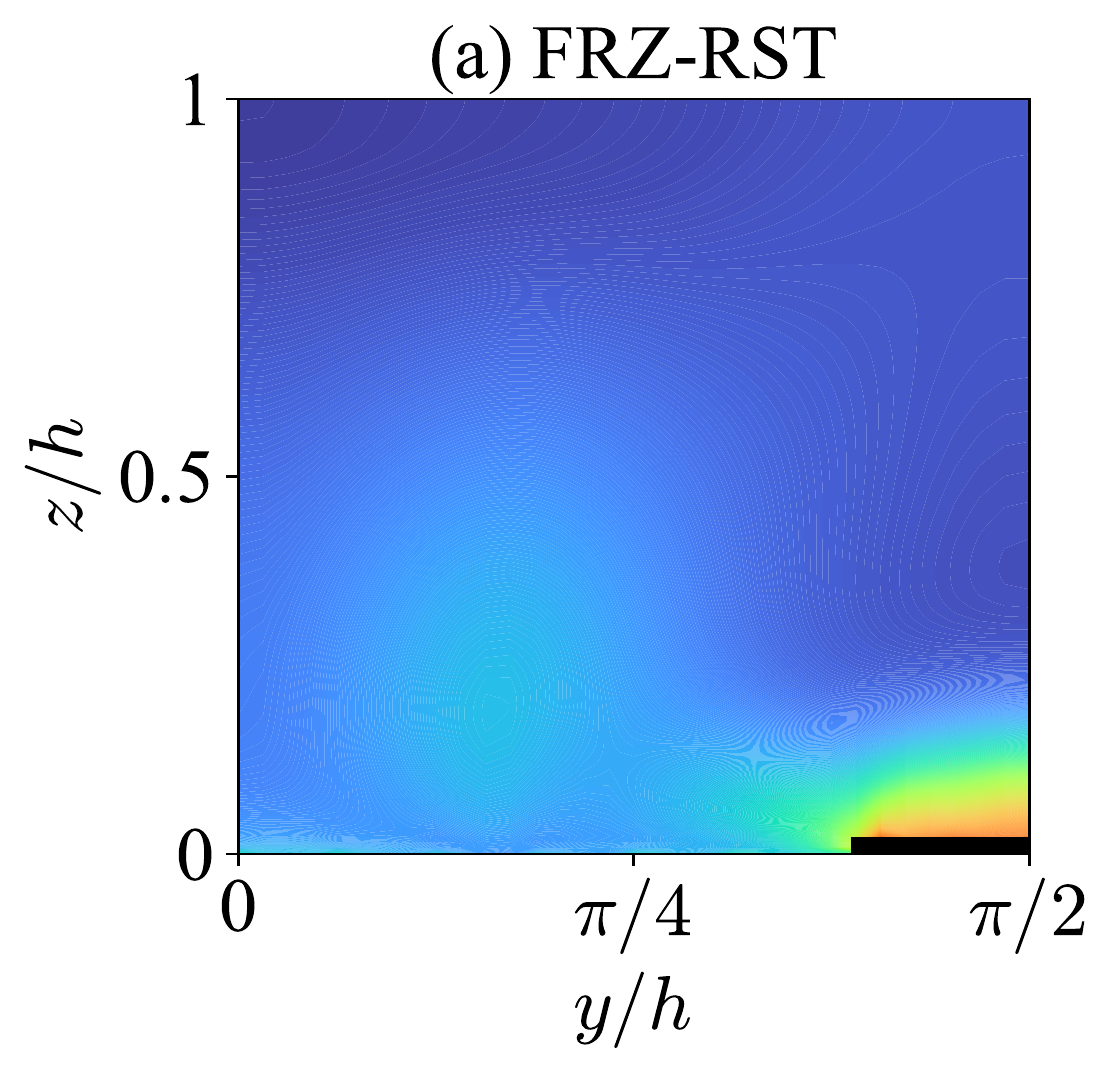}
\includegraphics[height=0.24\textwidth]{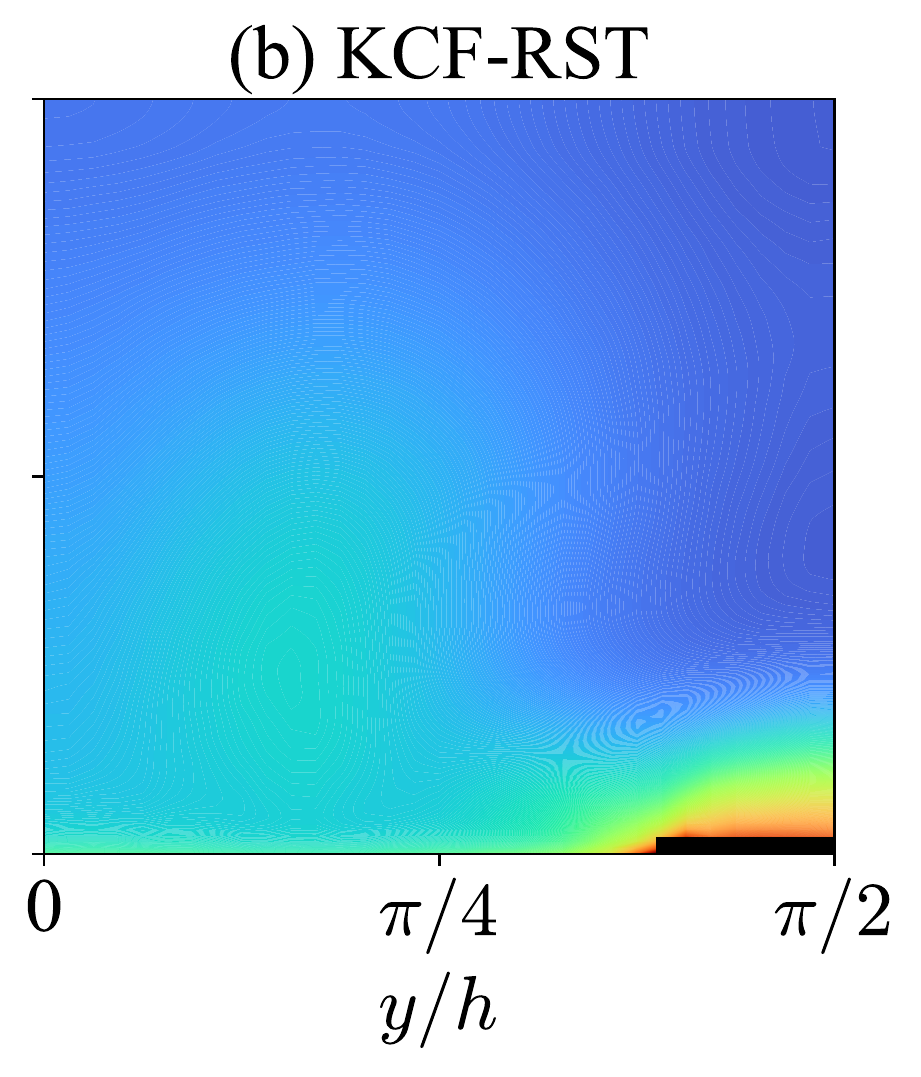}
\includegraphics[height=0.24\textwidth]{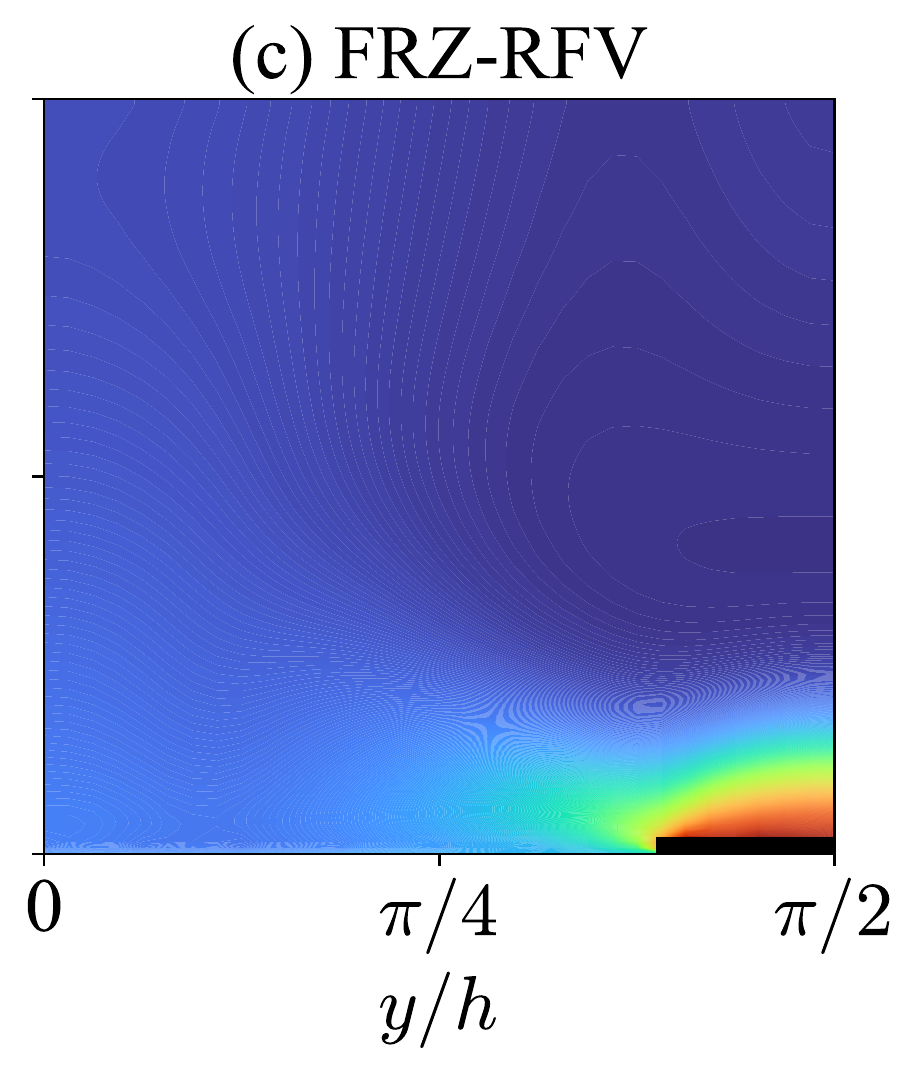}
\includegraphics[height=0.24\textwidth]{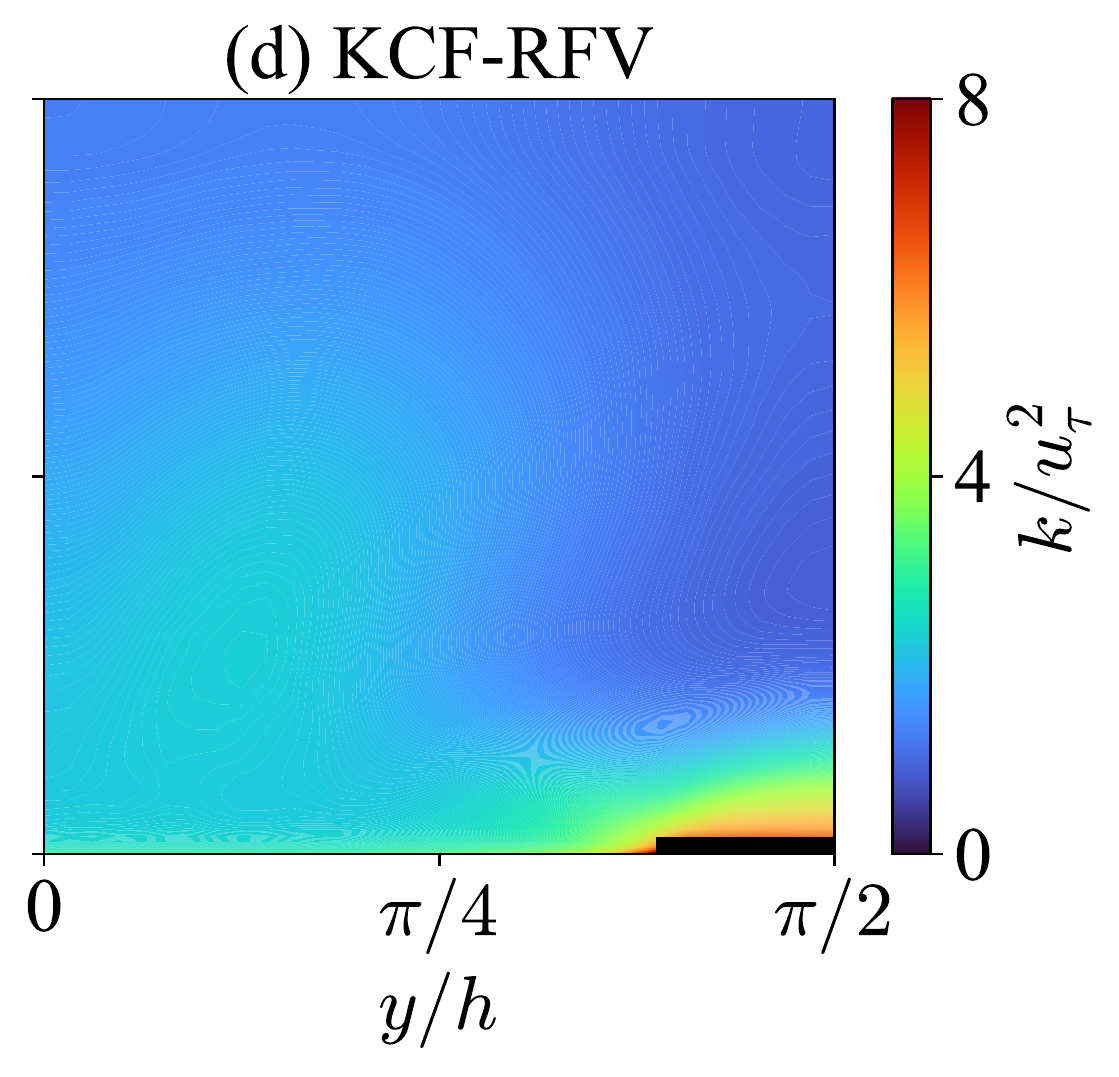}

\caption{\label{fig:RI_tke_frz} Contours of TKE for the case of roughness-induced secondary flow: (a) FRZ-RST, (b) KCF-RST, (c) FRZ-RFV, (d) KCF-RFV. The solid black line indicates the high-roughness patch.}%
\end{figure*}

The performance of the frozen treatments of both the RST and the RFV in modeling the TKE is compared in Fig.~\ref{fig:RI_tke_frz}. The k-corrective models (both KCF-RST and KCF-RFV) can reconstruct the TKE of the high-fidelity data (Fig.~\ref{fig:RI_rolls_imp}(c)). The frozen treatment of the RFV (FRZ-RFV) showed a remarkable performance in reproducing the velocity field (indicated in Table~\ref{tab:RO_errors}), but Fig.~\ref{fig:RI_tke_frz}(c) shows that it cannot reproduce the TKE similar to the LES data. Therefore, the KCF-RFV showed the best performance in the reconstruction of both the velocity field and the TKE field.

\subsection{\label{sec:BLresults} Sensitivity on the baseline model}
Since the frozen treatment of the RST and the RFV is showing lower errors in the propagation of information from the high-fidelity data, the sensitivity of this method on the baseline model will be evaluated here. It was mentioned before that predicting a proper value for the turbulent viscosity $\nu_t$ is necessary for successful propagation, and in the previous sections, the standard $k-\epsilon$ was used to find the $\nu_t$. Here, two other linear eddy-viscosity models are also tested to see how much the frozen treatments are sensitive to the baseline model. Tables~\ref{tab:SD_BL} and \ref{tab:RI_BL} show the effect of the baseline model on the area-averaged error of the velocity components for the square-duct and the roughness-induced secondary flows, respectively. They show that the propagation of the RFV results in less error propagation than the propagation of the RST regardless of the baseline model for both of the cases with low and high Reynolds numbers.

\begin{table}
\caption{\label{tab:SD_BL} \hltOn The area-averaged error propagation for the comparison of the baseline models in the case of the square-duct secondary flow. \hltOff}
\begin{ruledtabular}
\begin{tabular}{llccc}
Method &Baseline model &$E_{1}$ ($\%$)&$E_{2}$ ($\%$)& $E_{3}$ ($\%$)\\
\hline
\multirow{3}{*}{FRZ-RST}    &Standard $k-\epsilon$      & 0.141     & 0.741     & 0.803\\
                            &Realizable $k-\epsilon$    & 0.235     & 1.716     & 1.790\\
                            &$k-\omega$ SST             & 0.055     & 1.560     & 1.654\\
\multirow{3}{*}{FRZ-RFV}    &Standard $k-\epsilon$      & 0.009	    & 0.199	    & 0.203\\
                            &Realizable $k-\epsilon$    & 0.010     & 0.200     & 0.208\\
                            &$k-\omega$ SST             & 0.010     & 0.199     & 0.204\\
\end{tabular}
\end{ruledtabular}
\end{table}

\begin{table}
\caption{\label{tab:RI_BL} \hltOn The area-averaged error propagation for the comparison of the baseline models in the case of the roughness-induced secondary flow. \hltOff}
\begin{ruledtabular}
\begin{tabular}{llccc}
Method &Baseline model &$E_{1}$ ($\%$)&$E_{2}$ ($\%$)& $E_{3}$ ($\%$)\\
\hline
\multirow{3}{*}{FRZ-RST}    &Standard $k-\epsilon$      & 2.362     & 18.195    & 21.076\\
                            &Realizable $k-\epsilon$    & 1.500     & 11.919    & 12.780\\
                            &$k-\omega$ SST             & 2.752     & 16.759    & 18.242\\
\multirow{3}{*}{FRZ-RFV}    &Standard $k-\epsilon$      & 0.269     & 3.243     & 3.248\\
                            &Realizable $k-\epsilon$    & 0.211     & 2.500     & 2.652\\
                            &$k-\omega$ SST             & 0.461     & 5.656     & 5.296\\
\end{tabular}
\end{ruledtabular}
\end{table}
       
Table~\ref{tab:RI_BL} shows that the values of the $\nu_t$ modelled by the realizable $k-\epsilon$ end up in lower propagation error for the case of roughness-induced secondary flow. In the case of the square-duct secondary flow (Table~\ref{tab:SD_BL}), none of the baseline models show particularly better results. For both cases of low Reynolds number and high Reynolds number, a change of the baseline model changes the magnitude of the errors, but in the frozen treatment of the RFV for square duct flow, the error values are almost the same for all of the three baseline models. It can be inferred that the frozen treatment of the RFV is more robust to be used as a propagation technique regardless of the Reynolds number and the baseline model.

\section{\label{sec:Conclusion}Conclusions}
In this study, we compared a series of propagation methods for injecting the high-fidelity (i.e., DNS and LES) information into the RANS equations. Regarding the emerging field of data-driven RANS modeling, it is important that a reliable technique can be used for the propagation of data to the RANS equations. Since it has been established that a higher Reynolds number worsens the error propagation, we have chosen to include a case with a nominally infinite Reynolds number in our study. Two cases of Prandtl’s secondary flows of the second kind, which are known to not be captured in the RANS simulations with linear eddy-viscosity models\cite{nikitin2021prandtl}, were chosen to be studied: (1) the canonical case of the square duct with $Re = 3500$, and (2) the roughness-induced secondary flows with $Re_{\tau} = 1.3\times10^7$.

\hltOn \citet{wu2019reynolds} showed that the implicit treatment of RST has better conditioning for the RANS equation than the explicit treatment of RST. Here, we compared the frozen\cite{weatheritt2017development} and the k-corrective frozen\cite{schmelzer2020discovery} treatments of RST with the implicit and explicit treatments of RST. We showed that the frozen treatments have better performance and reduce the propagation error in comparison to the implicit method. \hltOff \hltOn In the case of square duct, the averaged error of the streamwise mean velocity was reduced from 0.416\% to 0.141\% by frozen treatment of RST instead of the implicit treatment of RST. \hltOff The k-corrective frozen method includes one extra term for the correction of the TKE equation which makes the RANS simulation capable of reproducing the TKE similar to the high-fidelity data. This conclusion was consistent in both cases at low Reynolds number (i.e., square-duct secondary flow) and high Reynolds number (i.e., roughness-induced secondary flow). 

\hltOn \citet{brener2021conditioning} showed that implicit treatment of RFV will result in less error propagation than the implicit treatment of RST. In this study, we showed that, regardless of treatment method, propagation of RFV will result in one order of magnitude lower errors compared to the propagation of RST in both cases at low and high Reynolds numbers. \hltOff \hltOn In the case of square duct, the averaged error of the streamwise mean velocity was reduced from 0.416\% to 0.015\% by implicit treatment of RFV instead of the implicit treatment of RST. \hltOff Therefore, we combined the strong points of the frozen technique and the RFV propagation, and we introduced two new propagation techniques: (1) the frozen treatment of the RFV, and (2) the k-corrective frozen treatment of the RFV. We showed that the frozen treatments of the RFV have less error propagation than the implicit treatment of the RFV in both cases of the square-duct and the roughness-induced secondary flows. \hltOn In the case of square duct, the averaged error of the streamwise mean velocity was reduced from 0.141\% to 0.009\% by frozen treatment of RFV instead of the frozen treatment of RST. In the case of roughness-induced secondary flow, the averaged error of the streamwise mean velocity was reduced from 2.362\% to 0.269\% by frozen treatment of RFV instead of the frozen treatment of RST \hltOff. The k-corrective frozen treatment of the RFV also can reproduce the TKE field similar to the high-fidelity data. 

Since the frozen treatments are using a baseline model to predict the turbulent viscosity, we used three different baseline models to evaluate the effect of the baseline model on the error propagation. It was found that, regardless of the baseline model, the frozen treatment of the RFV resulted in the least error propagation. In the case of roughness-induced secondary flow, the realizable $k-\epsilon$ showed a better performance but the same result was not seen for the case of the square-duct secondary flow.

In the case of very high Reynolds number, the explicit treatments of the RST and the RFV failed in convergence, the implicit treatment failed in reproducing the secondary flows correctly, and the frozen treatments showed a remarkable performance for reproducing the velocity fields; therefore, KCF-RFV technique has the potential to be used as a reliable propagation technique for cases with very high Reynolds number. Using KCF-RFV as a precursor, a well-trained data-driven technique that can predict three components of the RFV discrepancy and one correction term for the TKE equation can help RANS simulations to reproduce both the velocity field and the TKE field of the high-fidelity data.

Finally, the results showed that in the case of available high-fidelity data, the frozen propagation of RFV will result in lower errors than the other propagation methods. Therefore, a vector-basis prediction framework for the discrepancy of RFV has the potential to provide a proper correction for current eddy-viscosity models. This conclusion is valid in the case of developing a successful model for the prediction of RFV discrepancy which can be further studied in the future research.

\section*{Acknowledgment}
P.F. thanks the Aarhus University Research Foundation (AUFF) for the financial support. 
M.A. acknowledges the financial support from the Aarhus University Centre for Digitalisation, Big Data, and Data Analytics (DIGIT).

\section*{Conflict of interest}
The authors have no conflicts to disclose.

\section*{Data availability statement}
The data that support the findings of this study are available from the corresponding author upon reasonable request.

\bibliography{mainText.bib}

\end{document}